\documentclass[%
 reprint,
 amsmath,amssymb,
 aps,
 longbibliography,
pre,
]{revtex4-2}

\usepackage[caption=false]{subfig}
\usepackage{graphicx}
\usepackage{dcolumn}
\usepackage{bm}
\usepackage{xcolor}
\usepackage{comment}
\usepackage {hyperref}
\hypersetup{
    colorlinks = true,
    linkcolor = blue,
    anchorcolor = blue,
    citecolor = blue,
    filecolor = blue,
    urlcolor = blue
    }
\DeclareMathAlphabet{\mathpzc}{OT1}{pzc}{m}{it}
\DeclareMathOperator\erfc{erfc}
\begin{document}

\preprint{APS/123-QED}

\title{Understanding Degeneracy of Two-Point Correlation Functions via Debye Random Media}

\author{Murray Skolnick}
\affiliation{
    Department of Chemistry, Princeton University, Princeton, New Jersey 08544, USA
}%

\author{Salvatore Torquato}
\email{torquato@electron.princeton.edu}
\affiliation{
    Department of Chemistry, Department of Physics, Princeton Institute for the Science and Technology of Materials, and Program in Applied and Computational Mathematics, Princeton University, Princeton New Jersey 08544, USA
}%

\date{\today}

\begin{abstract}

It is well-known that the degeneracy of two-phase microstructures with the same volume fraction and two-point correlation function $S_2(\mathbf{r})$ is generally infinite. To elucidate the degeneracy problem explicitly, we examine \textit{Debye random media}, which are entirely defined by a purely exponentially decaying two-point correlation function $S_2(r)$. In this work, we consider three different classes of Debye random media. First, we generate the ``most probable" class using the Yeong-Torquato construction algorithm [Yeong and Torquato, \href{https://journals.aps.org/pre/abstract/10.1103/PhysRevE.57.495}{Phys. Rev. E, 57, 495 (1998)}]. A second class of Debye random media is obtained by demonstrating that the corresponding two-point correlation functions are effectively realized in the first three space dimensions by certain models of overlapping, polydisperse spheres. A third class is obtained by using the Yeong-Torquato algorithm to construct Debye random media that are constrained to have an unusual prescribed pore-size probability density function. We structurally discriminate these three classes of Debye random media from one another by ascertaining their other statistical descriptors, including the pore-size, surface correlation, chord-length probability density, and lineal-path functions. We also compare and contrast the percolation thresholds as well as the diffusion and fluid transport properties of these degenerate Debye random media. We find that these three classes of Debye random media are generally distinguished by the aforementioned descriptors and their microstructures are also visually distinct from one another. Our work further confirms the well-known fact that scattering information is insufficient to determine the effective physical properties of two-phase media. Additionally, our findings demonstrate the importance of the other two-point descriptors considered here in the design of materials with a spectrum of physical properties.

\end{abstract}

\maketitle


\section{\label{sec:intro}Introduction}

Two-phase disordered heterogeneous media in $d$-dimensional Euclidean space $\mathbb{R}^d$ are ubiquitous; examples include composites, porous media, polymer blends, colloids, complex fluids, and biological media \cite{Torq02,milt02,sahi03,zohd16,hris20,ashb99,yeomans98,ding16} among other synthetic and natural materials. Such two-phase media exhibit a rich range of complex structures that have varying degrees of disorder and intricate material properties \cite{torq97,zohd06,bros07}. 

To fully characterize the microstructure of a two-phase medium as well as its effective physical properties, an infinite set of $n$-point correlation functions are required in the infinite-volume limit \cite{Torq02}. A variety of different types of such correlation functions arise in rigorous theories that depend on the bulk physical property of interest \cite{Torq02}. For example, there is the standard $n$-point correlation function $S^{(i)}_n(\mathbf{x}_1,...,\mathbf{x}_n)$ which gives the probability that the position vectors $\mathbf{x}_1,...\mathbf{x}_n$ all lie in phase $i$ where $i=1,2$ for two-phase media (see Sec. \ref{sec:definitions} for details) \cite{Torq02,torq82}. Given that it is generally impossible to obtain the information contained in such an infinite set of correlation functions, their lower-order versions are often used as a starting point to characterize the structure and physical properties of a two-phase medium. 

For statistically homogeneous media, the one-point function is simply the volume fraction of the phase of interest, e.g., $S_1(\mathbf{x}_1)=\phi$, and hence position-independent. The two-point function $S_2(\mathbf{x}_1,\mathbf{x}_2)$, which is readily obtained from scattering experiments \cite{Torq02,deb57}, encodes information about pair separations, and depends only on the relative displacement ${\bf r}={\bf x}_2-{\bf x}_1$ for homogeneous media. The three-point function $S_3(\mathbf{x}_1,\mathbf{x}_2,\mathbf{x}_3)$ contains information about how these pair separations are assembled into triangles.

While $S_2$ contains important structural information, prior work has established that microstructures with a specific $S_1$ and $S_2$ are highly degenerate \cite{jiao10,jiaoPNAS09,gommesPRE,gommesPRL,jiao07}. Furthermore, the set of $S_1$- and $S_2$-degenerate microstructures is infinitely large in the thermodynamic limit. This degeneracy implies that the other microstructural descriptors of these two-phase systems will generally differ. There is a variety of descriptors that incorporate higher-order information that one could consider to differentiate $S_2$-degenerate microstructures \cite{Torq02}. At first glance, a natural higher-order function to include beyond $S_1$ and $S_2$ is the three-point function $S_3$. However, Jiao, Stillinger, and Torquato revealed that $S_3$ does not appreciably increase information content over pair statistics in systems that lack long-range order \cite{jiaoPNAS09}. 

\begin{figure}[t]
	\subfloat[\label{fig:manifolda}]{\includegraphics[width=0.45\linewidth]{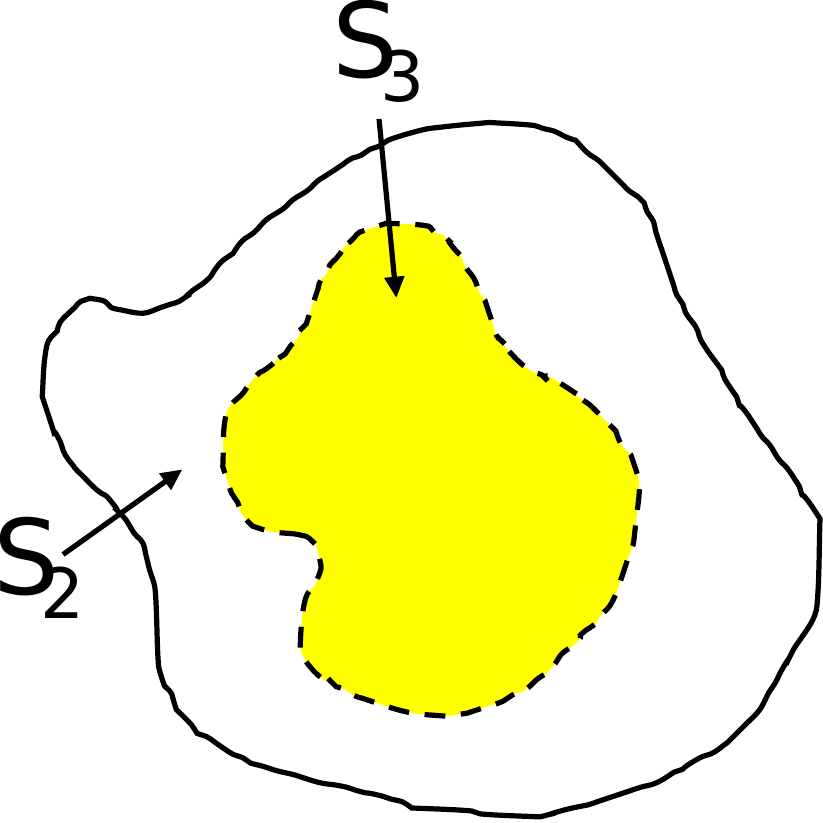}}~
	\subfloat[\label{fig:manifoldb}]{\includegraphics[width=0.45\linewidth]{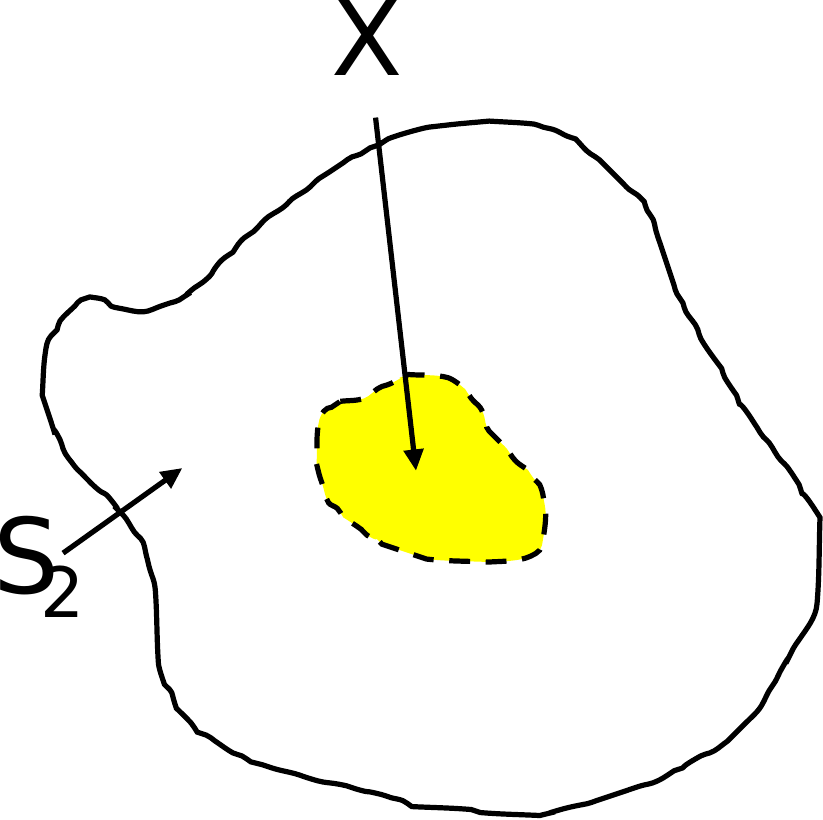}}
	\caption{The set of all microstructures associated with a particular $S_2$ is schematically shown as the region enclosed by the solid contour in (a) and (b). The shaded region in (a) shows the set of all microstructures associated with the same $S_2$ and $S_3$. The shaded and more restrictive region in (b) shows the set of all microstructures associated with the same $S_2$ and a superior set of two-point descriptors, $X$, which has a higher information content than $S_3$ does. This figure is adapted from Fig. 5 in Ref. \cite{jiaoPNAS09}.}
	\label{fig:manifold}
\end{figure}

In contrast, one can fruitfully increase information content by also incorporating superior two-point topological descriptors, such as the two-point cluster function $C_2(\mathbf{r})$ \cite{C2note,Beasley88,jiaoPNAS09}. It has been established that other two-point descriptors, which can be easier to compute than three-point statistics, also encode important higher-order nontrivial microstructural information \cite{yeong98,jiaoPNAS09,ma18}. Examples of such two-point quantities include the lineal-path function $L(z)$ and related chord-length probability density function $p(z)$ \cite{lu92}, the pore-size function $P(\delta)$ \cite{prag63}, surface-void correlation function $F_{sv}({\bf r})$, and the surface-surface correlation function $F_{ss}({\bf r})$  \cite{doi76,rube88,rube89} (see Sec. \ref{sec:definitions} for definitions). Figure \ref{fig:manifold} illustrates these ideas by schematically showing the relative sizes of the degenerate microstructures when $S_2$ and $S_3$ are used versus when $S_2$ and a set of superior two-point functions, $X$, are used.

So-called \textit{Debye random media} \cite{yeong98} are unique models of statistically isotropic and homogeneous two-phase media in that they are defined entirely by the two-point correlation function $S_2(r)$, namely,
\begin{equation}
    S_2^{(i)}(r) = \phi_i(1-\phi_i)e^{-r/a} + \phi_i^2\label{eqn:debS2First},
\end{equation}
where $r=|{\bf r}|$ is a radial distance, and $a$ is a positive constant that represents a characteristic length scale of the medium. Debye \textit{et al.} \cite{deb57} proposed the exponentially decaying two-point correlation function [Eq. \eqref{eqn:debS2First}] as a model of media with phases of ``fully random shape, size, and distribution." It is noteworthy that Debye random media are a good approximation of certain realistic two-phase media \cite{deb57}, including Fontainebleau sandstones \cite{dunsmuir96}. Given the aforementioned degeneracy associated with the same $S_1$ and $S_2$, there should exist a multitude of different \textit{classes} of Debye random media that are distinguished by other microstructural descriptors. Thus, such two-phase media provide a singular opportunity to study the degeneracy of a two-point correlation function.

In this paper, we examine three such classes of Debye random media. First, we consider Debye random media realized using the Yeong-Torquato stochastic (re)construction procedure \cite{yeong98} (see Sec. \ref{sec:YT} for details). These ``most probable" realizations of Debye random media, which we refer to as Yeong-Torquato Debye random media (YT-DRM), have been studied by Yeong and Torquato \cite{yeong98} and Ma and Torquato \cite{drm2020}. We obtain the second class of structures by demonstrating that certain systems of overlapping, polydisperse spheres with exponentially distributed radii effectively realize Debye random media in the first three space dimensions, i.e., for $d=1,2,$ and $3$ \cite{polydisperse91,lu92II,perm92}. Henceforth, we refer to this class of structures as overlapping-polydisperse-spheres Debye random media (OPS-DRM). To realize the third class, we use the Yeong-Torquato procedure to construct Debye random media constrained to have an unusual pore-size function $P(\delta)$ that has compact support (see Secs. \ref{sec:cpdrm} and \ref{sec:comp_porefuncs} for details). As such, we refer to this class as compact-pores Debye random media (CP-DRM).

We structurally discriminate these three classes of Debye random media from one another using various descriptors to characterize how the microstructures and physical properties of $S_2$-degenerate systems can vary. We determine $F_{sv}(r)$, $F_{ss}(r)$, $P(\delta)$, $L(z)$, and $p(z)$ for OPS-DRM analytically using the canonical correlation function formalism \cite{torquato86} (see Sec. \ref{sec:canonical}), for CP-DRM via empirical and semi analytical means, and subsequently compare these descriptors to their analogues for YT-DRM that were determined by Ma and Torquato \cite{drm2020}. Additionally, we compare and contrast the percolation and phase inversion symmetry properties of these three classes, both of which provide stringent tests for comparison (see Sec. \ref{sec:npoint} for definitions). Lastly, we treat these structures as porous media and compute bounds on their mean survival times, principal diffusion relaxation times, as well as bounds on and approximations of their fluid permeabilities. Our analysis considers these systems in 2D and in 3D for certain cases. Overall, we find that these degenerate Debye random media are generally differentiated by these descriptors to varying degrees.

The paper is organized as follows: in Sec. \ref{sec:definitions}, we provide definitions of and compare the microstructural descriptors used in this paper. In Sec. \ref{sec:YT}, we review the Yeong-Torquato (re)construction procedure. In Sec. \ref{sec:s2pds}, we derive the two-point correlation function for OPS-DRM in 1D, 2D, and 3D.  In Sec. \ref{sec:realization}, we demonstrate that our OPS systems are excellent models of Debye random media and possess effective phase inversion symmetry. In Sec. \ref{sec:cpdrm}, we describe CP-DRM. In Sec. \ref{sec:comparison}, we compare various two-point microstructural descriptors of YT-DRM, OPS-DRM, and CP-DRM in 2D and 3D. In Sec. \ref{sec:percolation}, we compare the percolation thresholds of these three classes of structures in 2D. In Sec. \ref{sec:properties}, we compare their diffusion properties in 2D and 3D as well as their fluid transport properties in 3D. In Sec. \ref{sec:conclusions}, we give concluding remarks and discuss possible future directions of research.

\section{\label{sec:definitions}Definitions of Microstructural Descriptors}

In this section, we briefly describe several microstructural descriptors that have been used to characterize two-phase random media and are particularly germane to the present study. To supplement this discussion, we briefly summarize the canonical correlation function formalism for overlapping monodisperse spheres to elucidate the nontrivial information contained in the various two-point descriptors described below.

\subsection{\label{sec:npoint}$n$-point correlation function}

A two-phase random medium is generally a domain of space $\mathcal{V}\subseteq\mathbb{R}^d$ that is partitioned into two disjoint regions: a region of phase 1, $\mathcal{V}_1$, and volume fraction $\phi_1$ as well as a region of phase 2, $\mathcal{V}_2$, of volume fraction $\phi_2$ \cite{Torq02}. The phase indicator function $\mathcal{I}^{(i)}(\mathbf{x})$ for a two-phase medium is defined as
\begin{equation}
    \mathcal{I}^{(i)}(\mathbf{x}) = \begin{cases}
        1, & \mathbf{x} \in \mathcal{V}_i,\\
        0, & \mathbf{x}_i \notin \mathcal{V}_i.
    \end{cases}
\end{equation}

The $n$-point correlation function $S_n^{(i)}$ for phase $i$ is defined as \cite{Torq02}
\begin{equation}
    S_n^{(i)}(\mathbf{x}_1,\mathbf{x}_2,...,\mathbf{x}_n) = \left\langle \prod_{i=1}^n \mathcal{I}^{(i)}(\mathbf{x}_i) \right\rangle,\label{eqn:defSn}
\end{equation}
where the angular brackets denote an ensemble average. The quantity $S_n^{(i)}(\mathbf{x}_1,\mathbf{x}_2,...,\mathbf{x}_n)$ can be interpreted as the probability of finding the ends of all vectors $\mathbf{x}_1,...,\mathbf{x}_n$ in phase $i$. Using relation \eqref{eqn:defSn}, the volume fraction of phase $i$ is the one-point correlation function
\begin{equation}
    S_1^{(i)} = \langle \mathcal{I}^{(i)}(\mathbf{x}) \rangle,
\end{equation}
which is equal to the volume fraction of phase $i$, $\phi_i$, for statistically homogeneous media. Similarly, the two-point correlation function is written as
\begin{equation}
    S_2^{(i)}(\mathbf{x}_1,\mathbf{x}_2) = \langle \mathcal{I}^{(i)}(\mathbf{x}_1)\mathcal{I}^{(i)}(\mathbf{x}_2) \rangle.
\end{equation}

A two-phase medium has phase-inversion symmetry if the morphology of phase 1 at volume fraction $\phi_1$ is statistically identical to that of phase 2 in the system where the volume fraction of phase 1 is $1-\phi_1$ \cite{Torq02}:
\begin{equation}
    S_n^{(1)}(\mathbf{x}^n;\phi_1,\phi_2)=S_n^{(2)}(\mathbf{x}^n;\phi_2,\phi_1), \textrm{    } n\geq2\label{eqn:phaseinv}.
\end{equation}
A notable property of such phase-inversion symmetric random media is that for $\phi_1=\phi_2=1/2$ it is possible to determine the odd-order probability functions $S^{(i)}_{2m+1}$ from $S^{(i)}_{2m},S^{(i)}_{2m-1},...,S^{(i)}_1$.

For statistically homogeneous systems, the two-point function depends only on the displacement vector $\mathbf{r}\equiv\mathbf{x}_2-\mathbf{x}_1$ and simplifies to $S_2(\mathbf{x}_1,\mathbf{x}_2)=S_2(\mathbf{r})$. If the medium is also statistically isotropic, the two-point function depends only on the magnitude of the displacement vector, simplifying as $S_2(\mathbf{r})=S_2(r)$. The two-point function $S_2^{(i)}(r)$ is related to the autocovariance function $\chi_{_V}(r)$ by subtracting its large-$r$ limit:
\begin{equation}
    \chi_{_V}(r) \equiv S_2^{(1)}(r) - \phi_1^2 = S_2^{(2)}(r) - \phi_2^2.
\end{equation}

Note the limits of the autocovariance function
\begin{equation}
    \lim_{r\to0}\chi_{_V}(r)=\phi_1\phi_2,\lim_{r\to\infty}\chi_{_V}(r)=0,
\end{equation}
where the later limit holds for systems that lack long-range order. Another important quantity is the spectral density which is the Fourier transform of the autocovariance function
\begin{equation}
    \tilde{\chi}_{_V}(\mathbf{k}) =\int \chi_{_V}(\mathbf{r}) e^{i\mathbf{k}\cdot\mathbf{r}} d\mathbf{r}\label{eqn:specDens}.
\end{equation}
The spectral density can be obtained from scattering experiments \cite{teub90,deb57}.

Debye and coworkers \cite{deb57} showed that the derivative of the two-point correlation function at the origin is proportional to the specific surface $s$ for 3D isotropic media. This property has been generalized to anisotropic media \cite{berry87} as well as $d$-dimensional media \cite{Torq02}, which is written as
\begin{equation}
    \frac{dS_2^{(i)}}{dr}\Big\rvert_{r=0} = -\frac{\omega_{d-1}}{\omega_d d}s\label{eqn:specsurf},
\end{equation}
where
\begin{equation}
    \omega_d = \frac{\pi^{d/2}}{\Gamma(1+d/2)}
\end{equation}
is the volume of a $d$-dimensional sphere of unit radius and $\Gamma(x)$ is the gamma function. For the first three spatial dimensions, the derivative in Eq. \eqref{eqn:specsurf} is $-s/2$, $-s/\pi$ and $-s/4$ which we employ in subsequent sections.

\subsection{\label{sec:surffuncs}Surface correlation functions}

Some important, but less well-known, descriptors are the two-point surface correlation functions which arise in rigorous bounds on transport properties of porous media \cite{Torq02,doi76}. The interface indicator function is defined as \cite{Torq02}

\begin{equation}
    \mathcal{M}(\mathbf{x})=|\nabla\mathcal{I}^{(1)}(\mathbf{x})|=|\nabla\mathcal{I}^{(2)}(\mathbf{x})|.
\end{equation}
The specific surface is the expected area of the interface per unit volume. For homogeneous media, $s$ is the ensemble average of the surface indicator function:
\begin{equation}
    s=\langle \mathcal{M}(\mathbf{x}) \rangle.
\end{equation}

The surface-void correlation function $F_{sv}(\mathbf{r})$ measures the correlation between one point on the interface and the other in the void phase. For homogeneous systems, it is defined as
\begin{equation}
    F_{sv}(\mathbf{r}) = \langle \mathcal{M}(\mathbf{x})\mathcal{I}^{(\textrm{void})}(\mathbf{x}+\mathbf{r}) \rangle.
\end{equation}
Henceforth, we will take phase 1 to be the void (matrix) phase and phase 2 to be the solid (inclusion) phase. For systems lacking long-range order, the surface-void correlation function has the large-$r$ limit
\begin{equation}
    \lim_{r\to\infty} F_{sv}(r) = s\phi_1.
\end{equation}
Ma and Torquato have shown that the derivative of $F_{sv}(r)$ can be related to the Euler characteristic $\chi$, a measure of phase connectivity, by the relation \cite{ma18}
\begin{equation}
    \frac{dF_{sv}(r)}{dr} \Big|_{r=0} = \frac{\chi}{V}\label{eqn:euler}.
\end{equation}
The right-hand side of relation \eqref{eqn:euler} can be interpreted as an intensive property or specific Euler characteristic. 

One may also measure the correlation of points on the phase interface using the surface-surface correlation function $F_{ss}(\mathbf{r})$. For homogeneous media, it is defined as
\begin{equation}
    F_{ss}(\mathbf{r}) = \langle \mathcal{M}(\mathbf{x})\mathcal{M}(\mathbf{x}+\mathbf{r})\rangle.
\end{equation}
It has been shown that $F_{ss}(r)$ diverges for small $r$ as $(d-1)\omega_{d-1}s/d\omega_d r$ \cite{ma18}. In the large-$r$ limit, we have
\begin{equation}
    \lim_{r\to\infty} F_{ss}(r) = s^2
\end{equation}
for systems with no long-range order.

\subsection{\label{sec:porefuncs}Pore-size function}

An important characterization of the pore (void) space is with the pore-size probability density function $P(\delta)$, which is defined by \cite{Torq02}
\begin{equation}
    P(\delta) = -\partial F(\delta)/\partial \delta\label{eqn:defFd},
\end{equation}
where $F(\delta)$ is the complementary cumulative distribution function that measures the probability that a randomly placed sphere of radius $\delta$ centered in the pore space $\mathcal{V}_1$ lies entirely in $\mathcal{V}_1$. We have that $F(0)=1$ and $F(\infty)=0$, and it immediately follows that $P(0)=s/\phi_1$ and $P(\infty)=0$. The $n$th moment of the pore-size probability density function is \cite{Torq02}
\begin{equation}
    \langle \delta^n \rangle \equiv \int_0^{\infty} \delta^n P(\delta) d\delta\label{eqn:mps}.
\end{equation}

These moments act as a measure of the characteristic length scale of the pore space and have been shown to be useful in the prediction of transport properties of random media \cite{prag61,avel91}. The first moment, the mean pore size $\langle\delta\rangle$, as well as the second moment $\langle\delta^2\rangle$ are of particular interest to us in this work.

\subsection{\label{sec:lpfunc}Lineal-path function}

An additional descriptor that we consider in this work is the lineal-path function $L^{(i)}(z)$ \cite{lu92}. The lineal-path function $L^{(i)}(z)$ is the probability that a line segment of length $z$ lies entirely in phase $i$. Thus, $L^{(i)}(z)$ contains degenerate connectedness information along a path in phase $i$. Naturally, it is a monotonically decreasing function with $L^{(i)}(0) = \phi_i$ and $L^{(i)}(z\to\infty)=0$.

\subsection{\label{sec:clpdf}Chord-length probability density function}

The chord-length density probability density function $p^{(i)}(z)$ is another descriptor and is related to the lineal-path function \cite{lu93,math75}. In this context, the chords are the line segments between intersections of an infinitely long line with the two-phase interface. For statistically isotropic media, $p^{(i)}(z)dz$ is the probability of finding a chord with length between $z$ and $z+dz$ in phase $i$. The chord-length density function often arises in the study of transport properties of porous media \cite{ho79,toku85,thom87}.

One can show that $p^{(i)}(z)$ is directly related to the second derivative of the lineal-path function $L^{(i)}(z)$ \cite{lu93},
\begin{equation}
    p^{(i)}(z) = \frac{\ell_C^{(i)}}{\phi_i} \frac{d^2L^{(i)}(z)}{dz^2}\label{eqn:pzdef}.
\end{equation}
Here, $\ell_C^{(i)}$ is the mean chord length for phase $i$ and thus the first moment of the chord-length probability density function.

\subsection{\label{sec:canonical}The Canonical Correlation Function $H_n$}

The canonical $n$-point correlation function $H_n$ developed by Torquato \cite{torquato86} provides a unified means to derive explicit closed-form expressions of any specific correlation function for various particle and cellular models of two-phase random media. This canonical function enables one to relate and compare the microstructural information contained in one descriptor to that of any other. For concreteness, we specialize the discussion of the $H_n$ for overlapping, $d$-dimensional, radius $R$ monodisperse spheres (phase 2) embedded in a matrix (phase 1).

The central idea employed by Torquato \cite{torquato86} to define and derive $H_n$ was to consider the space and surface that is available to a spherical ``test" particle that is inserted into the system. Following this principle, he derived
\begin{eqnarray}
    H_n(\mathbf{x}^m;\mathbf{x}^{p-m};\mathbf{r}^q)=(-1)^m \frac{\partial}{\partial a_1}...\frac{\partial}{\partial a_m}\Big\{ \rho^q\prod_{l=1}^q\prod_{k=1}^p \nonumber \\
    &&\hspace{-8cm} \Theta(|\mathbf{x}_k-\mathbf{r}_l| - a_k)\exp\left[-\rho v_p(\mathbf{x}^p;a_1,...,a_p) \right] \Big\}\label{eqn:Hn}.
\end{eqnarray}
Here, $H_n$ gives the probability of inserting $m$ test particles of radius $b=a-R$ whose centers $\mathbf{x}^m$ fall on the phase interface, inserting $p-m$ test particles of radius $b$ whose centers $\mathbf{x}^{p-m}$ fall in phase 1, and that the centers of any $q$ inclusions are given by $\mathbf{r}^q$. The function $v_p(\mathbf{x}^p;a_1,...,a_p)$ is the union volume of $p$, $d$-dimensional spheres of radii $a_1,..,a_p$, and $\rho$ is the number density. Also note the definition of the Heaviside step function
\begin{equation}
    \Theta(x) = \begin{cases}
    0, & x < 0, \\
    1, & x \geq 0.
    \end{cases}
\end{equation}

From here, one can use specific limits of Eq. \eqref{eqn:Hn} to derive key microstructural descriptors. All descriptors considered in this paper amount to placing different combinations of $p-m$ test particles into the matrix phase and $m$ test particles onto the phase interface, while placing no restriction on the centers of the spherical inclusions (i.e., $q=0$). For example, the $n$-point correlation function is derived using the following limit:
\begin{equation}
    S_n(\mathbf{x}^n) = \lim_{a_i\to R, \forall i} H_n(\emptyset;\mathbf{x}^n;\emptyset),
\end{equation} 
which clearly involves $p-m=n$ phase 1 test point-particles and $m=0$ interface test point-particles. From this expression, we can write the two-point correlation function as
\begin{equation}
    S_2(\mathbf{x}_1,\mathbf{x}_2) = \exp\left[ -\rho v_2(\mathbf{x}_1,\mathbf{x}_2;R) \right]\label{eqn:S2Hn}.
\end{equation}

For the surface-void and surface-surface correlation functions, we have the limits
\begin{eqnarray}
    F_{sv}(\mathbf{x}_1,\mathbf{x}_2) = &&\lim_{a_i\to R, \forall i} H_2(\mathbf{x}_1;\mathbf{x}_2;\emptyset) \nonumber \\
    &&\hspace{-1cm}=-\lim_{a_1\to R} \frac{\partial}{\partial a_1}\exp[-\rho v_2(r;a_1,R)],
\end{eqnarray}
and
\begin{eqnarray}
    F_{ss}(\mathbf{x}_1,\mathbf{x}_2) = &&\lim_{a_i\to R, \forall i} H_2(\mathbf{x}_1,\mathbf{x}_2;\emptyset;\emptyset) \nonumber \\
    &&\hspace{-2cm}=\lim_{a_1,a_2\to R} \frac{\partial}{\partial a_1}\frac{\partial}{\partial a_2} \exp[-\rho v_2(r;a_1,a_2)].
\end{eqnarray}
From these expressions, the extra information in the surface correlation functions is revealed: both $F_{sv}$ and $F_{ss}$ involve a product of Eq. \eqref{eqn:S2Hn} and a term related to surface area of the phase interface due to the partial derivatives.

The complementary pore-size distribution function $F(\delta)$ is related to the ``void" exclusion probability function $E_{_V}(r)$ which is defined in terms of $H_n$ as \cite{torquato86}
\begin{equation}
    E_{_V}(r) = H_1(\emptyset;\mathbf{x}_1;\emptyset).
\end{equation}
We see that higher-order microstructural information is incorporated into $F(\delta)=E_{_V}(\delta+R)/\phi_1$ by the requirement that the entire volume excluded by the radius $r$ test particle is devoid of phase 2. Lastly, Lu and Torquato found that the lineal-path function $L(z)$ is a special case of $E_{_V}(r)$ \cite{lu92} where a test line segment of length $z$ is inserted into the system. Thus, $L(z)$ incorporates functionals of higher-order information through the requirement that the entire test line is in phase 1 and not just its end points.

\section{\label{sec:YT}The Yeong-Torquato Reconstruction Algorithm}

\begin{figure*}
	    \subfloat[\label{fig:YT2Dstructsa}]{\includegraphics[width=0.33\textwidth]{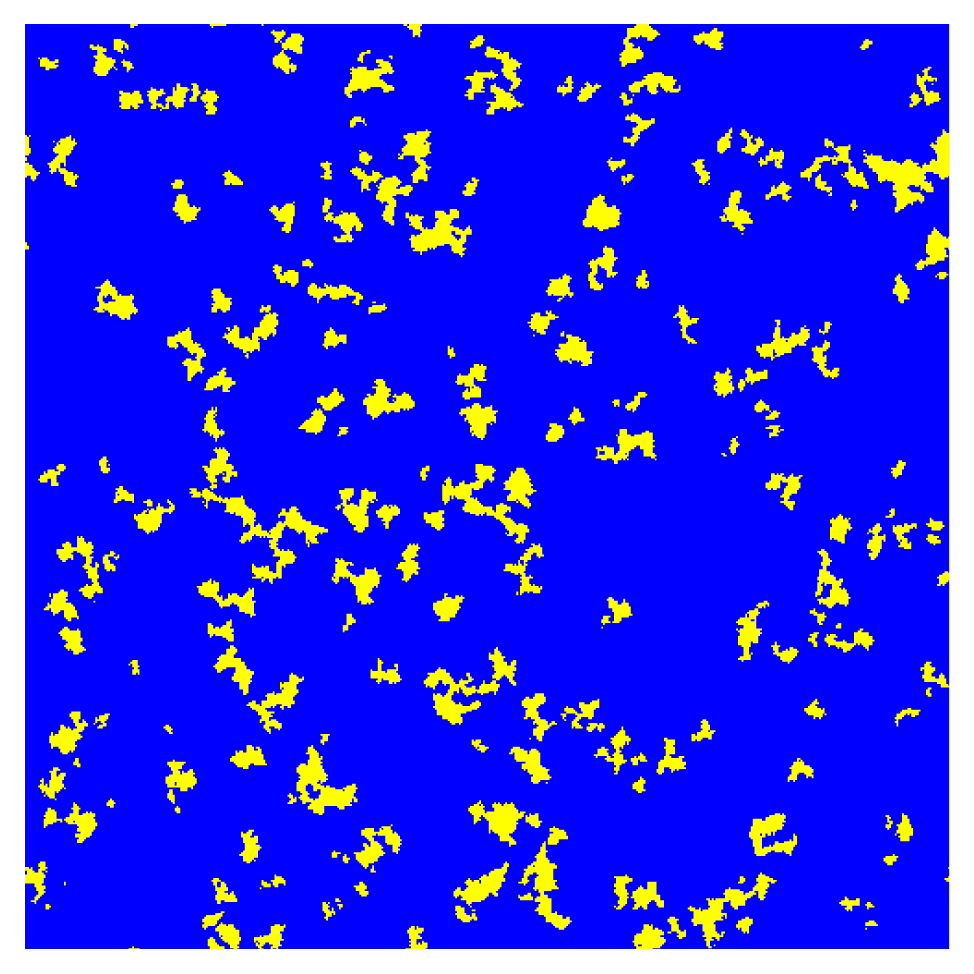}}
	    \subfloat[\label{fig:YT2Dstructsb}]{\includegraphics[width=0.33\textwidth]{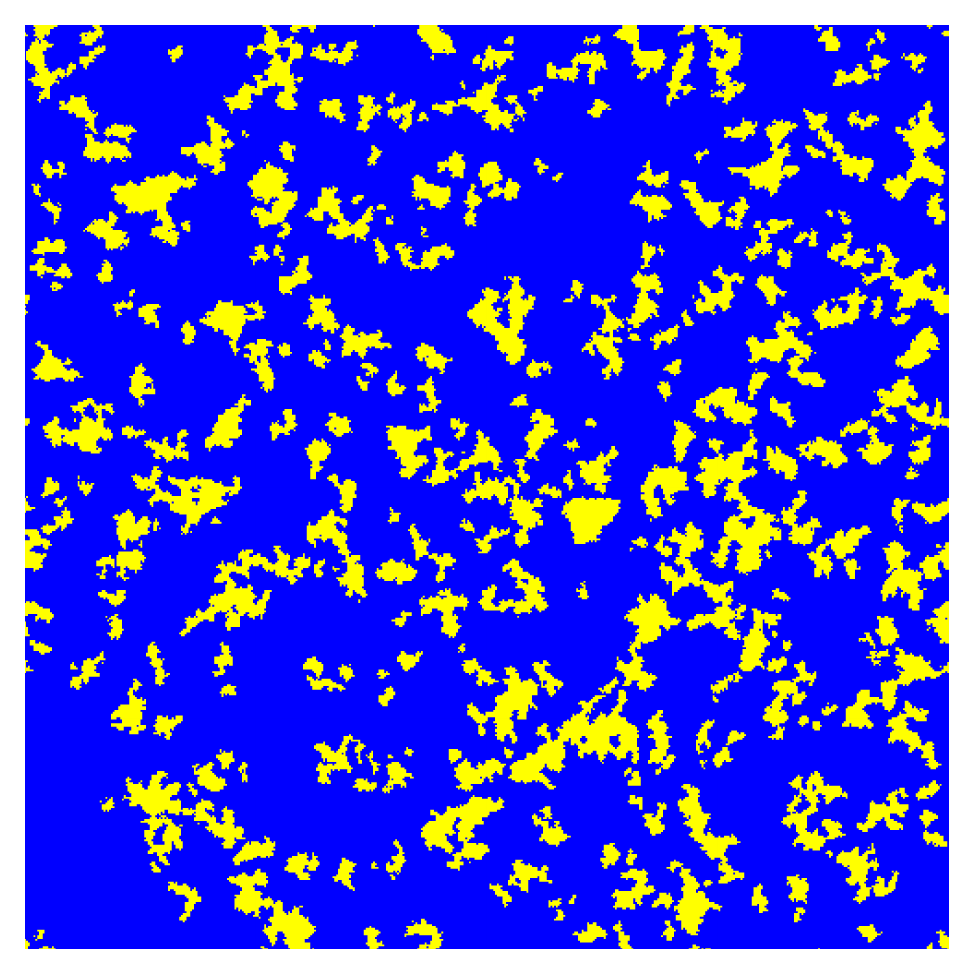}}
	    \subfloat[\label{fig:YT2Dstructsc}]{\includegraphics[width=0.33\textwidth]{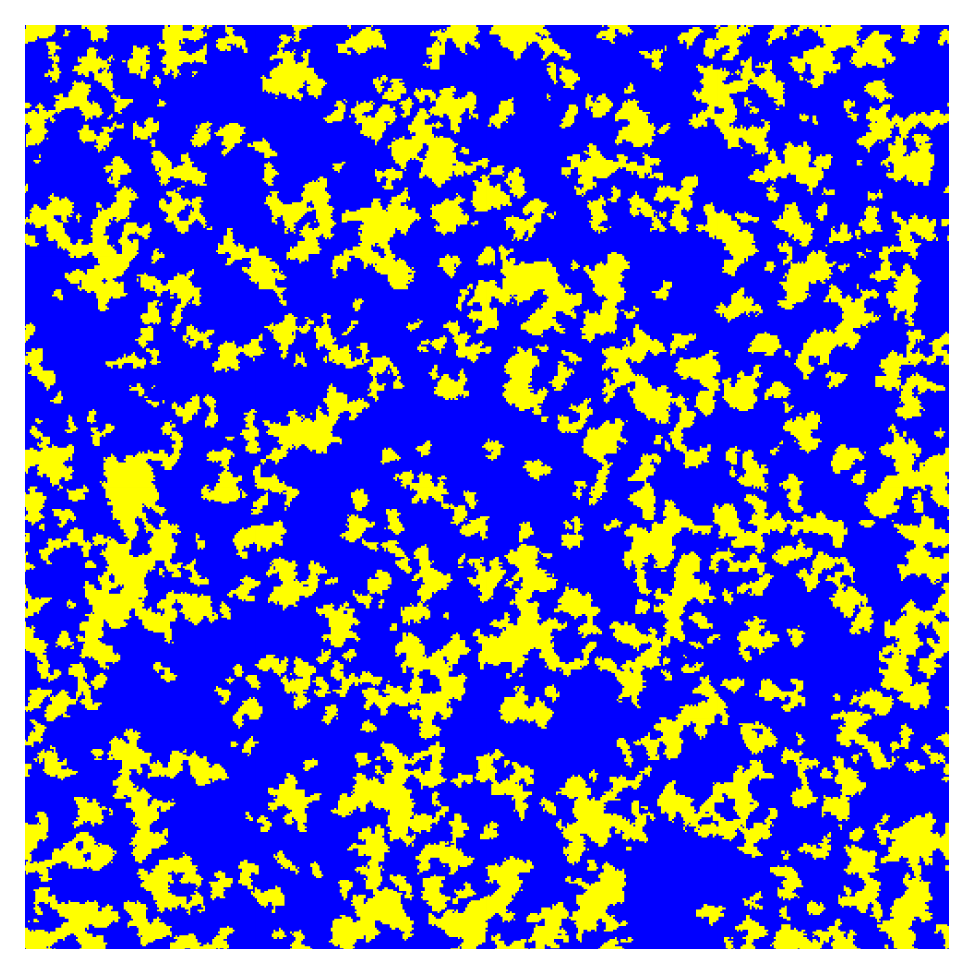}}
	    \hfill
    	\subfloat[\label{fig:YT2Dstructsd}]{\includegraphics[width=0.33\textwidth]{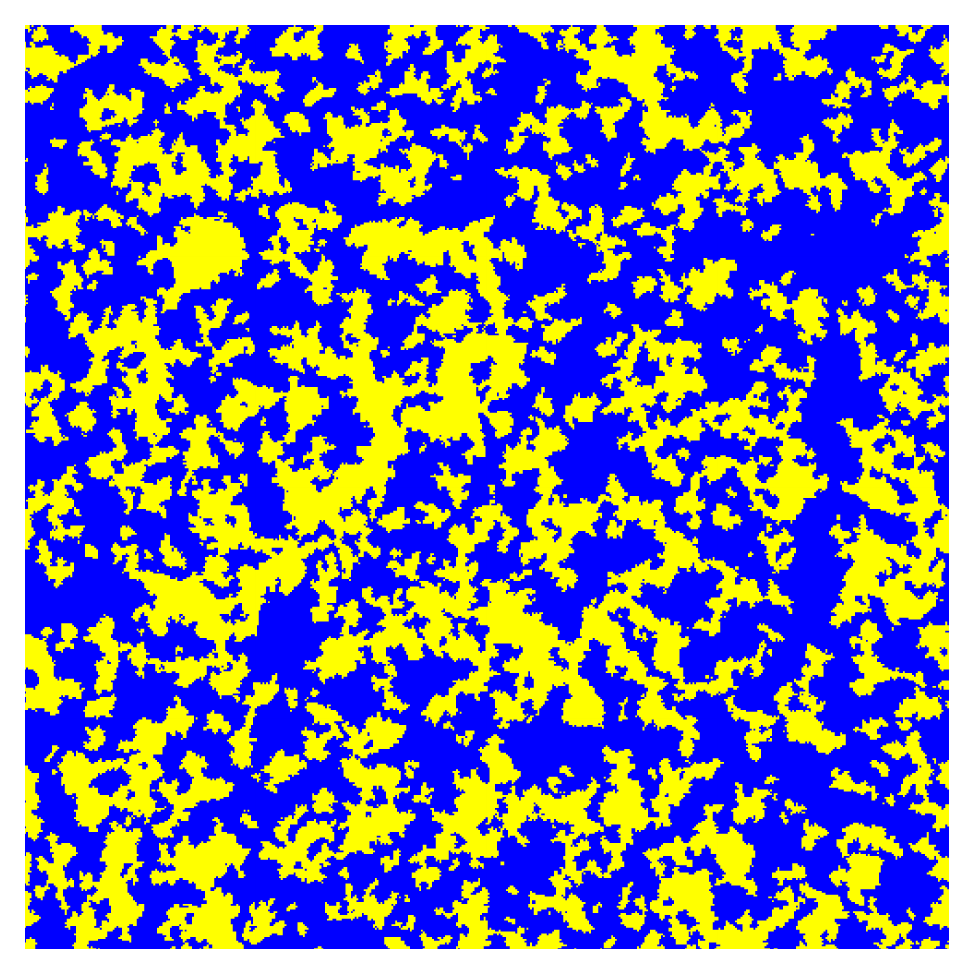}}
    	\subfloat[\label{fig:YT2Dstructse}]{\includegraphics[width=0.33\textwidth]{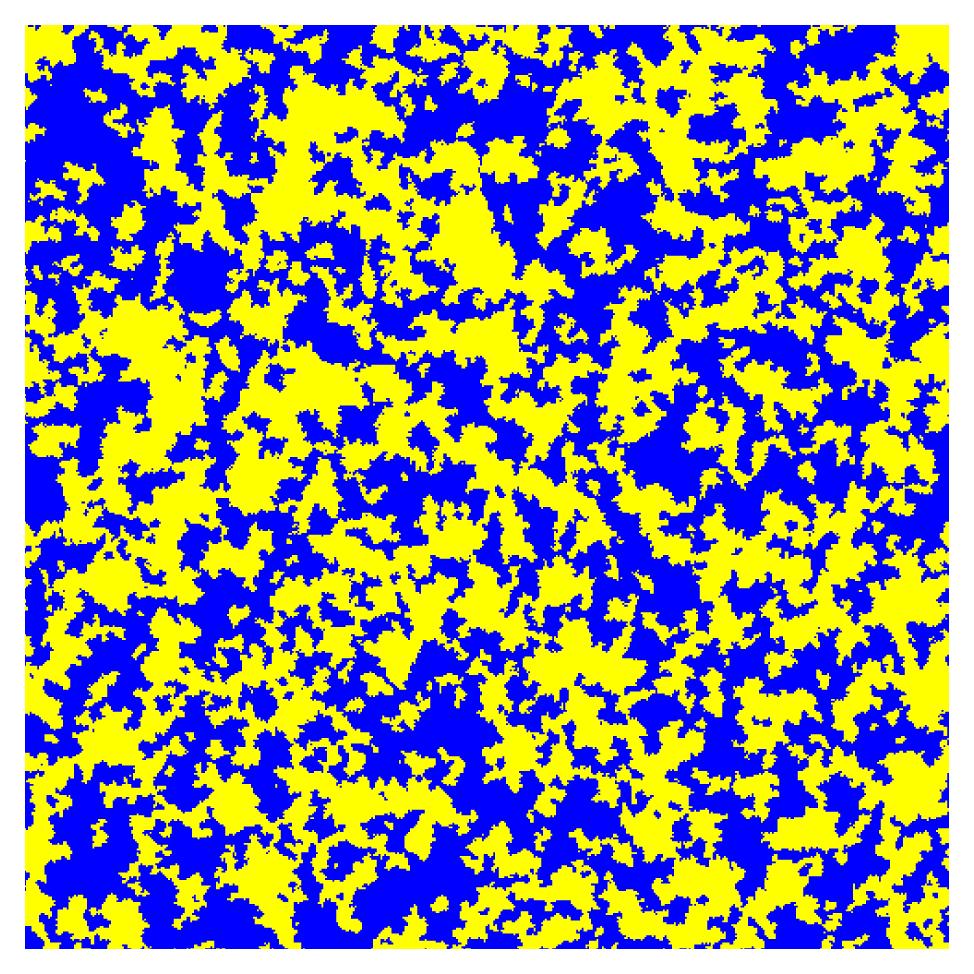}}
    	\subfloat[\label{fig:YT2Dstructsf}]{\includegraphics[width=0.33\textwidth]{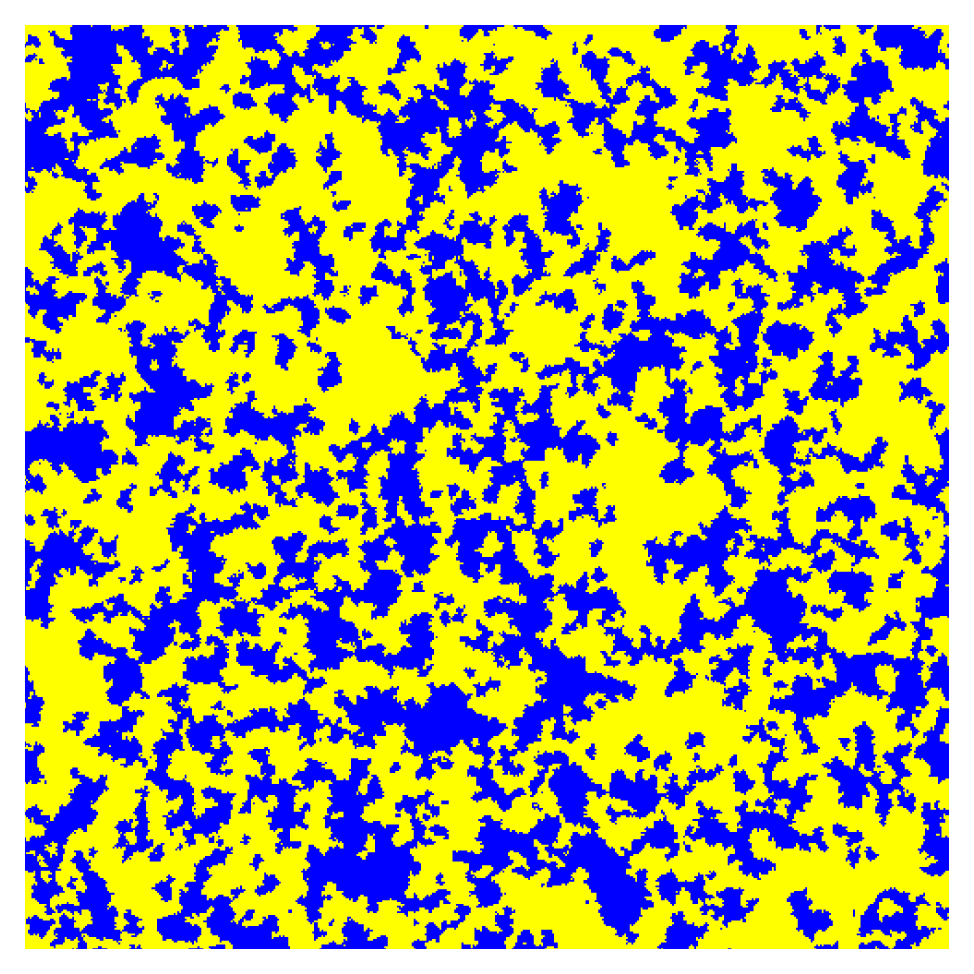}}
    	\hfill
    	\subfloat[\label{fig:YT2Dstructsg}]{\includegraphics[width=0.33\textwidth]{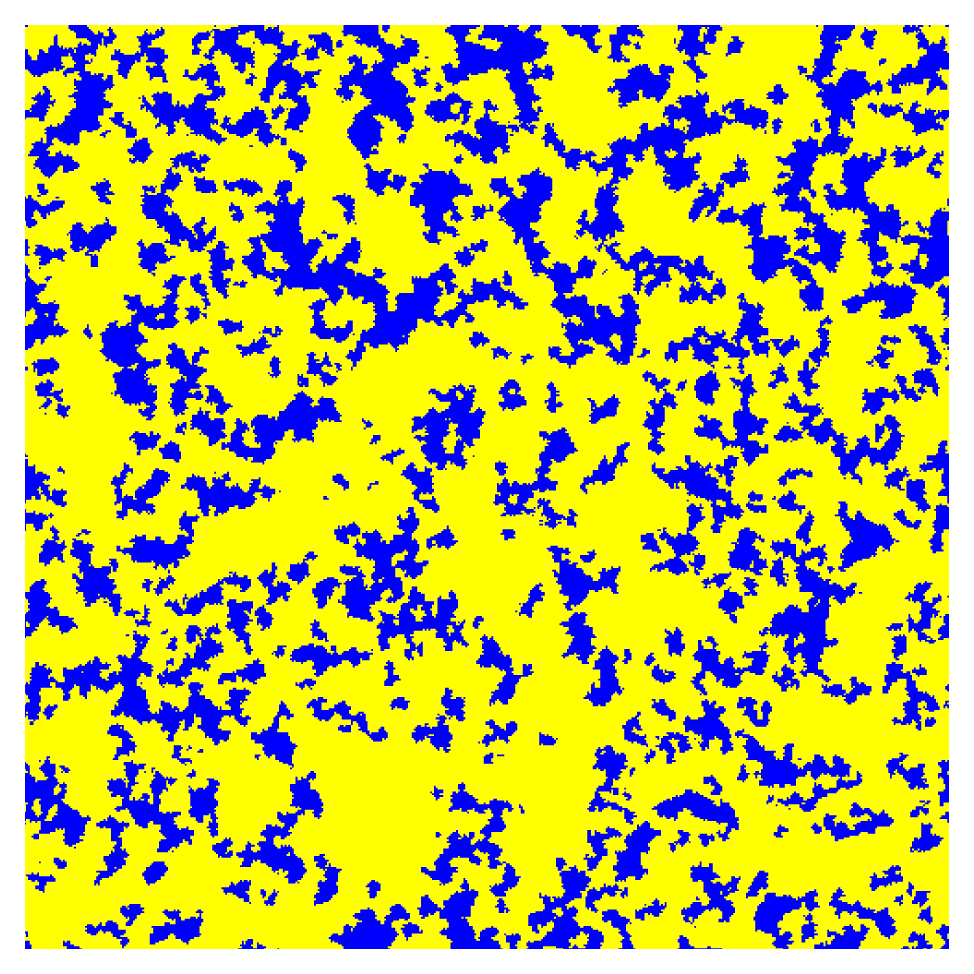}}
    	\subfloat[\label{fig:YT2Dstructsh}]{\includegraphics[width=0.33\textwidth]{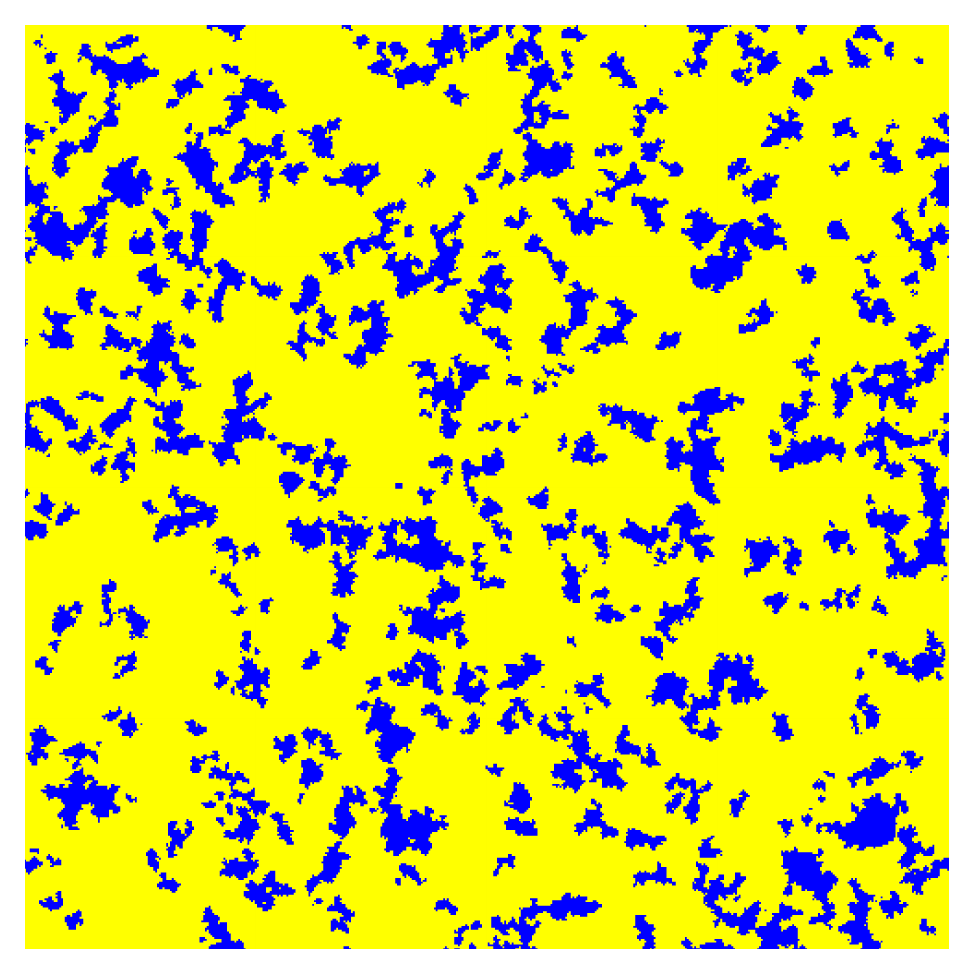}}
    	\subfloat[\label{fig:YT2Dstructsi}]{\includegraphics[width=0.33\textwidth]{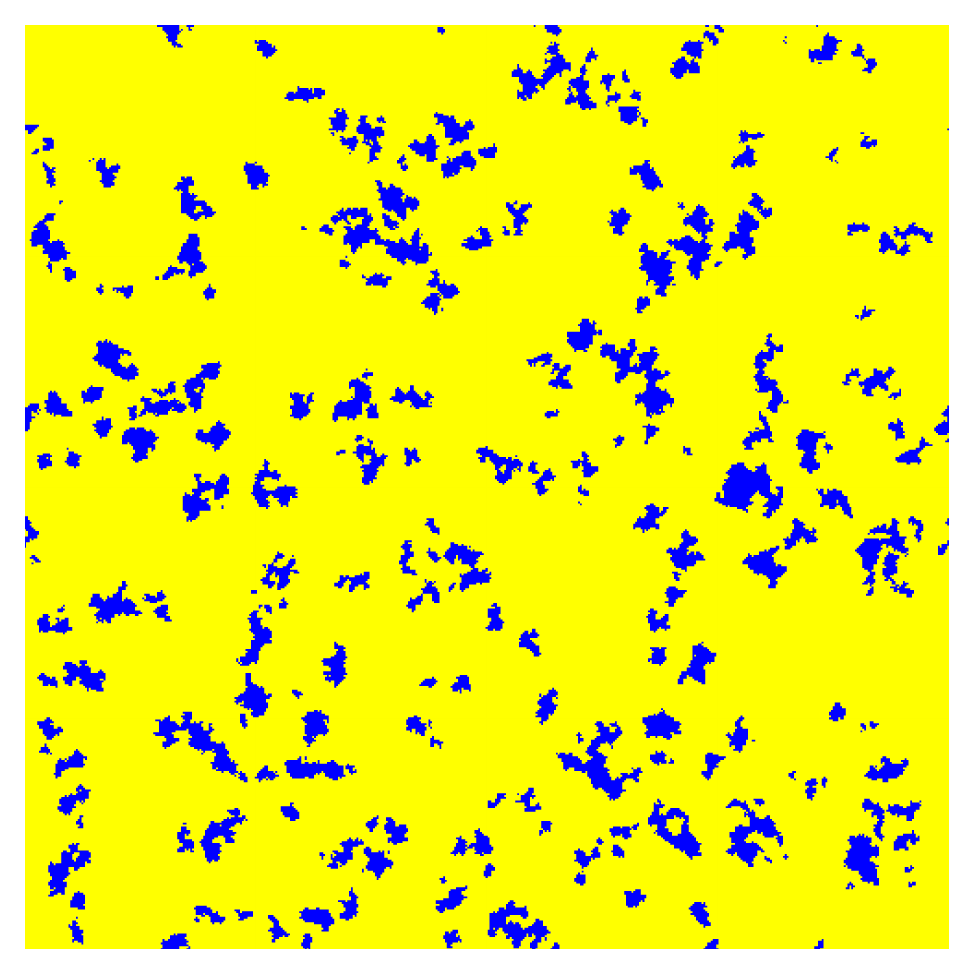}}
	\caption{Realizations of 2D YT-DRM. Images (a)-(i) correspond to void phase (yellow) volume fraction $\phi_1=0.1-0.9$ and inclusion phase (blue) volume fraction $\phi_2=0.9-0.1$, respectively. Following Ma and Torquato \cite{drm2020}, these microstructures are $501\times501$ pixels with characteristic length $a=5$, and a cutoff $l_c=10a$ was used for sampling $S_2^{(1)}(r)$.}\label{fig:YT2Dstructs}
\end{figure*}

The Yeong-Torquato optimization procedure is a popular algorithm that has been used by various groups to construct or reconstruct microstructures that realize a set of prescribed correlation functions \cite{jiao07,jiaoPNAS09,gerke18,capek18,chen18,secanell15,katsman19}. Here, we briefly describe the Yeong-Torquato algorithm. For the 2D reconstructions employed in this work, the two-phase system is represented as a square grid of pixels that is subject to periodic boundary conditions. This square has side length $L$ and contains $N^2$ pixels which can represent phase 1 or 2. The Yeong-Torquato procedure treats the task of transforming this grid into the desired microstructure as an energy-minimization problem that it solves by simulated annealing. 

The ``energy" is defined as
\begin{equation}
    E = \sum_{\alpha} w_{\alpha} E_{\alpha} = \sum_{\alpha} \sum_{\mathbf{x}} w_{\alpha}[f_n^{\alpha}(\mathbf{x}) - \hat{f}_n^{\alpha}(\mathbf{x})]^2\label{eqn:YTenergy}
\end{equation}
and measures how close the current system is to realizing the prescribed, target statistical descriptors: the volume fraction and some set $\hat{f}_n^{1}(\mathbf{x}),\hat{f}_n^{2}(\mathbf{x}),...$ where $\hat{f}_n^\alpha$ is an $n$-point correlation function of type $\alpha$ and $\mathbf{x}\equiv\mathbf{r}_1,\mathbf{r}_2,...$ denotes position vectors in the medium. Note that $f_n^{1}(\mathbf{x}),f_n^{2}(\mathbf{x}),...$ is the set of correlations measured from the system that is being reconstructed and the number $w_\alpha$ is a weight for target descriptor $\hat{f}_n^{\alpha}$. The microstructure of the system is evolved using volume fraction conserving pixel swapping moves which are accepted according to the Metropolis rule while a fictitious temperature is lowered which has the effect of reducing the acceptance probability. For more details on the Yeong-Torquato procedure and simulated annealing, see Ref. \cite{yeong98}.

In this work, we employ an accelerated implementation of the Yeong-Torquato construction algorithm developed by Ma and Torquato \cite{drm2020}. In this scheme, relatively large 2D systems ($501\times501$ pixels) are more easily realized by using a cutoff $l_c<L$ when sampling $S_2^{(i)}(r)$. For $S_2^{(i)}(r)$ like Eq. \eqref{eqn:debS2First} that decay to their asymptotic value ($\phi_i^2$) rapidly, the use of a cutoff is valid as long as it is sufficiently larger than the characteristic length of the system. Notably, the computational cost of the accelerated scheme scales as $\mathcal{O}(N^d)$; an improvement over the $\mathcal{O}(N^{2d})$ scaling of the original Yeong-Torquato implementation. Moreover, in this work, we found that the accelerated scheme frees sufficient computational resources to facilitate the construction of Debye random media with a specific pore-size probability density function (see Sec. \ref{sec:cpdrm}).

The implementation of the Yeong-Torquato procedure used here employs a pixel refinement phase where, after a fraction of the total Monte Carlo steps, only pixels at the phase interface are selected for trial swaps. This refinement phase has the net effect of eliminating small isolated ``islands" of one phase embedded in a ``sea" of the other phase. Lastly, $S_2$ is sampled in all directions (as described in Ref. \cite{jiao07}) which contrasts the original scheme used by Yeong and Torquato wherein two-point correlations were sampled only along orthogonal directions \cite{yeong98}.

Samples of Debye random media realized with the Yeong-Torquato procedure for various volume fractions in 2D are presented in Fig. \ref{fig:YT2Dstructs}. Note how, at lower $\phi_1$, the void phase consists of islands with a spectrum of sizes and shapes. As $\phi_1$ is increased, the islands continually merge until phases 1 and 2 are statistically indistinguishable at $\phi_1=1/2$. Due to the phase inversion symmetry that is manifest in Eq. \eqref{eqn:debS2First}, realizations of YT-DRM for $\phi_1=0.6-0.9$ are identical to those with $\phi_1=0.4-0.1$, which is evident in Fig. \ref{fig:YT2Dstructs}.

\section{\label{sec:s2pds}Two-point correlation function for Overlapping, polydisperse spheres}

In this section, we derive the two-point correlation function for systems of polydisperse, \textit{totally penetrable} spheres in the first three dimensions following the approach in Refs. \cite{Torq02,polydisperse91,lu92II,perm92}. We take sphere radii $R$ to follow the normalized probability density $f(R)$. The average of any $R$-dependent function is thus computed as
\begin{equation}
    \langle w(R) \rangle = \int_0^{\infty} w(R)f(R)dR.
\end{equation}
As in prior work, \cite{Torq02,polydisperse91,lu92II,perm92} we define a reduced density to be
\begin{equation}
    \eta = \rho \langle v_1(R) \rangle,
\end{equation}
where the average volume of the spheres is $\langle v_1(R) \rangle = \omega_d \langle R^d \rangle$. Following Torquato, \cite{Torq02,polydisperse91,lu92II,perm92}, we consider the Schulz distribution \cite{schulz39} 
\begin{equation}
    f(R) = \frac{1}{\Gamma(m+1)}\left( \frac{m+1}{\langle R \rangle} \right)^{m+1}R^m e^{-(m+1)R/\langle R \rangle} \label{eqn:schulz},
\end{equation}
where $\langle R \rangle$ is the mean radius of the distribution, and $m$ is restricted to integer values in the interval $[0,\infty)$. Increasing the parameter $m$ lowers the variance of the distribution and the monodisperse limit is recovered when $m\to\infty$, i.e., $f(R)\to\delta(R-\langle R \rangle)$. In this work, we take $m=0$, which corresponds to an exponential distribution where many particles have small radii.

The two-point correlation function for the void-phase of these systems is \cite{Torq02}
\begin{equation}
    S_2^{(1)}(r) = \exp\left[ \ln\phi_1 \frac{\langle v_2(r;R) \rangle}{\langle v_1(R) \rangle} \right]\label{eqn:S2PDS},
\end{equation}
where $v_2(r;R)$ is the union volume of two $d$-dimensional spheres of radius $R$, which is given for $d=1,2$ and $3$ in Ref. \cite{Torq02}. Using the union volume formulas in Eq. \eqref{eqn:S2PDS}, we find that the two-point probability function for the first three dimensions has the form
\begin{equation}
    S_2^{(1)}(r) = \exp\left[\ln\phi_1 h(r;\langle R \rangle) \right],\label{eqn:OPSs2}
\end{equation}
where for $d=1,2$ and $3$, respectively,
\begin{equation}
    h(r;\langle R \rangle) = 2 - e^{-r/2\langle R \rangle}\label{eqn:hfunc1D},
\end{equation}
\begin{eqnarray}
    h(r;\langle R \rangle) = &&2 + \frac{r^2}{4\pi\langle R \rangle^2} K_1\left( \frac{r}{2\langle R \rangle} \right) - \nonumber \\
    &&\frac{2}{\pi} G_{2,4}^{4,0}\left(\frac{r^2}{16\langle R \rangle^2}\Big|\begin{array}{c} 1,1 \\ 0,\frac{1}{2},\frac{3}{2},2 \\\end{array}\right)\label{eqn:hfunc2D},
\end{eqnarray}
\begin{equation}
    h(r;\langle R \rangle) = \frac{8\langle R \rangle - e^{-r/2\langle R \rangle}(r + 4\langle R \rangle)}{4\langle R \rangle}\label{eqn:hfunc3D}.
\end{equation}
For $h(r,\langle R \rangle)$ of 2D systems, $K_1(x)$ is the first order, modified Bessel function of the second kind, and $G_{m,n}^{p,q}\left(z\Big|\begin{array}{c} a_1,...,a_p \\ b_1,...,b_q \\\end{array}\right)$ is the Meijer-G function.

\section{\label{sec:realization}Realizing Debye Random Media with Overlapping, Polydisperse Spheres}

\begin{figure*}
    \subfloat[\label{fig:S2plotsa}]{\includegraphics[width=0.5\linewidth]{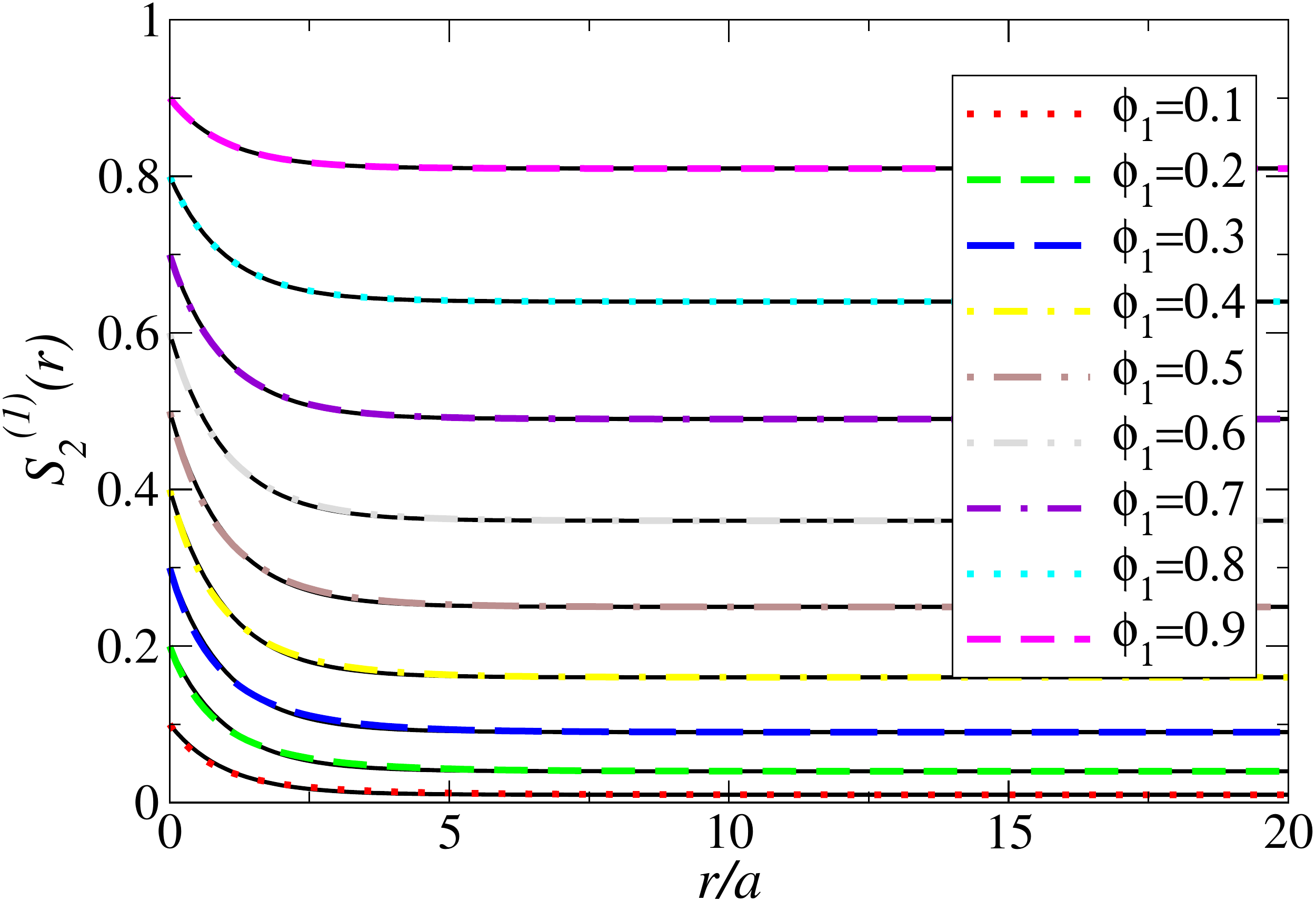}}
    \subfloat[\label{fig:S2plotsb}]{\includegraphics[width=0.5\linewidth]{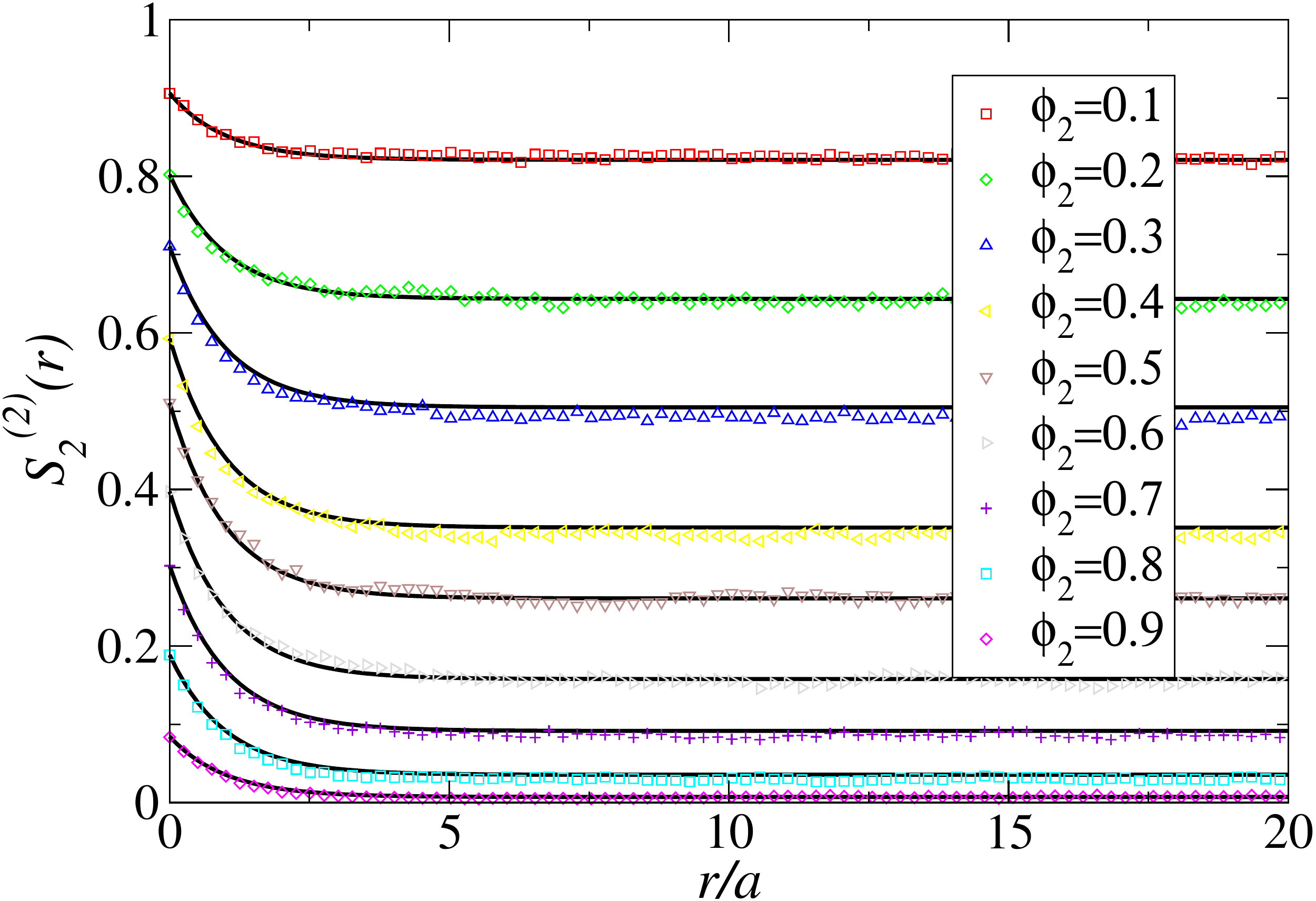}}
    \hfill
    \subfloat[\label{fig:S2plotsc}]{\includegraphics[width=0.5\linewidth]{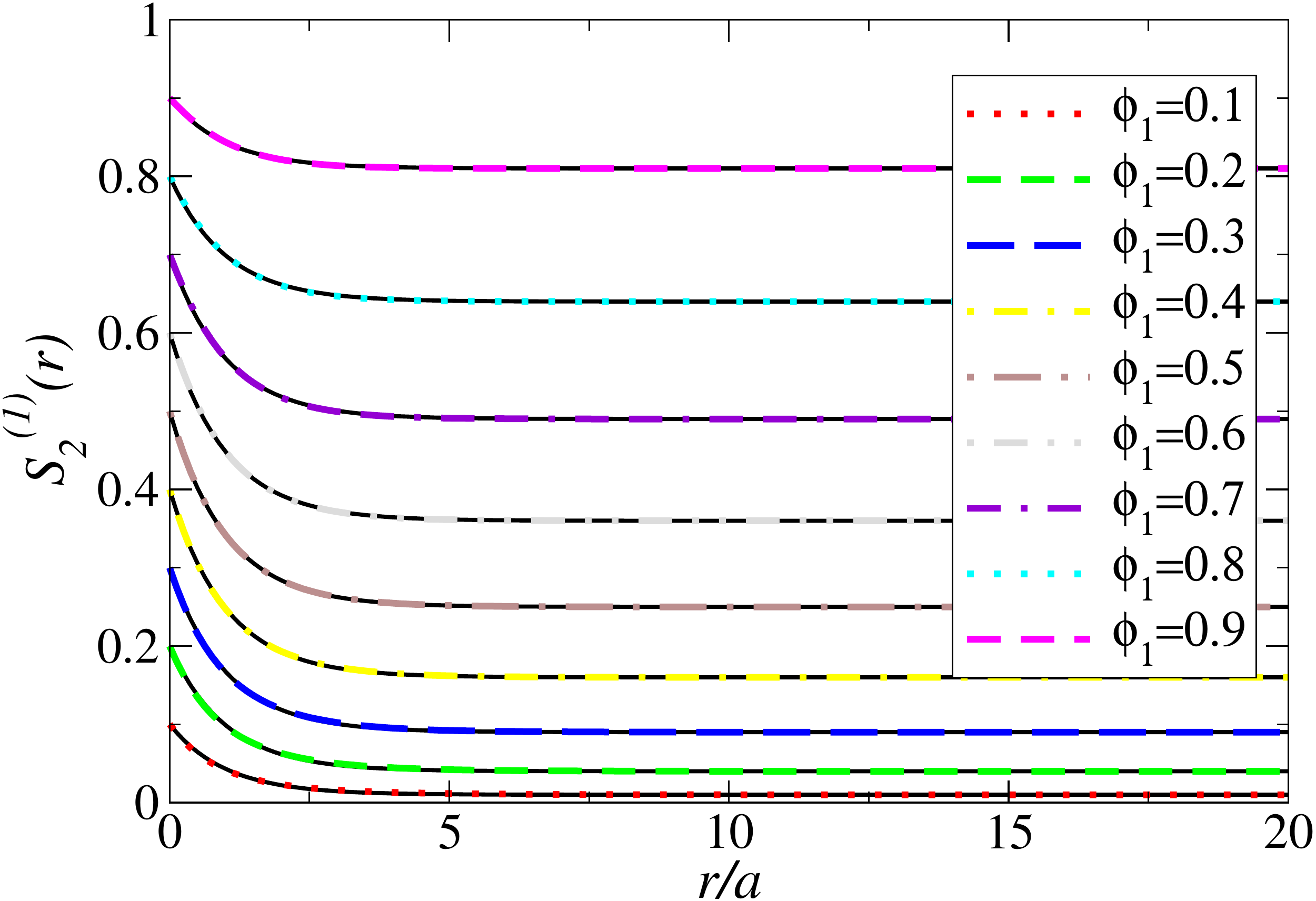}}
    \subfloat[\label{fig:S2plotsd}]{\includegraphics[width=0.5\linewidth]{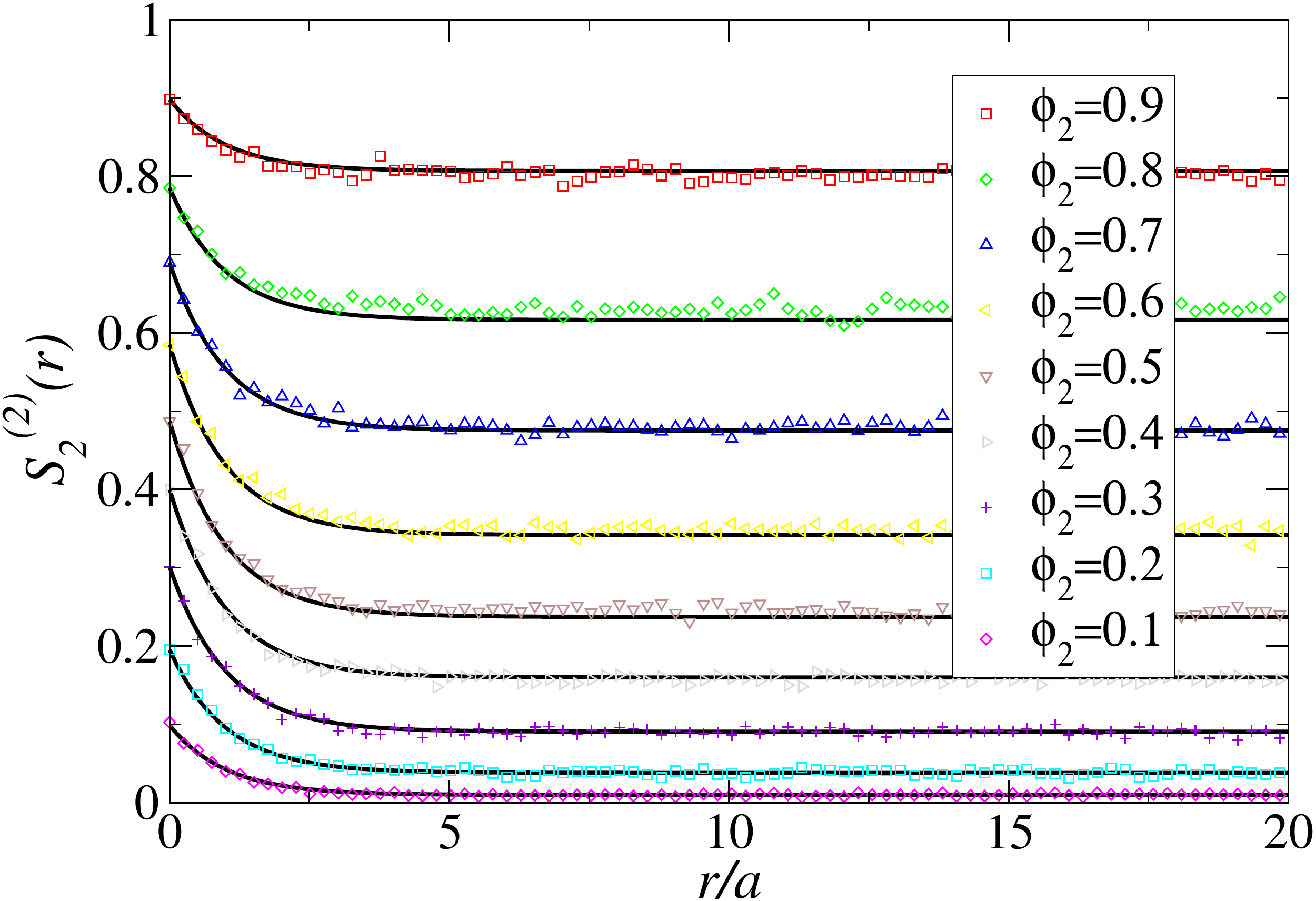}}
    \hfill
    \subfloat[\label{fig:S2plotse}]{\includegraphics[width=0.5\linewidth]{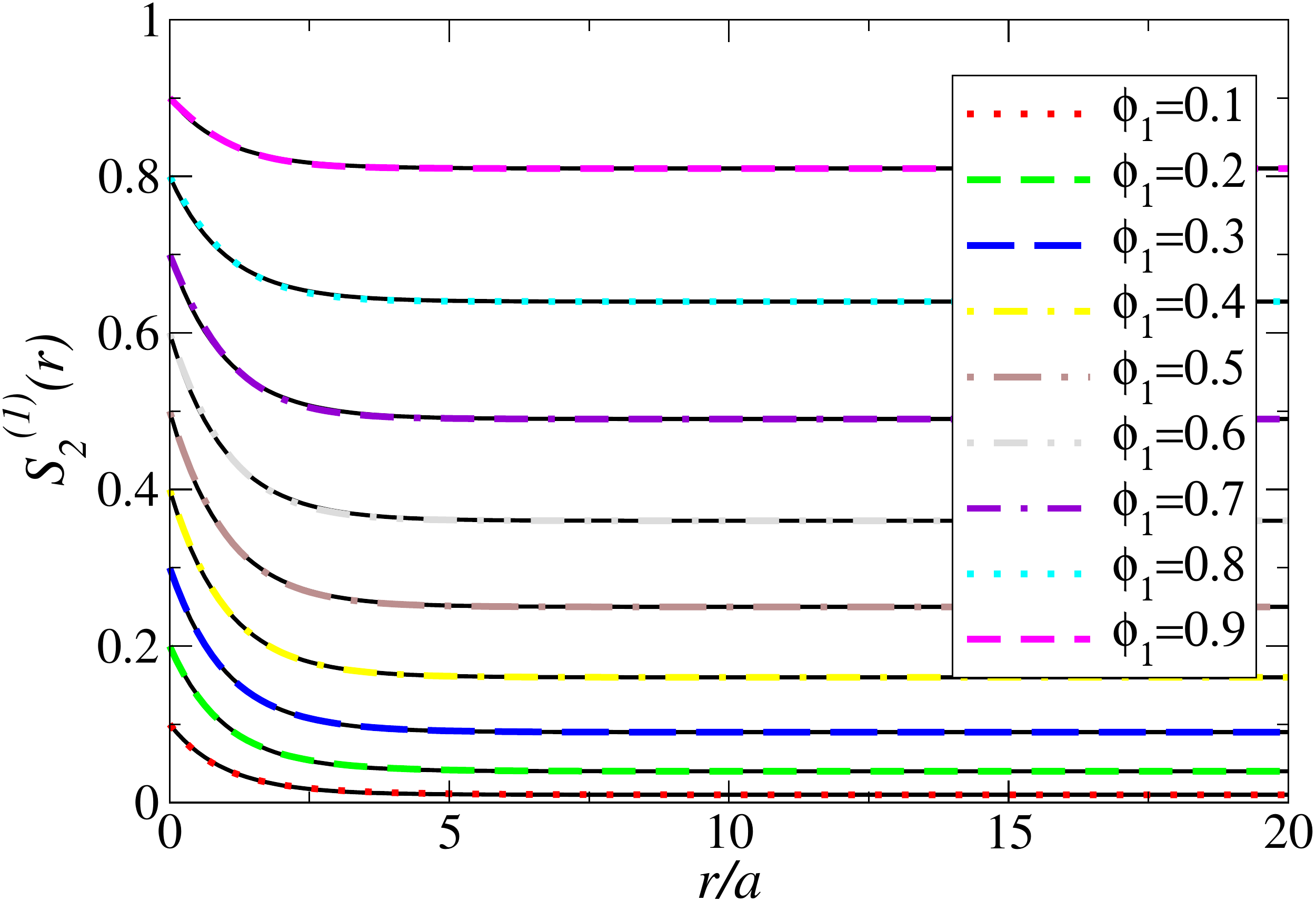}}
    \subfloat[\label{fig:S2plotsf}]{\includegraphics[width=0.5\linewidth]{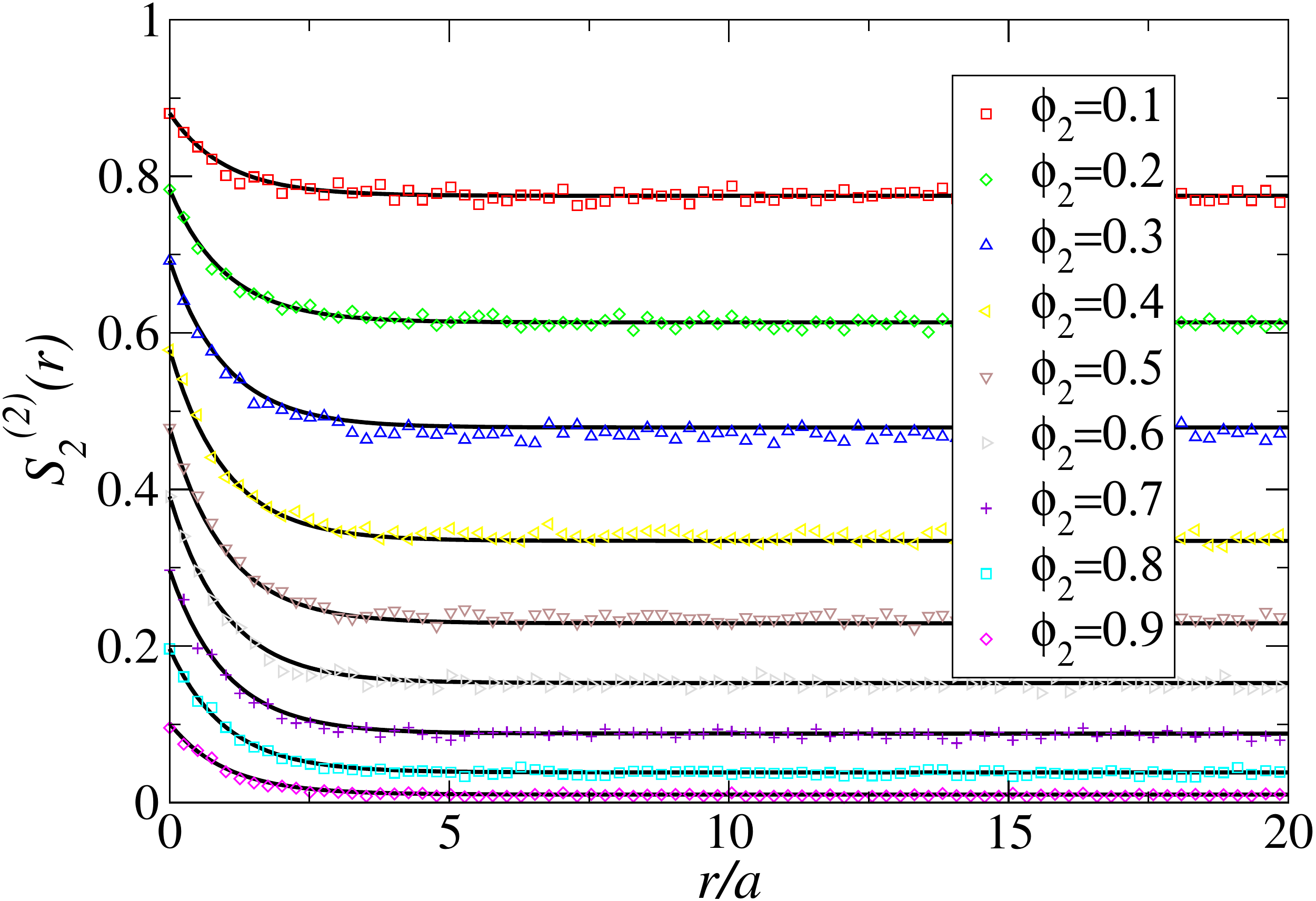}}
    \hfill
    \caption{The plots in the left column [(a), (c), and (e)] are for the two-point function of the void phase of OPS-DRM, where the colored lines are of function \eqref{eqn:OPSs2} with $h(r;\langle R \rangle)$ given by \eqref{eqn:hfunc1D} for 1D (a), \eqref{eqn:hfunc2D} for 2D (c), and \eqref{eqn:hfunc3D} for 3D (e). For $h(r;\langle R \rangle)$, the value of mean radius $\langle R \rangle$ is given by Eq. \eqref{eqn:eplw}. The plots in the right column [(b), (d), and (f)] are for the two-point function of the disk phase, where the colored markers are for $S_2^{(2)}(r)$ that was sampled numerically from realizations of OPS-DRM. Panel (b) is from 1D systems, (d) is from 2D systems, and (f) is from 3D systems. Note that in all cases, $S_2^{(i)}(r=0)=\phi_i$ and $S_2^{(i)}(r\to\infty)=\phi_i^2$. The black lines are all given by \eqref{eqn:debS2First} with the appropriate volume fractions. \label{fig:S2plots}}
\end{figure*}

\begin{figure*}
	    \subfloat[\label{fig:OPS2Dstructsa}]{\includegraphics[width=0.33\textwidth]{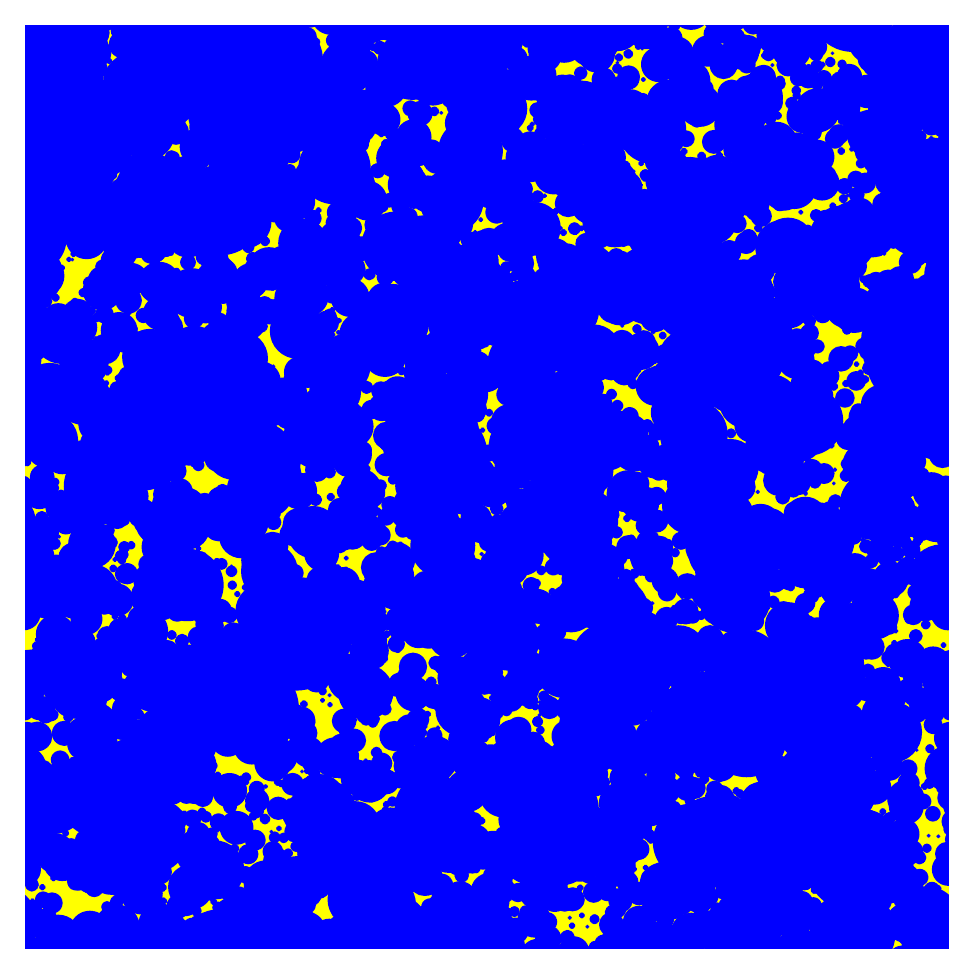}}
	    \subfloat[\label{fig:OPS2Dstructsb}]{\includegraphics[width=0.33\textwidth]{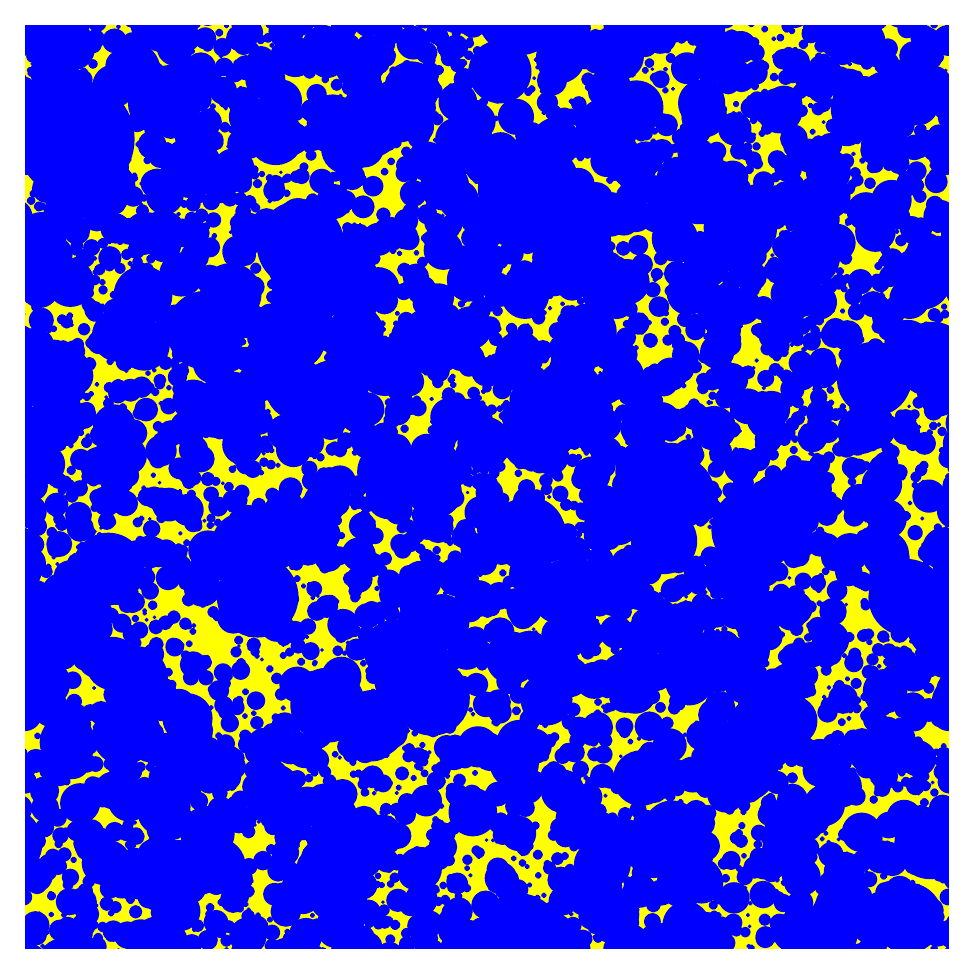}}
	    \subfloat[\label{fig:OPS2Dstructsc}]{\includegraphics[width=0.33\textwidth]{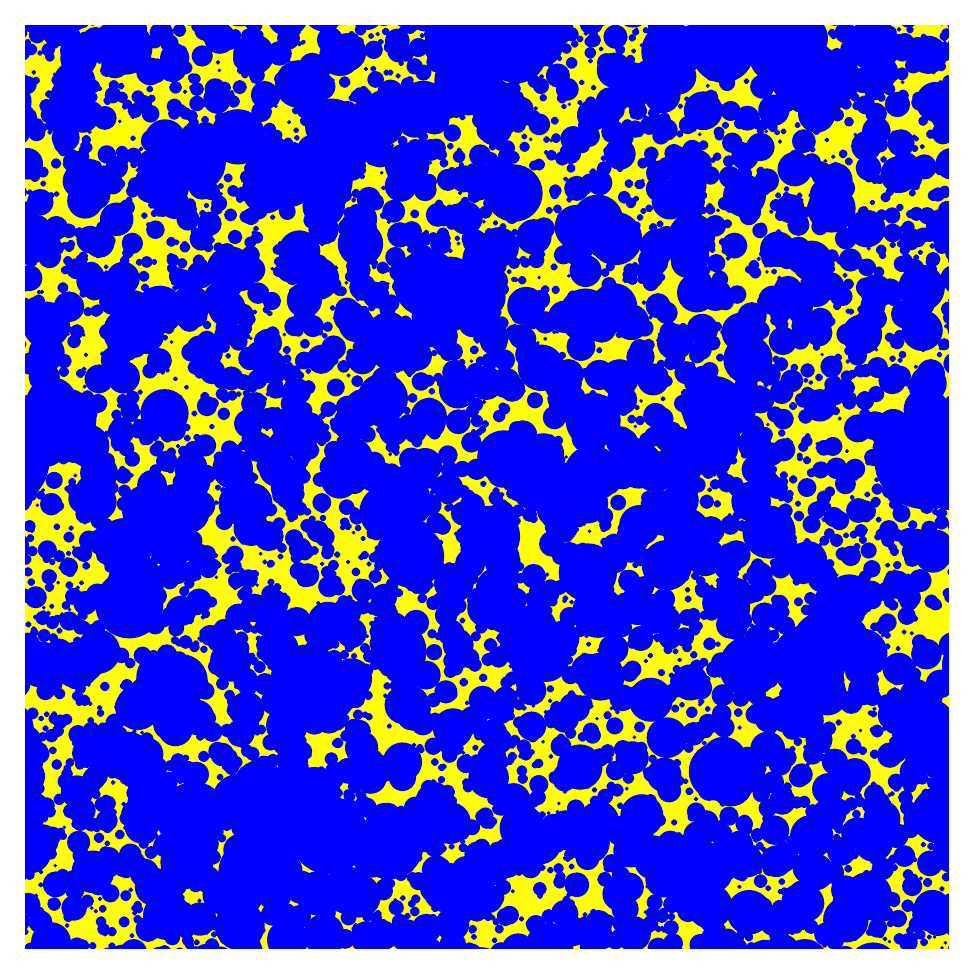}}
	    \hfill
    	\subfloat[\label{fig:OPS2Dstructsd}]{\includegraphics[width=0.33\textwidth]{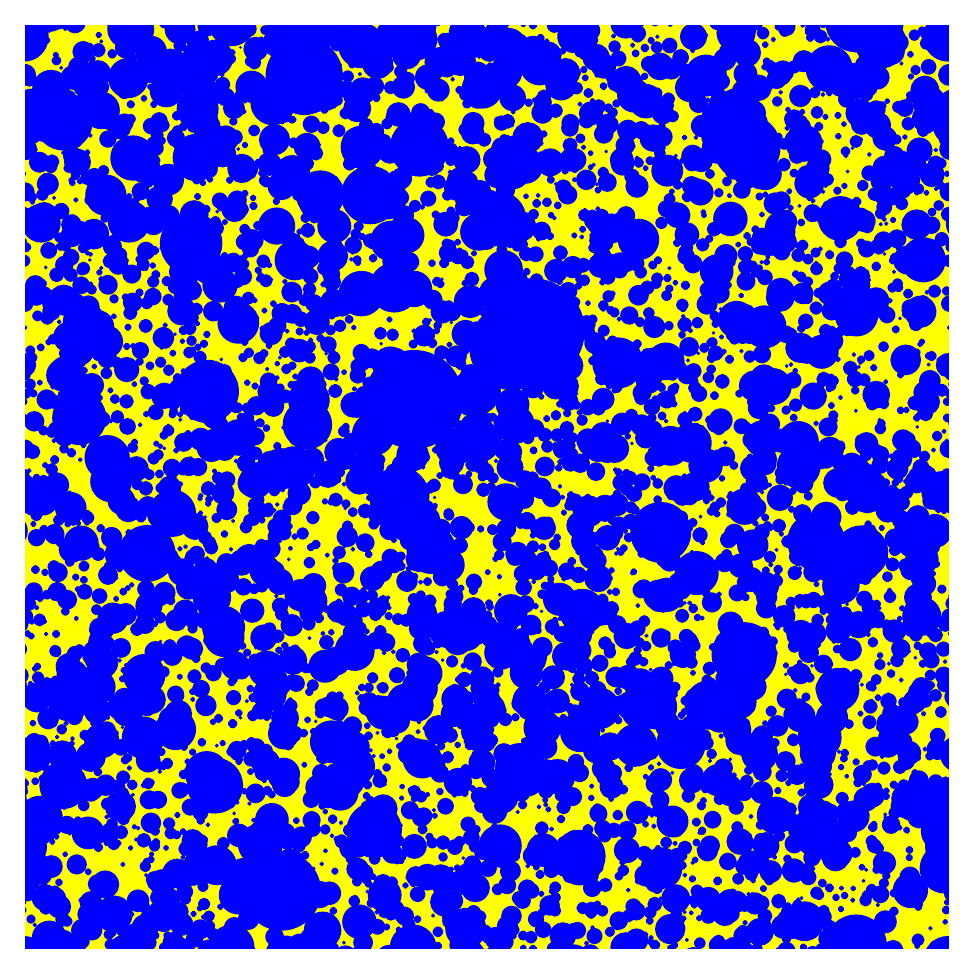}}
    	\subfloat[\label{fig:OPS2Dstructse}]{\includegraphics[width=0.33\textwidth]{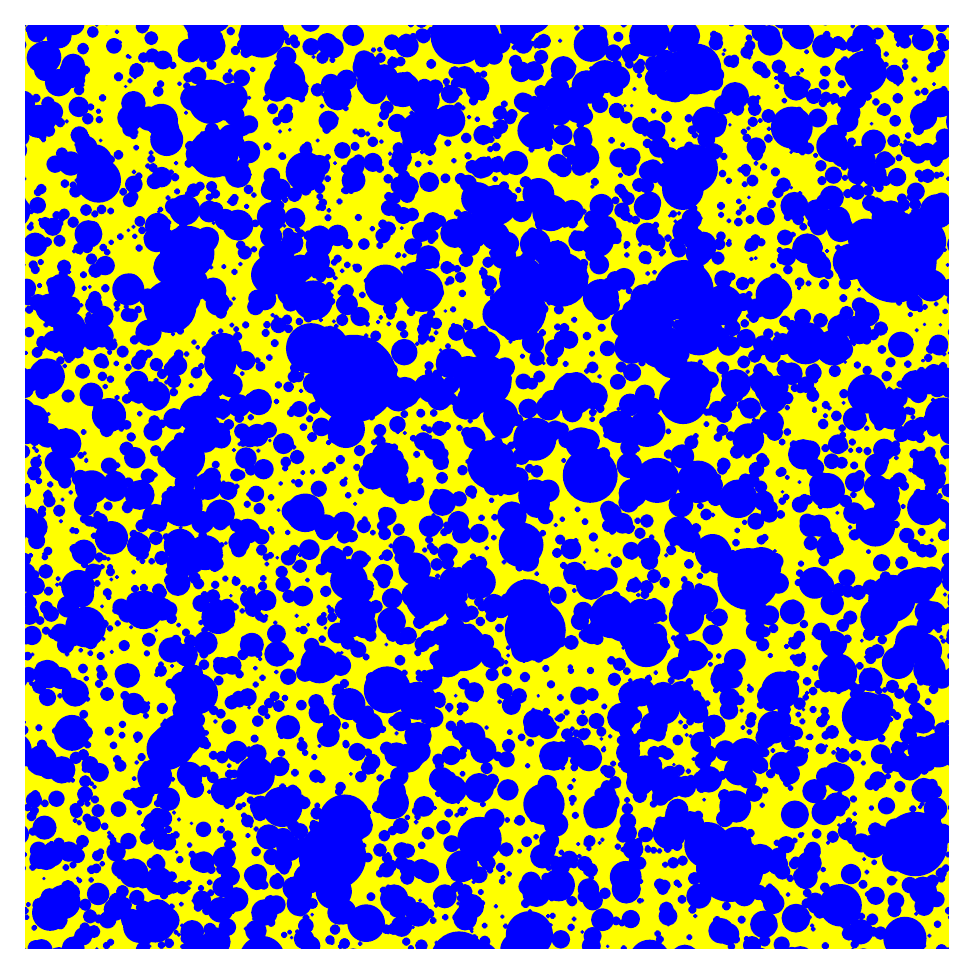}}
    	\subfloat[\label{fig:OPS2Dstructsf}]{\includegraphics[width=0.33\textwidth]{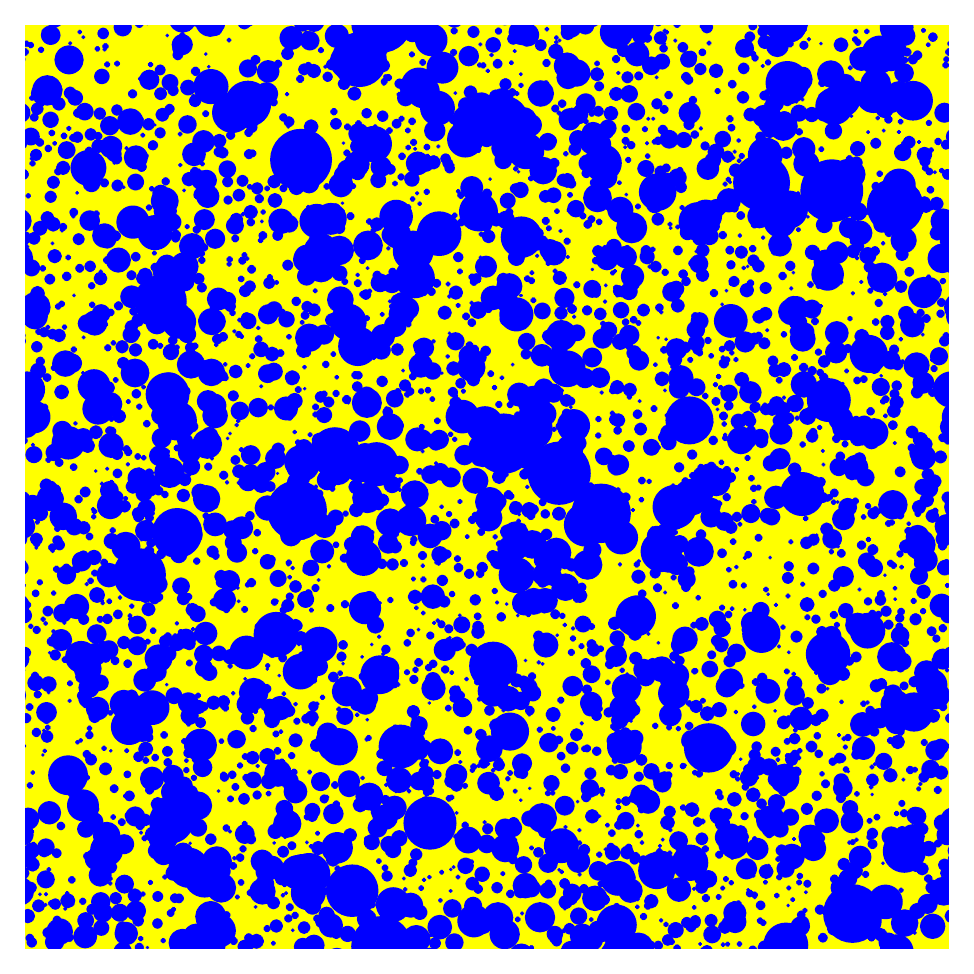}}
    	\hfill
    	\subfloat[\label{fig:OPS2Dstructsg}]{\includegraphics[width=0.33\textwidth]{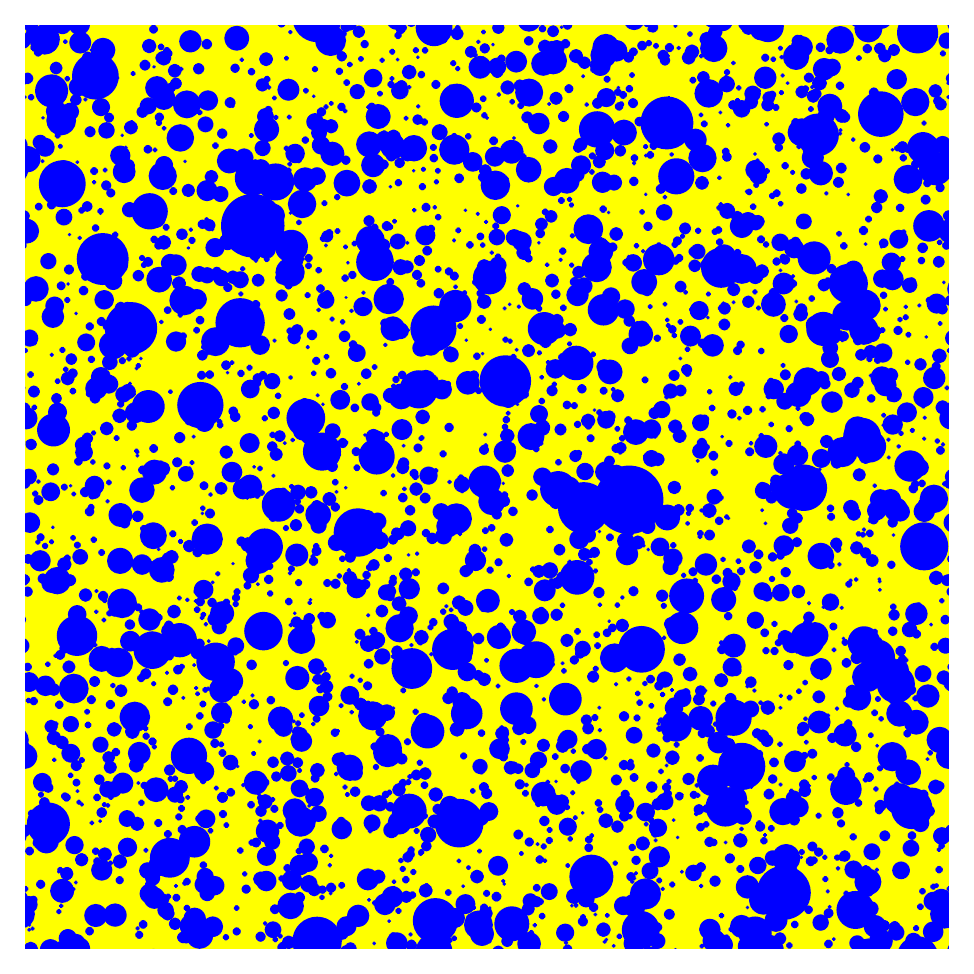}}
    	\subfloat[\label{fig:OPS2Dstructsh}]{\includegraphics[width=0.33\textwidth]{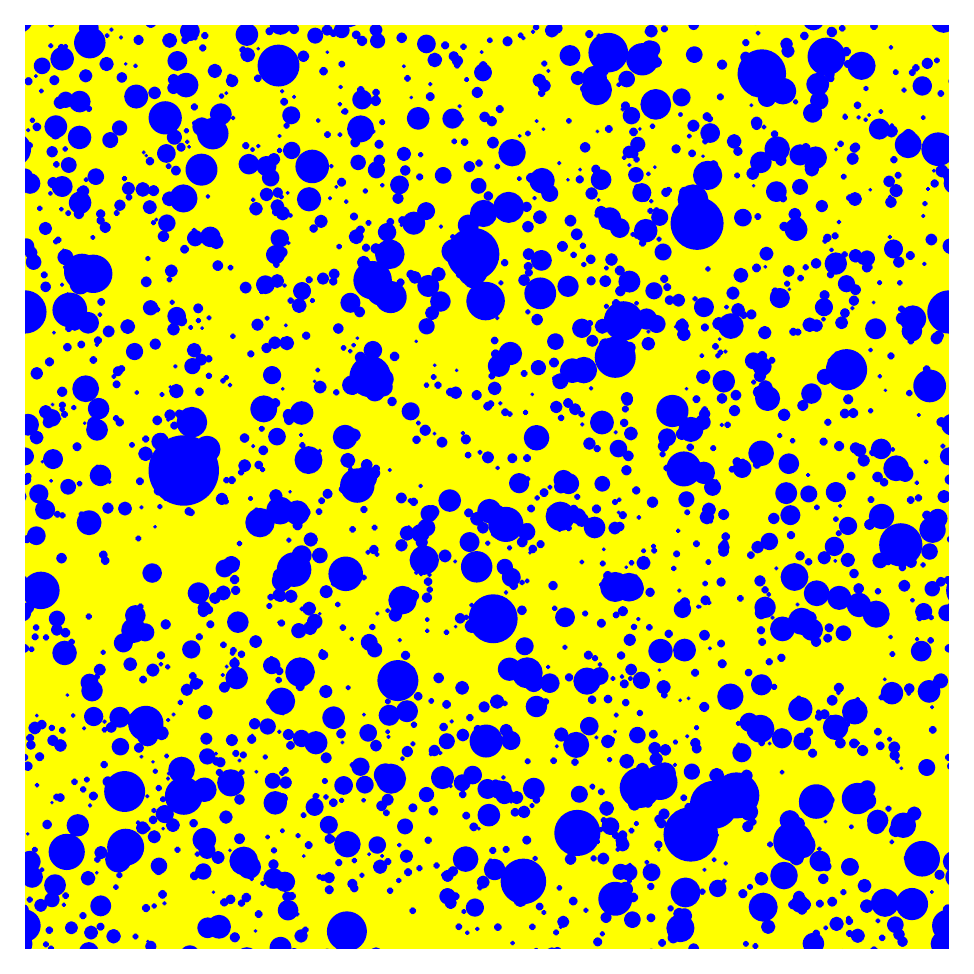}}
    	\subfloat[\label{fig:OPS2Dstructsi}]{\includegraphics[width=0.33\textwidth]{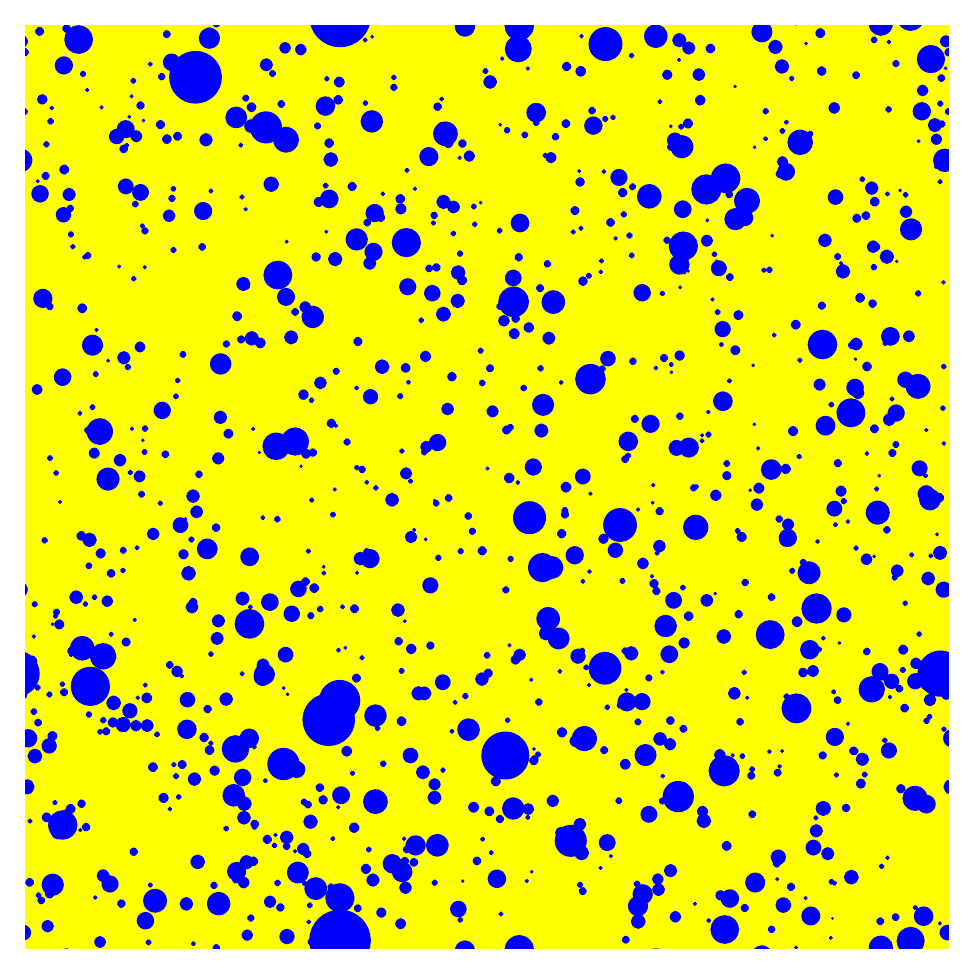}}
	\caption{Realizations of 2D OPS-DRM. Images (a)-(i) correspond to void phase (yellow) volume fraction $\phi_1=0.1-0.9$ and disk phase (blue) volume fraction $\phi_2=0.9-0.1$, respectively.}\label{fig:OPS2Dstructs}
\end{figure*}

In this section, we show that $S_2^{(1)}(r)$ for overlapping, polydisperse spheres with exponentially distributed radii [described by Eq. \eqref{eqn:OPSs2}] is an excellent approximation of the exponentially decaying $S_2(r)$ of Debye random media, defined by Eq. \eqref{eqn:debS2First}, across the first three space dimensions. The analytically known two-point correlation function for the void phase of OPS-DRM in dimensions $1,2,$ and $3$ are plotted in Figs. \ref{fig:S2plots}\subref{fig:S2plotsa}, \ref{fig:S2plots}\subref{fig:S2plotsc}, and \ref{fig:S2plots}\subref{fig:S2plotse}, respectively. Analogous results for the numerically sampled two-point correlation function for the sphere phase of OPS-DRM are plotted in Figs. \ref{fig:S2plots}\subref{fig:S2plotsb}, \ref{fig:S2plots}\subref{fig:S2plotsd}, and \ref{fig:S2plots}\subref{fig:S2plotsf}.

We measure the discrepancies between the $S_2^{(i)}(r)$ for OPS-DRM and Eq. \eqref{eqn:debS2First} using the following error estimate generalized from Ref. \cite{DeltaS208}:
\begin{equation}
    \Delta f_2(r) = \frac{1}{N_L} \sum_r | \delta f(r) |,
\end{equation}
where $f_2(r)$ is a two-point descriptor, $N_L$ is the number of sampling bins, and $\delta f(r)$ is the difference between the two functions being compared. We specifically found that $10^{-5}<\Delta S_2^{(1)}(r)<10^{-4}$ and $10^{-4}<\Delta S_2^{(2)}(r)<10^{-3}$, which are both sufficiently small \cite{DeltaS208,drm2020}. In summary, these results indicate that, while Eq. \eqref{eqn:OPSs2} is not mathematically symmetric under $\phi_1\to\phi_2$, OPS-DRM has effective phase inversion symmetry at the two-point level.

Recall from Sec. \ref{sec:YT} that actual Debye random media has phase inversion symmetry at the two-point level. Also note that because the forms of Eqs. \eqref{eqn:debS2First} and \eqref{eqn:S2PDS} are distinct, a fitting procedure must be employed to determine the mean radius $\langle R \rangle$, which yields an OPS system with effective characteristic length $a$ for a given $\phi_1$. These values of $\langle R \rangle$ were determined using a least-squares optimization scheme and then fitted to the exponentially damped power law
\begin{equation}
    \langle R \rangle(\phi_1) = a_1 e^{-a_2\phi_1}\phi_1^{-a_3} + a_4\label{eqn:eplw}
\end{equation}
in order to interpolate values of $\langle R \rangle $ for $\phi_1\in[0,1]$. The parameters $a_1,a_2,a_3,$ and $a_4$ for dimensions $1,2,$ and $3$ are listed in Table \ref{tab:Rmvals}. Samples of OPS-DRM microstructures in 2D for different volume fractions are presented in Fig. \ref{fig:OPS2Dstructs}. Interestingly, for $\phi_1<1/2$ the void space of OPS-DRM is filamentous while that of YT-DRM consists of more compact regions. For $\phi_1>1/2$, we see that OPS-DRM has a wide range of inclusion sizes, whereas those in YT-DRM are more uniformly distributed in size; see Figs. \ref{fig:OPS2Dstructs}\subref{fig:OPS2Dstructsi} and \ref{fig:YT2Dstructs}\subref{fig:YT2Dstructsi}, respectively.

\begin{table}
\caption{\label{tab:Rmvals}%
Parameters for Eq. \eqref{eqn:eplw} for dimensions $1,2,$ and $3$. These values were computed for OPS-DRM with characteristic length $a=0.2$ and system side-length $L=20$.}
\begin{ruledtabular}
\begin{tabular}{cccc}
$a_n$&
$d=1$&
$d=2$&
$d=3$\\
\colrule
$a_1$ & 0.05063 & 0.04399 & 0.03015\\
$a_2$ & 1.23419 & 1.96841 & 1.19292\\
$a_3$ & 0.41971 & 0.33559 & 0.39660\\
$a_4$ & 0.08573 & 0.07204 & 0.05611
\end{tabular}
\end{ruledtabular}
\end{table}

\section{\label{sec:cpdrm}{Debye Random Media with Compact Pores}}

In this section, we introduce a class of Debye random media whose pore-size probability density function is constrained to have compact support as follows:
\begin{equation}
    P(\delta) = (A - m\delta)\Theta\left( \frac{A}{m} - \delta \right)\label{eqn:CP_Pd}.
\end{equation}
The parameter $A$ must be equal to $s/\phi_1$ from the condition that $P(0)=s/\phi_1$ where the specific surface $s$ for $d$-dimensional Debye random media is given by
\begin{equation}
    s = \frac{\omega_d d \phi_1\phi_2}{\omega_{d-1}a}.\label{eqn:drmss}
\end{equation}
The slope $m$ must equal $\phi_1^2/(2s^2)$ per the normalization condition on $P(\delta)$. The complementary cumulative distribution function corresponding to \eqref{eqn:CP_Pd} is given by
\begin{equation}
    F(\delta) = \left( \frac{s\delta - 2\phi_1}{2\phi_1} \right)^2 \Theta\left( \frac{2\phi_1}{s} - \delta \right)\label{eqn:CP_Fd}
\end{equation}
using relation \eqref{eqn:defFd}. Also note that the $n$th moment of \eqref{eqn:CP_Pd} is given by
\begin{equation}
    \langle \delta^n \rangle = \frac{2^{n+1}}{2+3n+n^2}\left( \frac{\phi_1}{s} \right)^n\label{eqn:CP_dn}
\end{equation}
from relation \eqref{eqn:mps}. The critical feature of $P(\delta)$ and $F(\delta)$ for this class of Debye random media is that they are equal to zero for $\delta>\Lambda$ where the pore-size cutoff $\Lambda=2\phi_1/s$. Moreover, this cutoff makes the pore regions of such structures more compact (see Sec. \ref{sec:comp_porefuncs}).

We realized CP-DRM in 2D using the accelerated Yeong-Torquato procedure with $S_2^{(1)}(r)$ constrained to be Eq. \eqref{eqn:debS2First} and $F(\delta)$ constrained to be Eq. \eqref{eqn:CP_Fd}. For the simulated annealing energy function \eqref{eqn:YTenergy}, the weight $w_{F(\delta)}$ was chosen such that $w_{F(\delta)}E_{F(\delta)}=E_{S_2(r)}$ for the initial configuration. Ten configurations of CP-DRM were made for each volume fraction $\phi_1=0.1,0.2,...,0.9$. The ``pixel-refinement phase" was also utilized for all constructions. To sample $F(\delta)$, we treated every pixel as a pore center to ensure that the pore space of the final structure was completely consistent with Eq. \eqref{eqn:CP_Fd}. Additionally, when updating $F(\delta)$, we considered only pores that included the swapped pixels to improve performance.

\begin{figure}
	    \centering
	    \subfloat[\label{fig:CPVERIFa}]{\includegraphics[width=0.9\columnwidth]{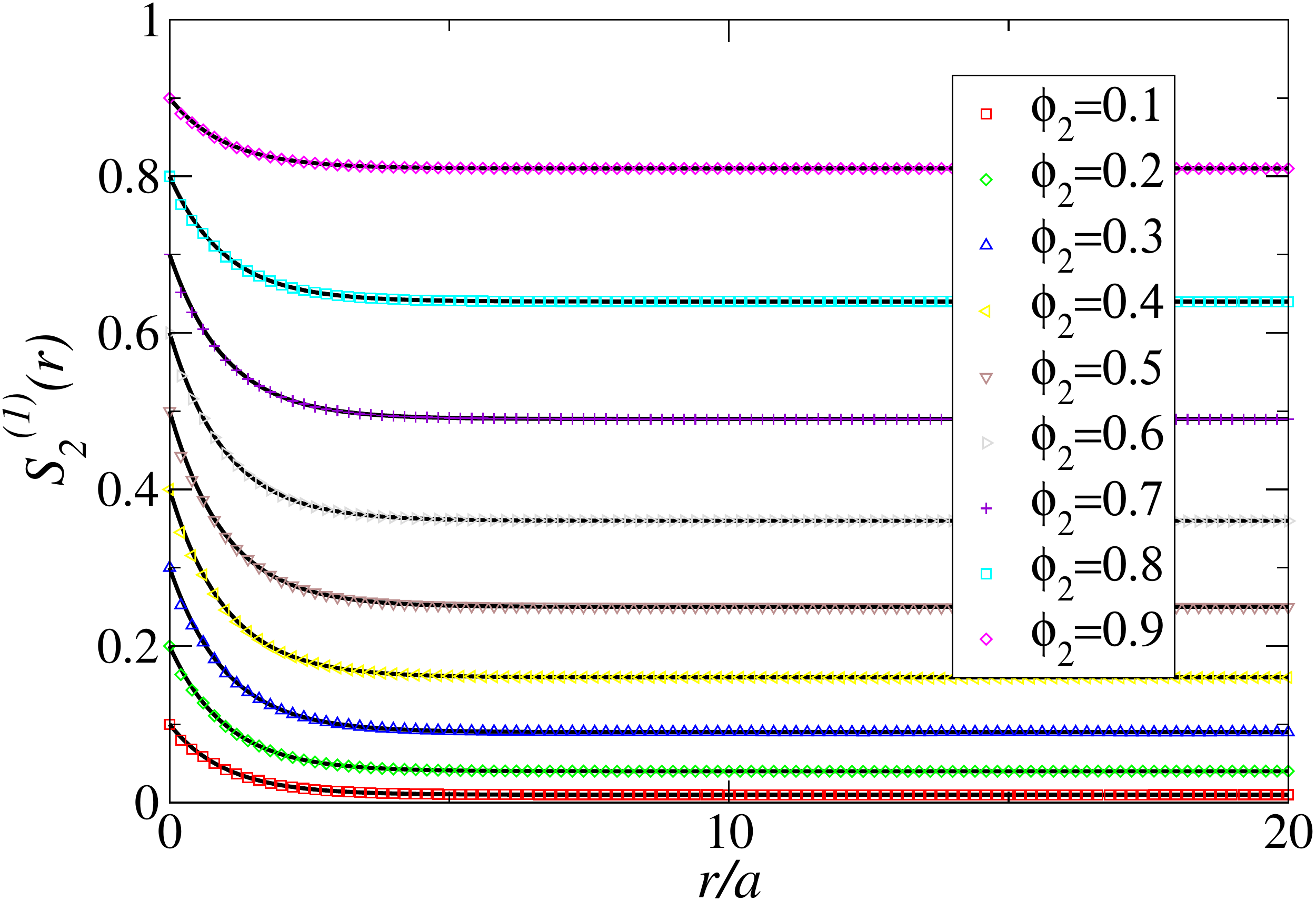}}
	    \hfill
	    \subfloat[\label{fig:CPVERIFb}]{\includegraphics[width=0.9\columnwidth]{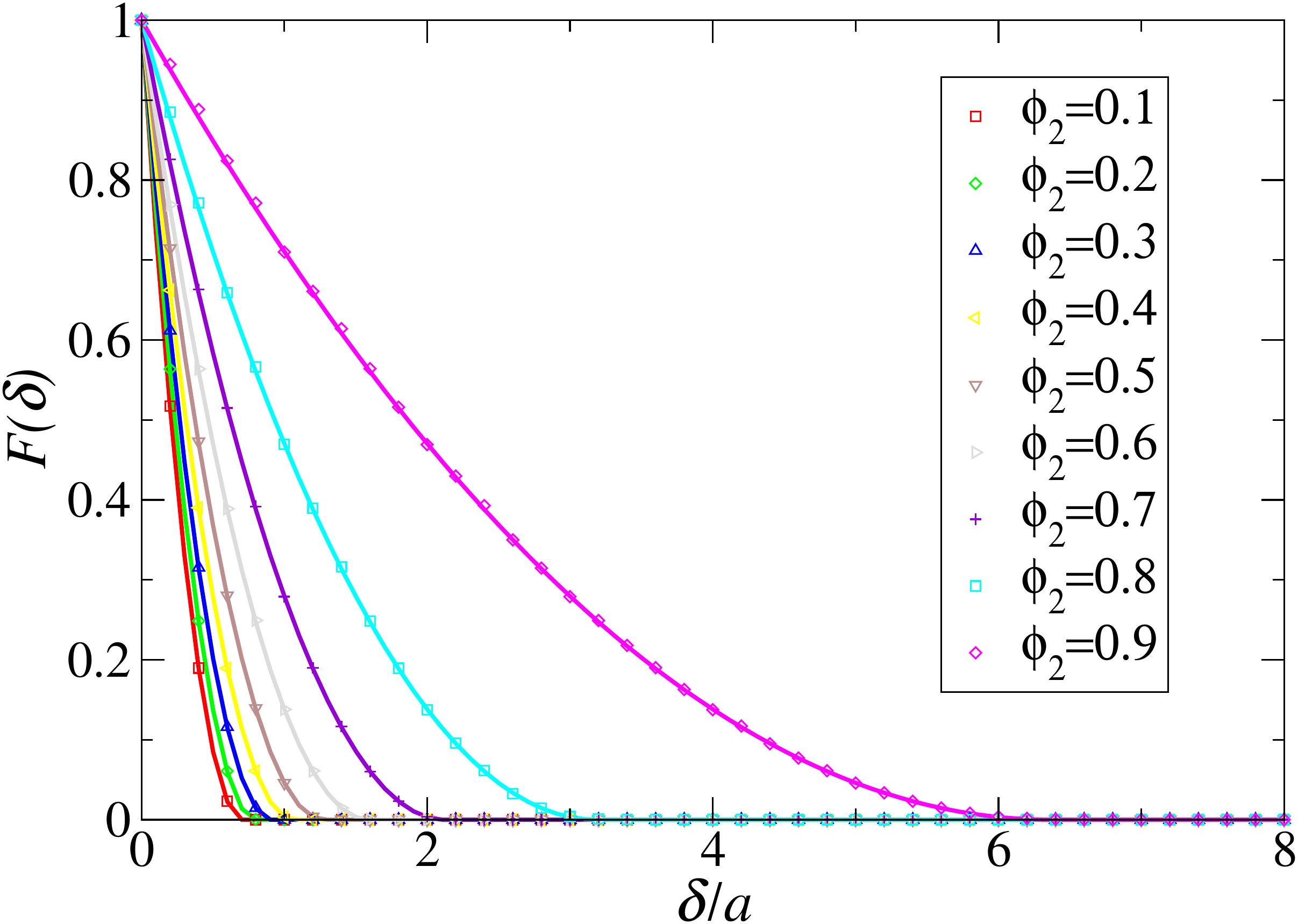}}
	    \caption{(a) Plots of $S_2^{(1)}(r)$ for several $\phi_1$ where the scatter plots are numerically sampled from realizations of 2D CP-DRM and the solid lines are of Eq. \eqref{eqn:debS2First}. (b) Analogous plots of $F(\delta)$, but the solid lines are of Eq. \eqref{eqn:CP_Fd}. \label{fig:CPVERIF}} 
\end{figure}

\begin{figure*}
	    \subfloat[\label{fig:LIN2Dstructsa}]{\includegraphics[width=0.33\textwidth]{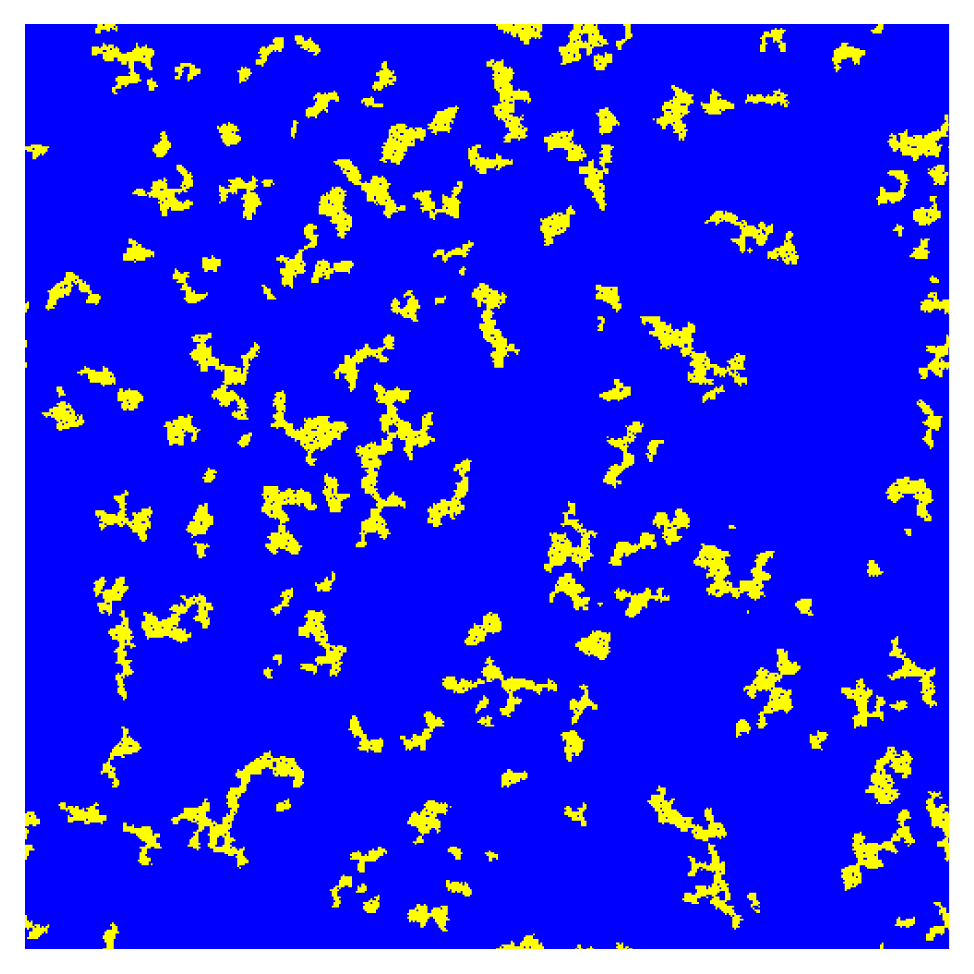}}
	    \subfloat[\label{fig:LIN2Dstructsb}]{\includegraphics[width=0.33\textwidth]{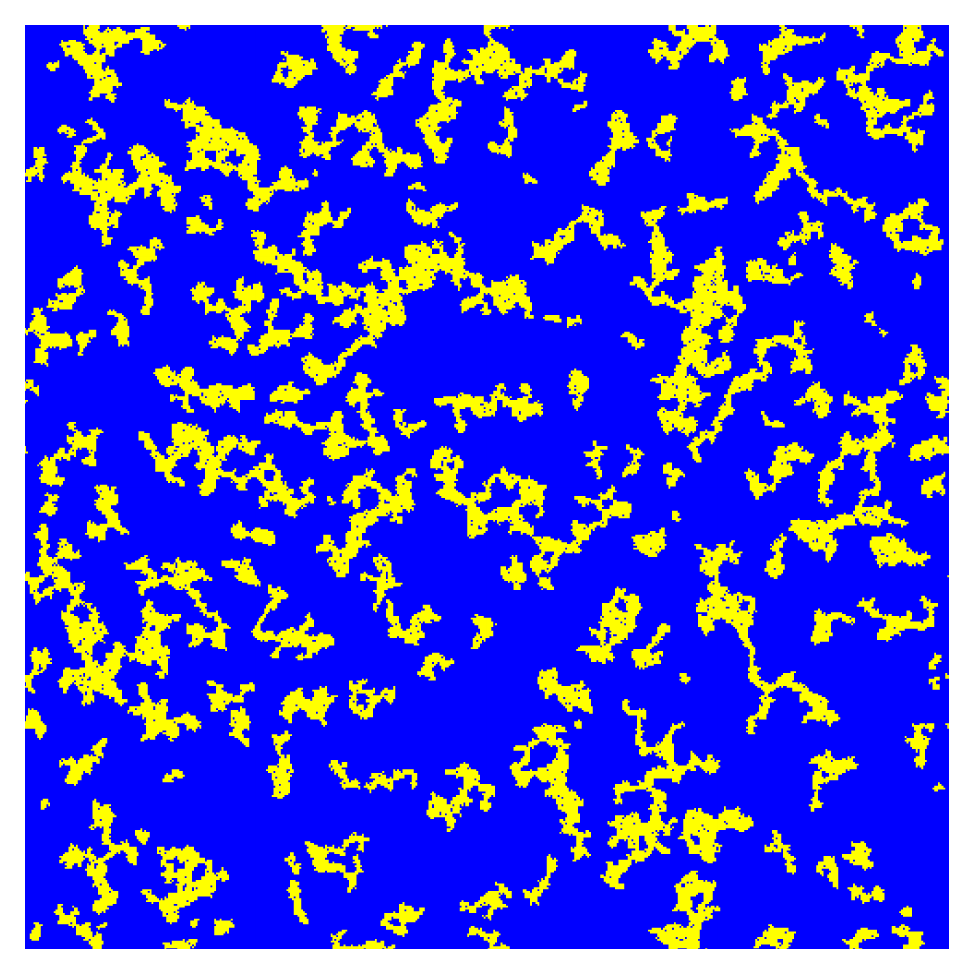}}
	    \subfloat[\label{fig:LIN2Dstructsc}]{\includegraphics[width=0.33\textwidth]{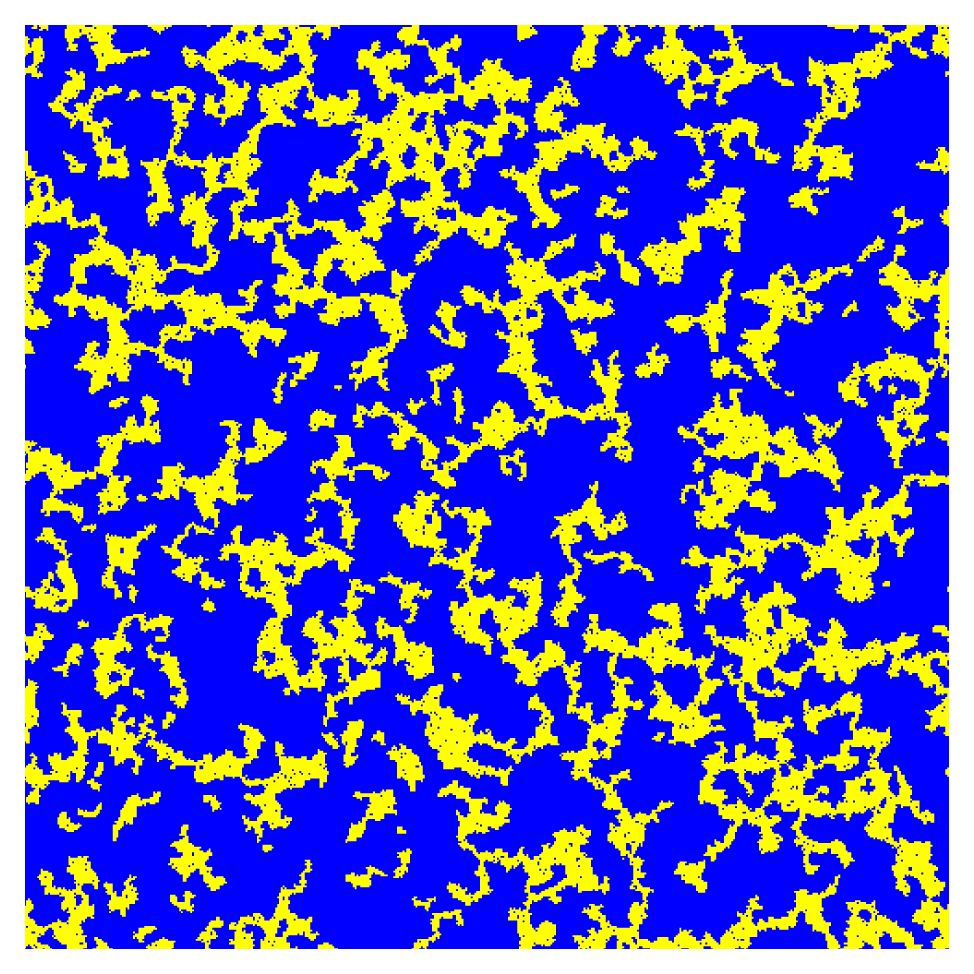}}
	    \hfill
    	\subfloat[\label{fig:LIN2Dstructsd}]{\includegraphics[width=0.33\textwidth]{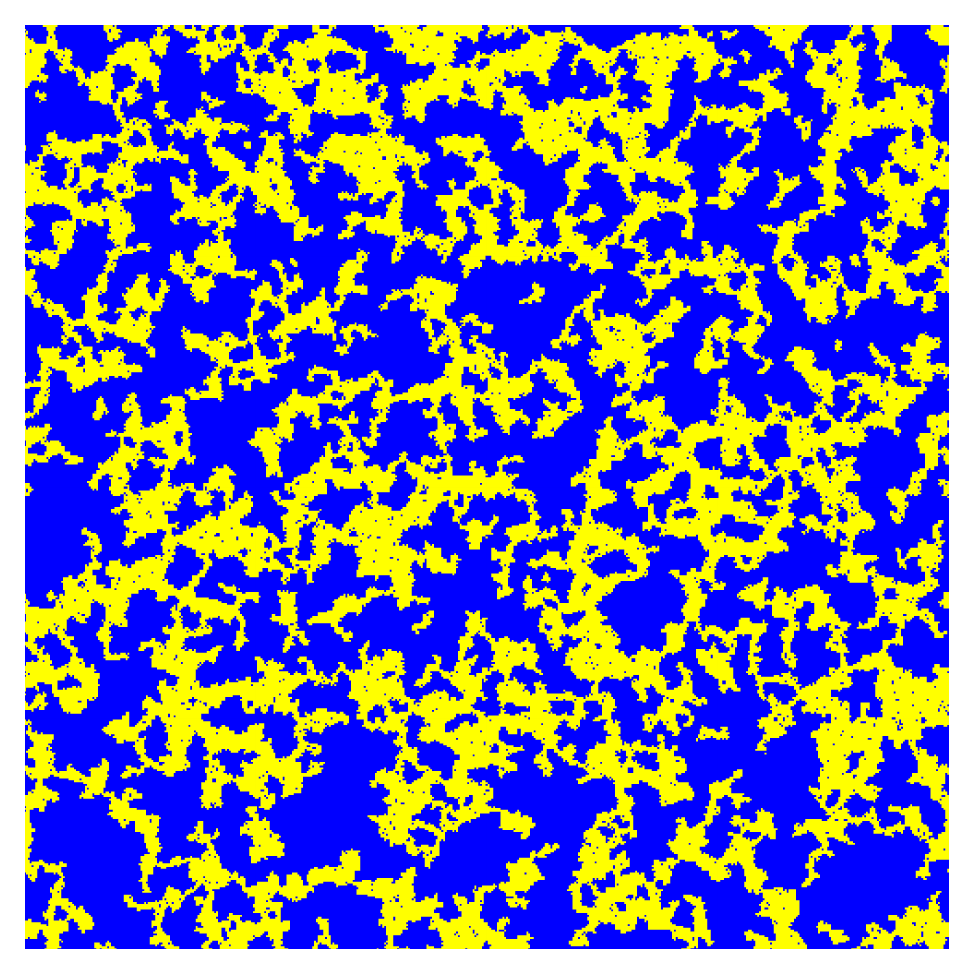}}
    	\subfloat[\label{fig:LIN2Dstructse}]{\includegraphics[width=0.33\textwidth]{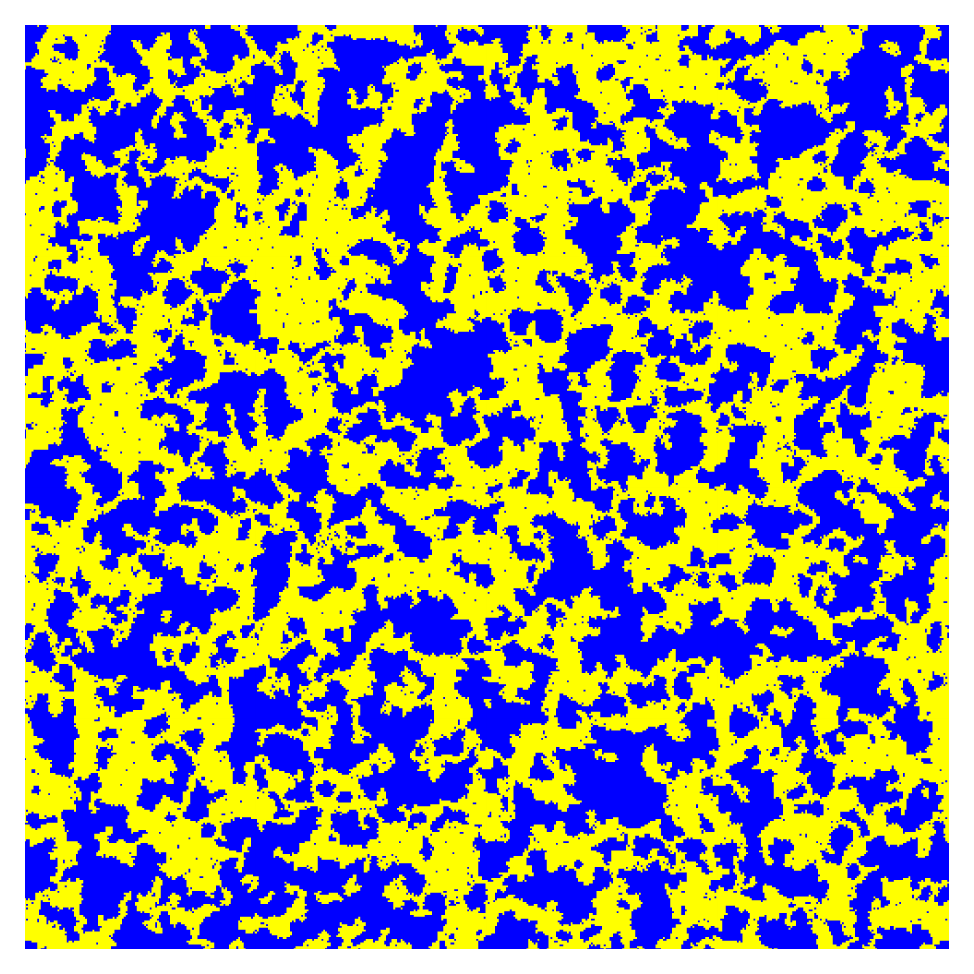}}
    	\subfloat[\label{fig:LIN2Dstructsf}]{\includegraphics[width=0.33\textwidth]{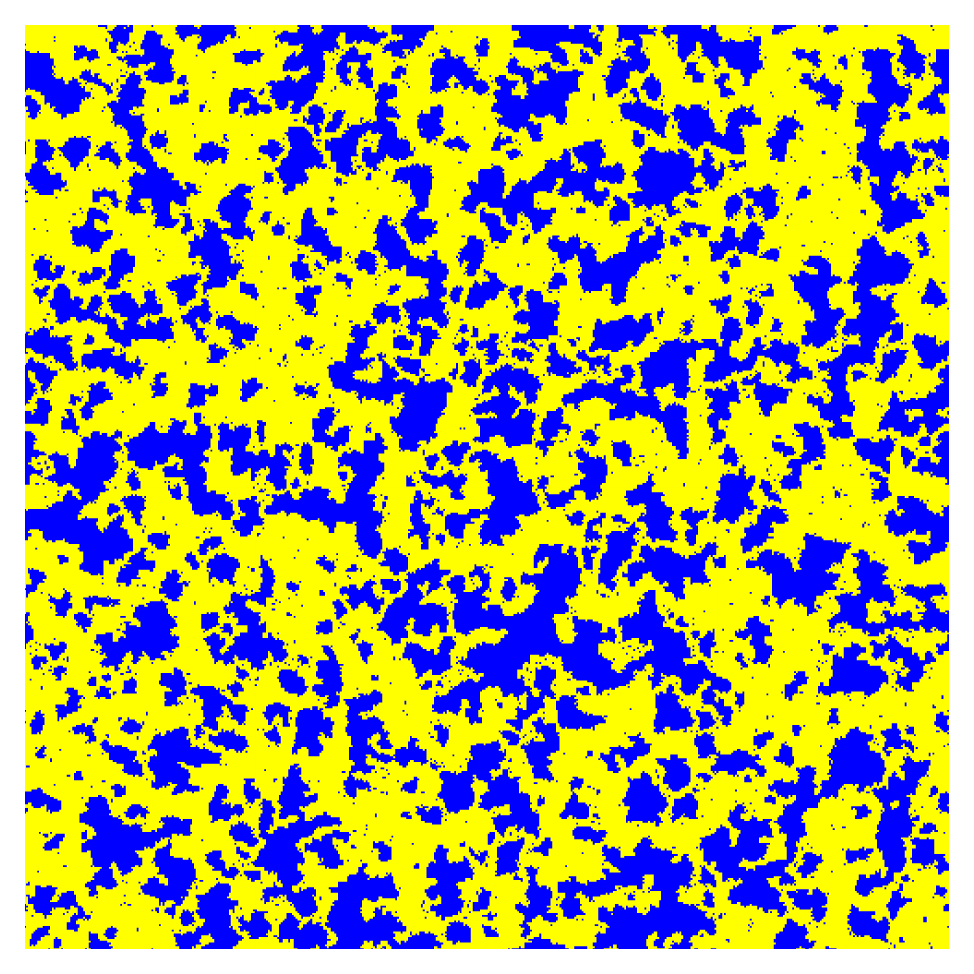}}
    	\hfill
    	\subfloat[\label{fig:LIN2Dstructsg}]{\includegraphics[width=0.33\textwidth]{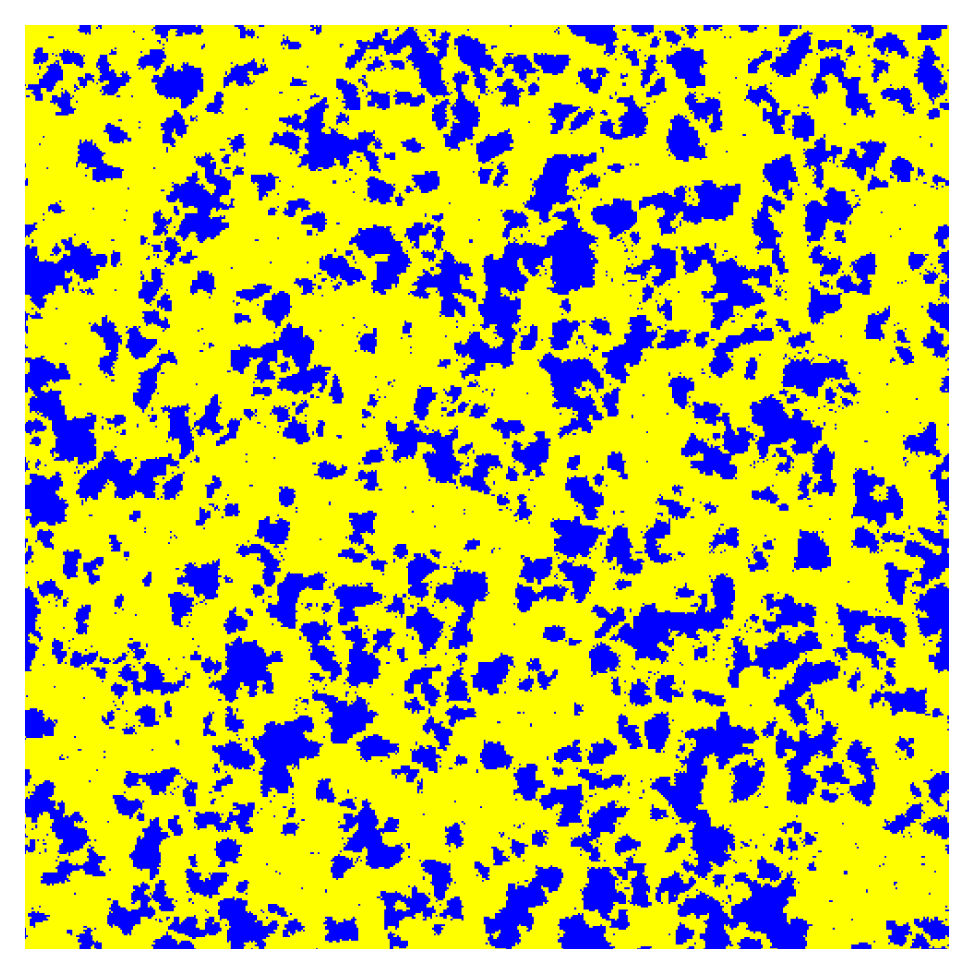}}
    	\subfloat[\label{fig:LIN2Dstructsh}]{\includegraphics[width=0.33\textwidth]{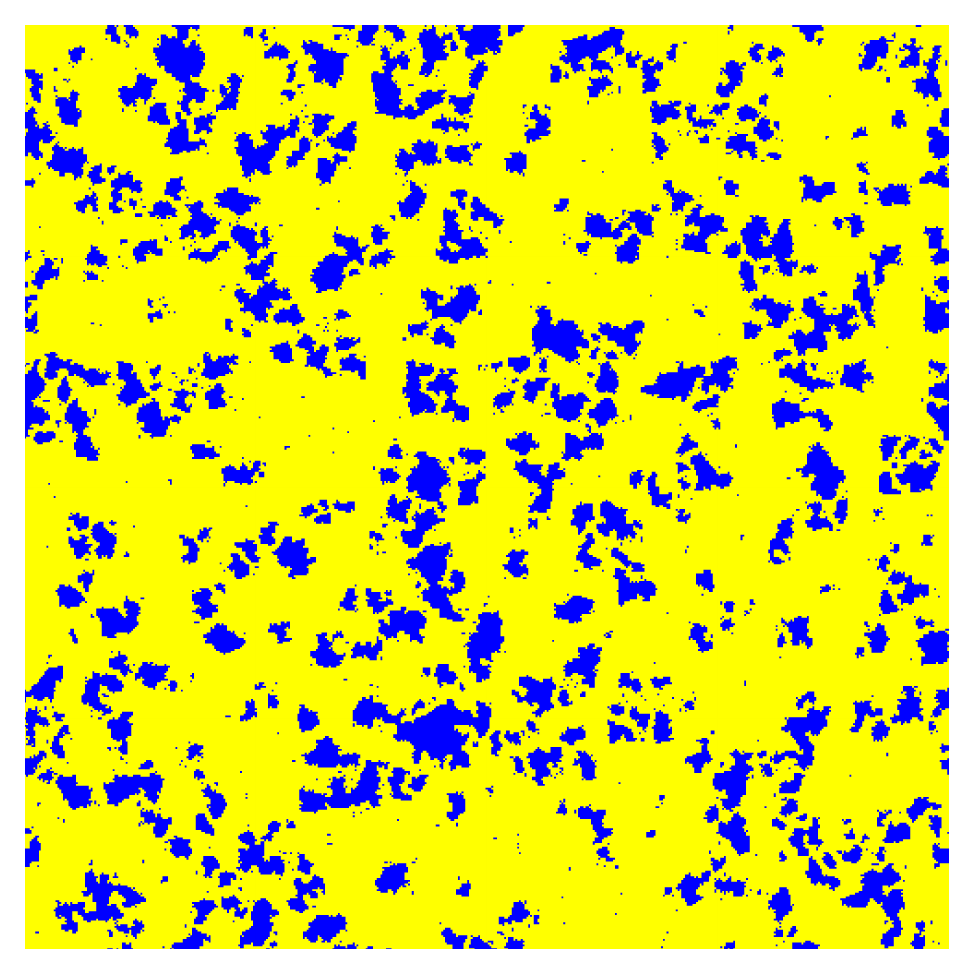}}
    	\subfloat[\label{fig:LIN2Dstructsi}]{\includegraphics[width=0.33\textwidth]{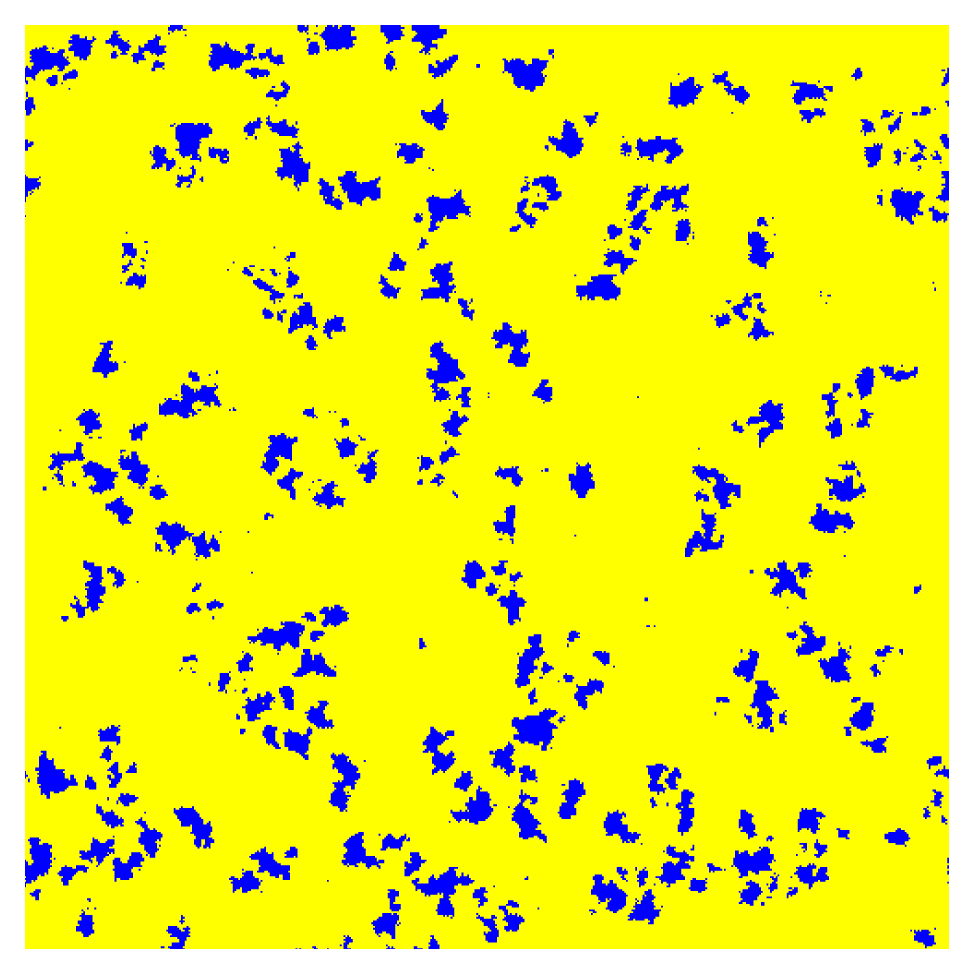}}
	\caption{Realizations of 2D CP-DRM. Images (a)-(i) correspond to void phase (yellow) volume fraction $\phi_1=0.1-0.9$ and inclusion phase (blue) volume fraction $\phi_2=0.9-0.1$, respectively. Once again following Ma and Torquato \cite{drm2020}, our configurations are $501\times501$ pixels with characteristic length $a=5$, and a cutoff $l_c=10a$ was used for sampling $S_2^{(1)}(r)$.}\label{fig:LIN2Dstructs}
\end{figure*}

Comparison of $S_2^{(1)}(r)$ sampled from our constructed CP-DRM to Eq. \eqref{eqn:debS2First} for various volume fractions in Fig. \ref{fig:CPVERIF}\subref{fig:CPVERIFa} confirms that these systems are in fact Debye random media [$10^{-5}<\Delta S_2(r)<10^{-4}$]. In Fig. \ref{fig:CPVERIF}\subref{fig:CPVERIFb}, analogous plots of sampled $F(\delta)$ against Eq. \eqref{eqn:CP_Fd} indicate that our constructed systems completely satisfy the prescribed pore-size statistics [$10^{-5}<\Delta F(\delta)<10^{-4}$]. Selected constructed configurations of this class of Debye random media for different volume fractions are presented in Fig. \ref{fig:LIN2Dstructs}. For $\phi_1<0.4$, note how the void spaces of these microstructures are more elongated and channel-like. We observe similar features in OPS-DRM but not in YT-DRM; see Figs. \ref{fig:OPS2Dstructs} and \ref{fig:YT2Dstructs}, respectively.

\section{\label{sec:comparison}Comparison of YT-DRM and OPS-DRM Microstructures}

In order to probe how $S_2$-degenerate two-phase media differ in their other microstructural statistics, we compute and compare a set of alternative two-point descriptors [i.e., $F_{ss}(r),F_{sv}(r),P(\delta),L(z)$ and $p(z)$] for YT-DRM, OPS-DRM and CP-DRM in 2D. In 3D, while we know all of these descriptors for OPS-DRM, we know only $F_{sv}(r)$ and $F_{ss}(r)$ for YT-DRM and $P(\delta)$ for CP-DRM. As such, our exploration of the effect of dimension on the degeneracy problem is limited to these descriptors.

\subsection{\label{sec:comp_svfuncs}Surface correlation functions}

In 2D and 3D, the specific surface for OPS-DRM is given by
\begin{equation}
    s = \frac{\eta\phi_1}{\langle R \rangle},
\end{equation}
which contrasts $s$ for YT-DRM and CP-DRM which is equal to $\pi\phi_1\phi_2/a$ in 2D and $4\phi_1\phi_2/a$ in 3D [see Eq. \eqref{eqn:drmss}]. For the surface-void and surface-surface correlation functions for the overlapping polydisperse sphere systems, we employ the canonical correlation function formalism \cite{torquato86} and find that for 2D structures
\begin{equation}
    F_{sv}(r) = 2\pi\rho\left\langle R - \frac{R}{\pi}\cos^{-1}\left( \frac{r}{2R} \right)\Theta(2R-r) \right\rangle S_2^{(1)}(r)\label{eqn:opsdrmfsv2D}
\end{equation}
and
\begin{eqnarray}
    F_{ss}(r) = &&\Bigg[ \left\langle 2\rho R \left( \pi - \cos^{-1}\left(\frac{r}{2R}\right) \Theta(2R-r) \right) \right\rangle^2  \nonumber + \\ 
    &&\hspace{-1cm} \left\langle \frac{2\rho R \Theta(2R-r)}{r\sqrt{1-(r/2R)^2}} \right\rangle \Bigg] 
    S_2^{(1)}(r)\label{eqn:opsdrmfss2D}.
\end{eqnarray}
Note that these average value integrals must be computed numerically. For 3D, Lu and Torquato derived expressions for $F_{sv}(r)$ and $F_{ss}(r)$ for overlapping, polydisperse spheres for a general distribution of radii \cite{perm92}. Here, we evaluate these expressions for the $m=0$ Schulz distribution and find that
\begin{equation}
    F_{sv}(r) = \pi \langle R \rangle \rho \left[ 8 \langle R \rangle - e^{-r/2\langle R \rangle} (r + 4\langle R \rangle) \right] S_2^{(1)}(r)\label{eqn:opsdrmfsv3D}
\end{equation}
and
\begin{eqnarray}
    F_{ss}(r) = &&\frac{\pi\rho}{2}\Bigg[ \frac{e^{-r/2\langle R \rangle}}{r} \left( r^2 + 4r\langle R \rangle  + 8 \langle R \rangle^2 \right) + \nonumber \\
    &&\hspace{-2cm} 2\pi\rho\langle R \rangle^2 \left( e^{-r/2\langle R \rangle}(r+4\langle R \rangle) - 8\langle R \rangle \right)^2 \Bigg] S_2^{(1)}(r)\label{eqn:opsdrmfss3D}. 
\end{eqnarray}

For general-dimensional Debye random media realized via stochastic reconstruction, Ma and Torquato \cite{drm2020} proposed the following semi empirical forms for the surface-correlation functions:
\begin{equation}
    F_{sv}(r) = \frac{s}{\phi_1} \frac{1}{1 + e^{-r/a}}S_2^{(1)}(r)\label{eqn:ytdrmfsv},
\end{equation}
and
\begin{equation}
    F_{ss}(r) = s^2 + \frac{(d-1)\phi_1\phi_2}{ar}e^{-r/a} + \frac{|\phi_2-\phi_1|}{2a^2}\frac{e^{-r/a}}{1 + e^{-r/a}}.\label{eqn:ytdrmfss}
\end{equation}
These functions were originally fit using statistics sampled from realizations of 2D YT-DRM and then generalized to dimension $d$ using theoretical arguments presented in \cite{ma18}. We use the method developed by Ma and Torquato \cite{ma18} to sample $F_{sv}(r)$ and $F_{ss}(r)$ for 2D CP-DRM.

\begin{figure*}
    \centering
	\subfloat[\label{fig:Fsv2DCOMPa}]{\includegraphics[width=1.0\columnwidth]{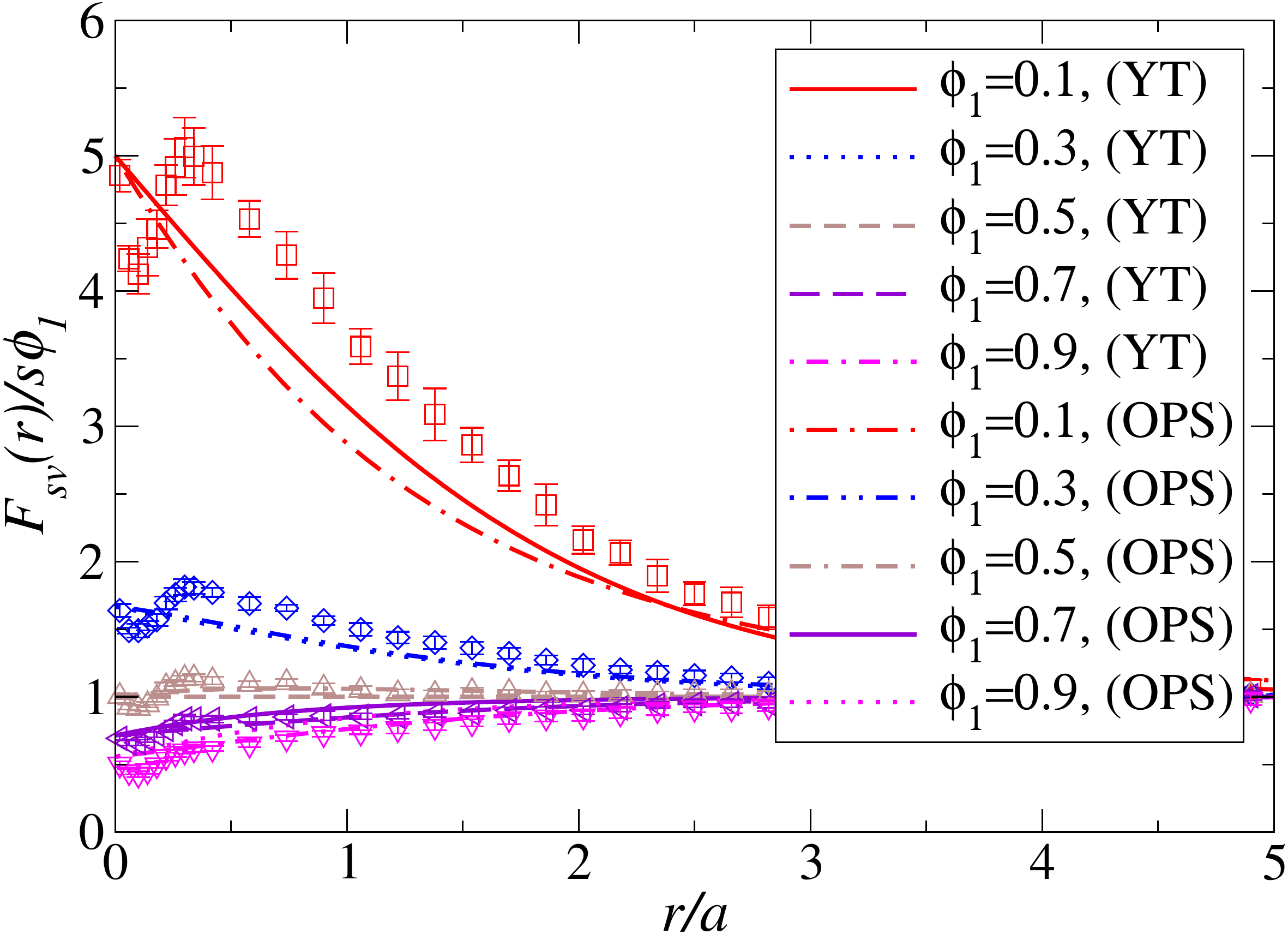}}
	\subfloat[\label{fig:Fsv2DCOMPb}]{\includegraphics[width=1.0\columnwidth]{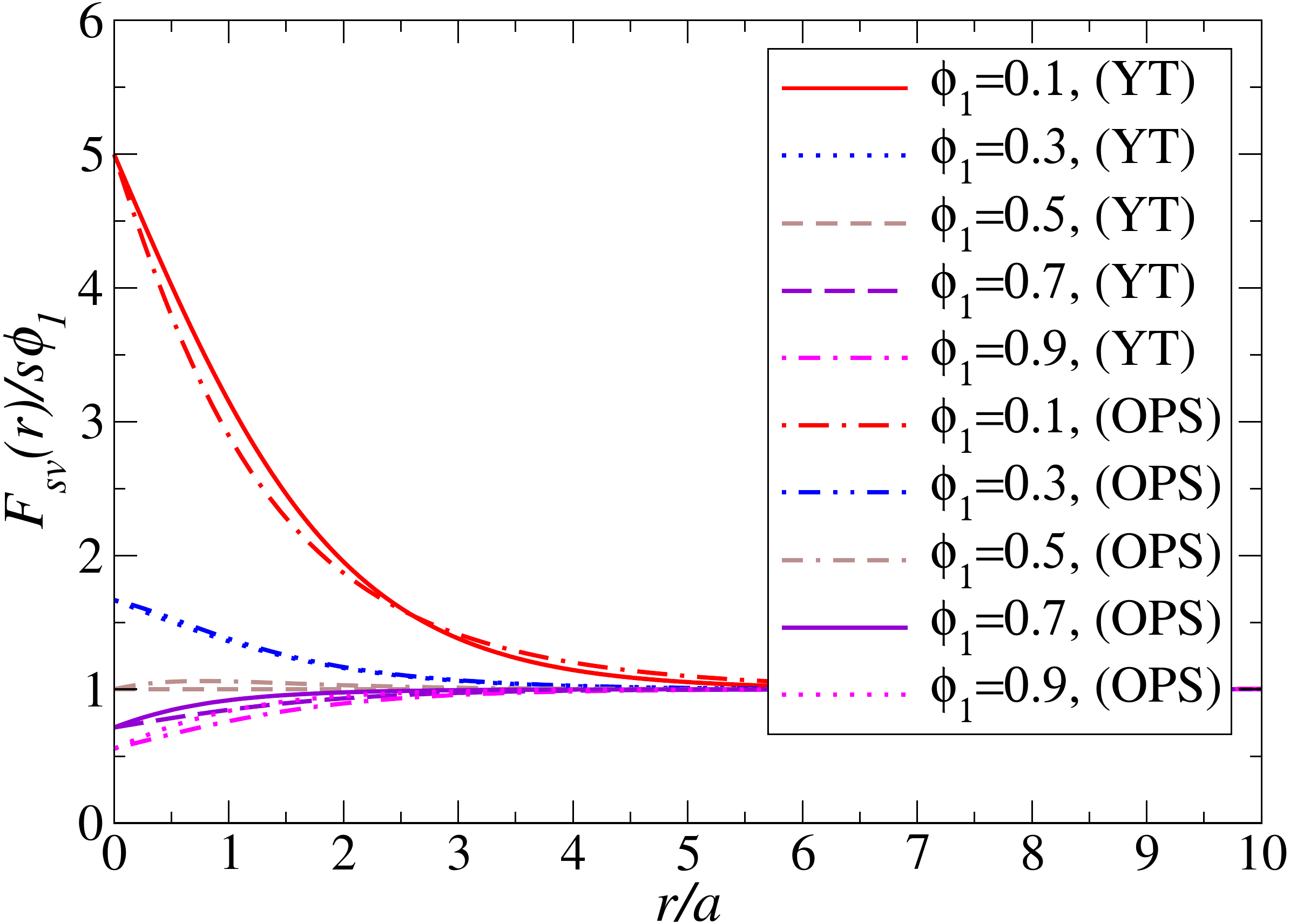}}
	\hfill
	\centering
	\subfloat[\label{fig:Fsv2DCOMPc}]{\includegraphics[width=1.0\columnwidth]{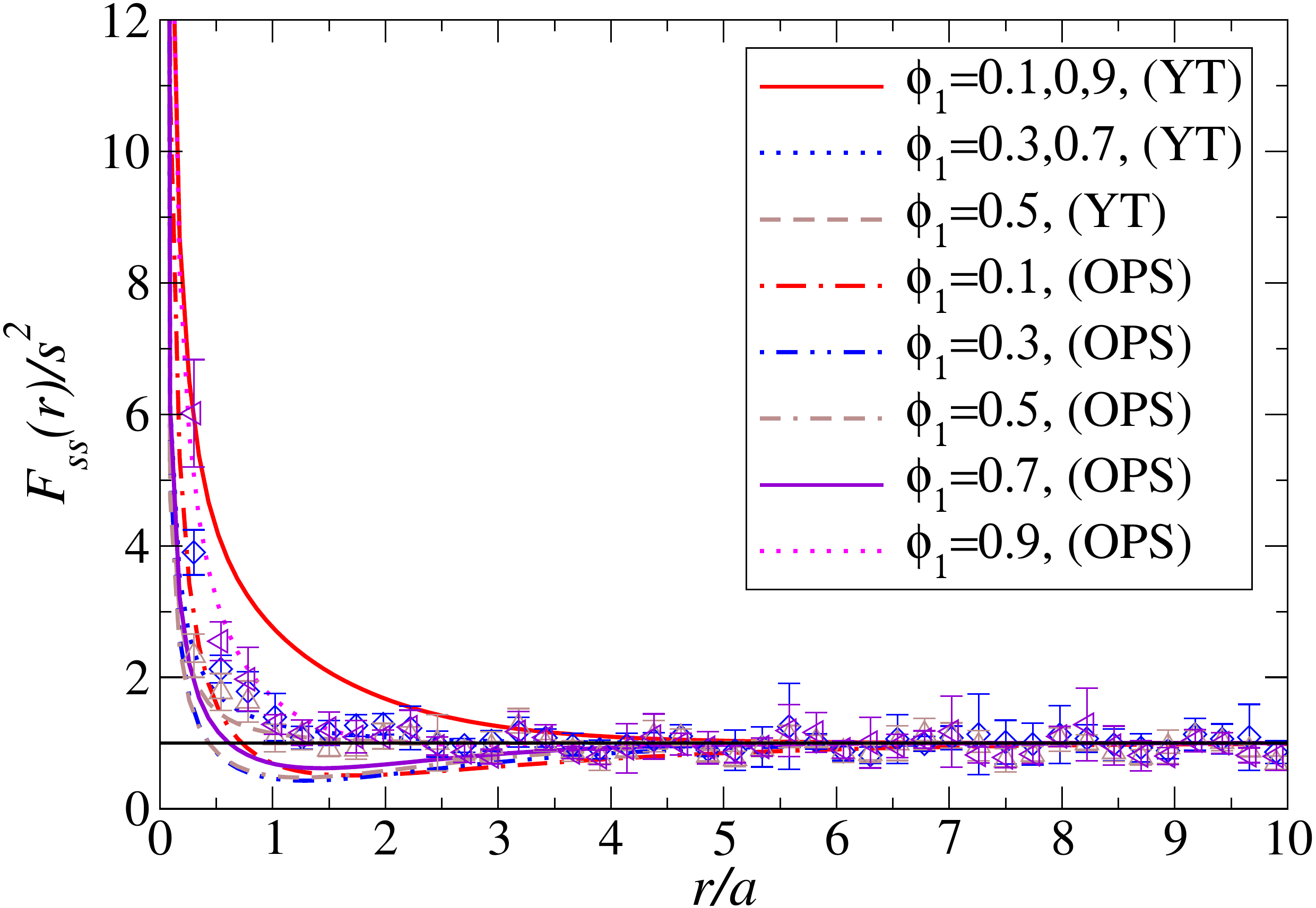}}
	\subfloat[\label{fig:Fsv2DCOMPd}]{\includegraphics[width=1.0\columnwidth]{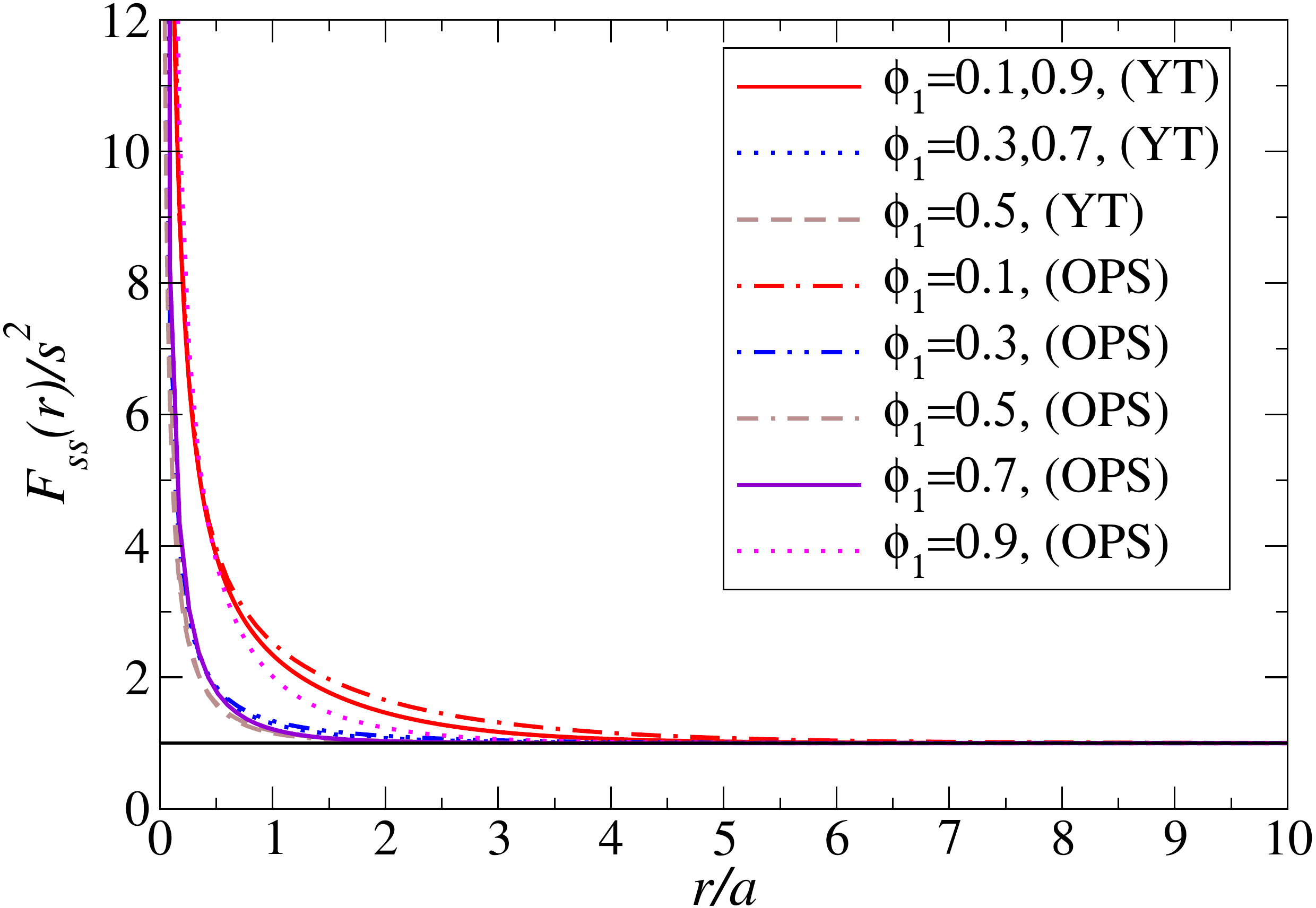}}
	\caption{Plots of the surface-void for 2D (a) and 3D (b) and surface-surface for 2D (c) and 3D (d) correlation functions for the three different classes of Debye random media. For YT-DRM, $F_{sv}(r)$ and $F_{ss}(r)$ are given by Eqs. \eqref{eqn:ytdrmfsv} and \eqref{eqn:ytdrmfss}, respectively. $F_{sv}(r)$ and $F_{ss}(r)$ for the OPS class in 2D are given by Eqs. \eqref{eqn:opsdrmfsv2D} and \eqref{eqn:opsdrmfss2D}, respectively, and by Eqs. \eqref{eqn:opsdrmfsv3D} and \eqref{eqn:opsdrmfss3D} for 3D OPS-DRM. Open symbols are the results for CP-DRM and are color-coded according to volume fraction. In the plots of $F_{ss}(r)$, the horizontal black line is meant to aid in visualization of the asymptotic value. In (a), the range of $r$ values is half of that used in the other subplots in order to facilitate viewing the fine features of $F_{sv}(r)$ for CP-DRM around the origin. In (c), for the sake of clarity, the results for $\phi_1=0.1,0.9$ have been omitted due to larger error bars that obscure the curves for the other volume fractions.}\label{fig:Fsv2DCOMP}
\end{figure*}

In Figs. \ref{fig:Fsv2DCOMP}\subref{fig:Fsv2DCOMPa} and \ref{fig:Fsv2DCOMP}\subref{fig:Fsv2DCOMPc}, plots of $F_{sv}(r)$ and $F_{ss}(r)$ \cite{Fssnote} for the three different classes of 2D Debye random media are shown. We see that $F_{sv}(r)$ is a monotonically decreasing function of $r$ only for YT-DRM, but that it is otherwise similar to $F_{sv}(r)$ for OPS-DRM. Additionally, $F_{sv}(r)$ is flat only for YT-DRM at $\phi_1=1/2$, and thus the Euler characteristic for this class of Debye random media is equal to zero when $\phi_1=\phi_2$ [see relation \eqref{eqn:euler}]. This behavior is related to the percolation threshold of YT-DRM (see Sec. \ref{sec:percolation}). Most notably, $F_{sv}(r)$ for CP-DRM has a negative slope at the origin and a local minimum for each volume fraction considered. Note also that $F_{ss}(r)$ is monotonically decreasing and symmetric under the transformation $\phi_1\to(1-\phi_1)$ for both YT-DRM and CP-DRM, whereas it has a minimum and no such symmetry for OPS-DRM. Lastly, the large error bars on the plot of $F_{ss}(r)$ for CP-DRM suggest that these structures posses a high degree of variability in their surface geometries. In Figs. \ref{fig:Fsv2DCOMP}\subref{fig:Fsv2DCOMPb} and \ref{fig:Fsv2DCOMP}\subref{fig:Fsv2DCOMPd}, plots of $F_{sv}(r)$ and $F_{ss}(r)$ for 3D OPS-DRM and YT-DRM are shown. For OPS-DRM, the $F_{sv}(r)$ curves are extremely similar to their 2D versions, while the $F_{ss}(r)$ curves become monotonically decreasing. It should be noted that in all plots, $F_{sv}(r)$ is scaled by $s\phi_1$ and $F_{ss}(r)$ by $s^2$ to bring their large-$r$ asymptotic values to unity.

\subsection{\label{sec:comp_porefuncs}Pore-size function}

Here, we compare the pore statistics of the three classes of Debye random media in 2D and 3D. Following Torquato \cite{Torq02,lu92pores}, we find that the pore-size probability density function $P(\delta)$ for Debye random media approximated by overlapping, polydisperse spheres in 2D is given by

\begin{eqnarray}
    P(\delta) = &&\frac{2\pi\rho}{\phi_1}(\langle R \rangle + \delta) \times \nonumber \\
    &&\hspace{-1cm}\exp\left[ -\pi\rho(\delta^2 + 2\delta\langle R \rangle  + 2\langle R \rangle^2) \right]\label{eqn:PdOPS2D}.
\end{eqnarray}
We use Eq. \eqref{eqn:mps} to compute the first and second moments of this distribution and find that they are
\begin{equation}
    \langle\delta\rangle = \frac{e^{-l^2}\erfc(l)}{2\phi_1\sqrt{\rho}}\label{eqn:mdeltaOPS2D},
\end{equation}
and
\begin{equation}
    \langle\delta^2\rangle = \frac{e^{-2l^2}\left( 1 - e^{l^2}\pi\langle R \rangle\sqrt{\rho}\erfc(l)\right)}{\phi_1\pi\rho},\label{eqn:delta2OPS2D}
\end{equation}
respectively, where $\erfc(x)$ is the complementary error function and $l=\sqrt{\pi\rho}\langle R\rangle$. Using a similar approach, we find that the pore-size probability density function for 3D OPS-DRM is given by
\begin{eqnarray}
    P(\delta) = &&\frac{4\pi\rho}{3\phi_1}(3\delta^2 + 6\delta\langle R \rangle^2 + 6\langle R \rangle^3) \times \nonumber \\
    &&\hspace{-1cm}\exp\left[ -\frac{4\pi\rho}{3}(\delta^3 + 3\delta^2\langle R \rangle + 6\delta\langle R \rangle^2 + 6\langle R \rangle^3) \right]\label{eqn:PdOPS3D}.
\end{eqnarray}
Numerical integration must be used to find $\langle\delta^n\rangle$ of Eq. \eqref{eqn:PdOPS3D} for $n\geq1$.

Ma and Torquato, guided by the scaled-particle theory \cite{Torq02,drm2020}, proposed the following form of $P(\delta)$ for Debye random media realized with the Yeong-Torquato procedure:
\begin{equation}
    P(\delta) = \left(\frac{\pi\phi_2}{a} + 2p_1\delta \right)\exp\left(-p_1\delta^2 - \frac{\pi\phi_2}{a}\delta\right) \label{eqn:PdYT2D}.
\end{equation}
Here, $p_1=(1.05\phi_2 - 2.41\phi_2^2 + 4.16\phi_2^3)/a^2$ is a free parameter whose value was determined by a fitting Eq. \eqref{eqn:PdYT2D} to simulated data. The first and second moments of Eq. \eqref{eqn:PdYT2D} are
\begin{equation}
    \langle\delta\rangle = \frac{1}{2} \sqrt{\frac{\pi}{p_1}} e^{k^2} \erfc\left( k \right)\label{eqn:mdeltaYT2D},
\end{equation}
and
\begin{equation}
    \langle\delta^2\rangle = \frac{1}{p_1} - \frac{\phi_2}{2a} \left( \frac{\pi}{p_1} \right)^{3/2} e^{k^2} \erfc(k) \label{eqn:delta2YT2D}
\end{equation}
respectively, where $k=\phi_2\pi/(2a\sqrt{p_1})$. 

\begin{figure}
    \centering
	\subfloat[\label{fig:pores2DCOMPa}]{\includegraphics[width=1.0\columnwidth]{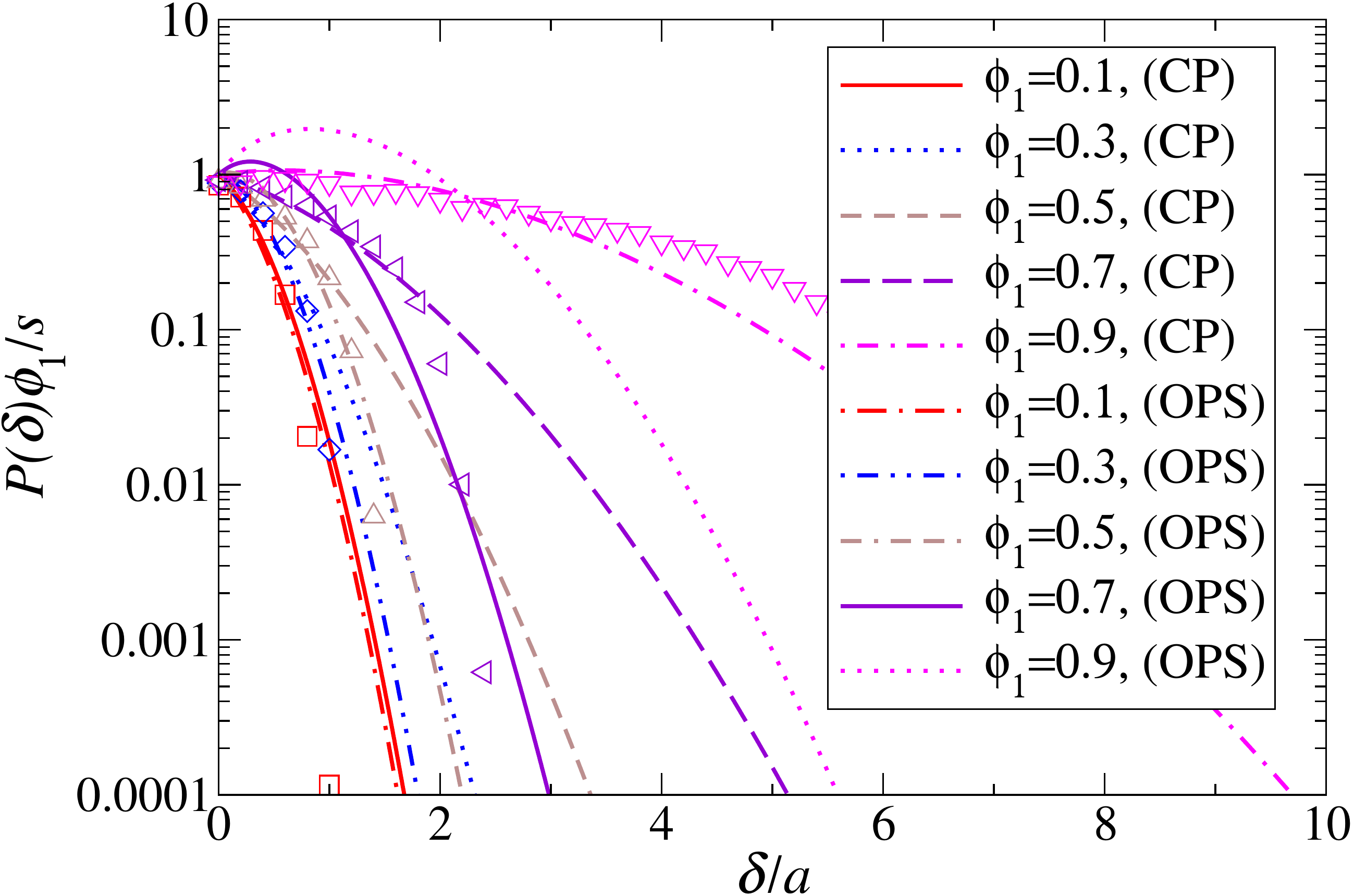}}
	\hfill
	\centering
	\subfloat[\label{fig:pores2DCOMPb}]{\includegraphics[width=1.0\columnwidth]{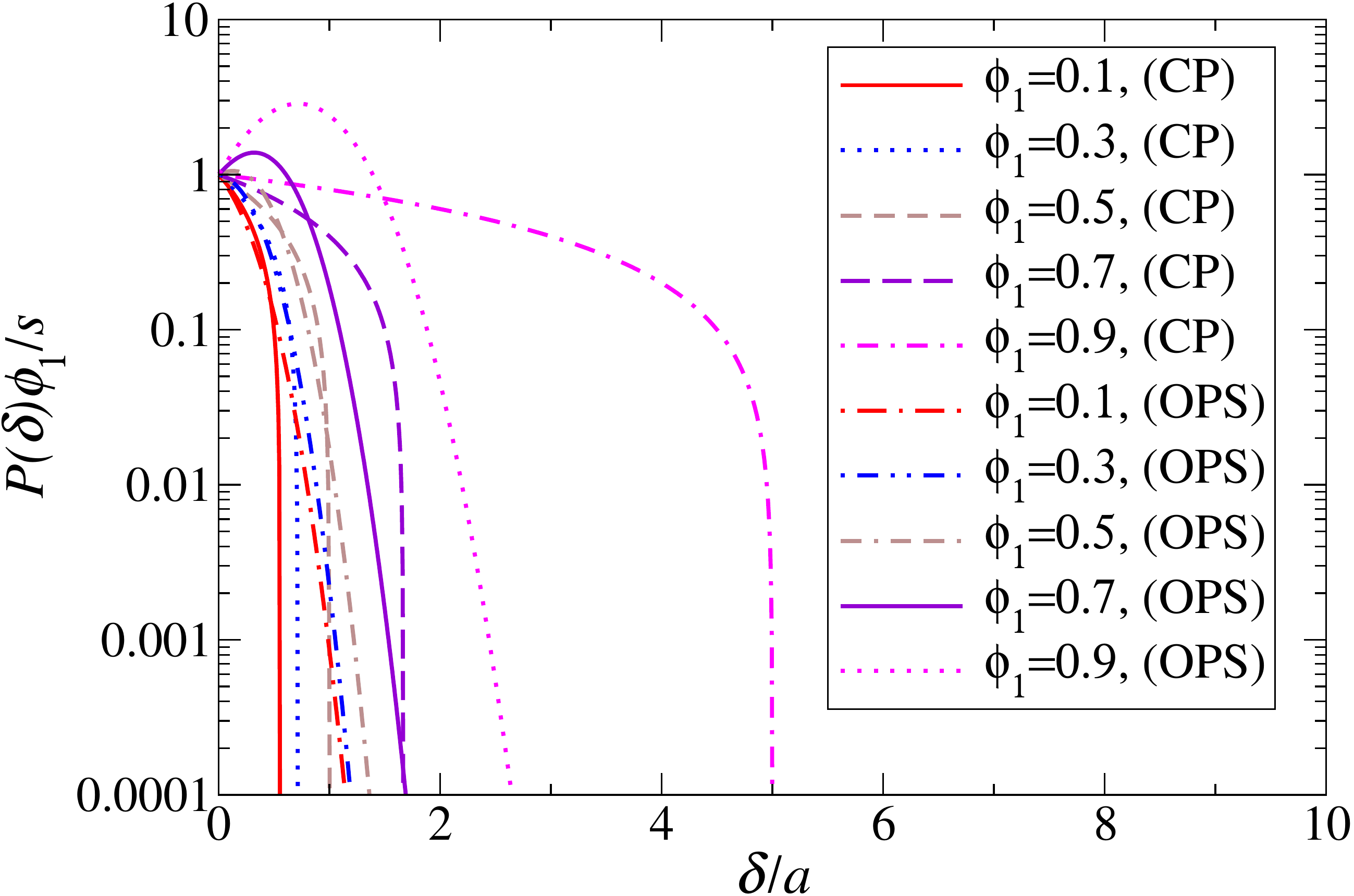}}
	\hfill
	\centering
	\subfloat[\label{fig:pores2DCOMPc}]{\includegraphics[width=1.0\columnwidth]{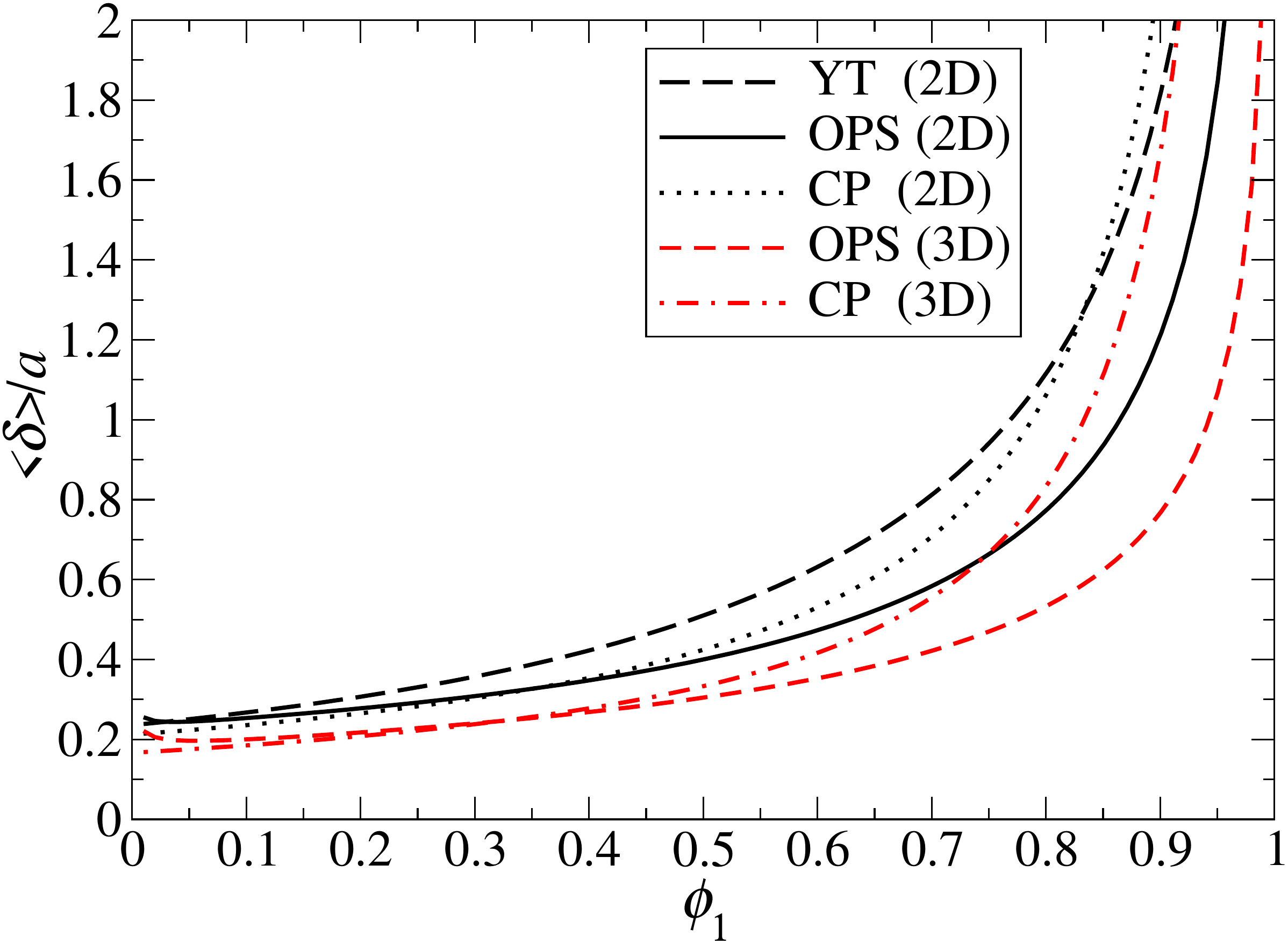}}
	\caption{Plots of the pore-size probability density function $P(\delta)$ for 2D (a) and 3D (b), as well as the mean pore size $\langle\delta\rangle$ as a function of $\phi_1$ (c). For YT-DRM, $P(\delta)$ is given by Eq. \eqref{eqn:PdYT2D} and $\langle\delta\rangle$ by Eq. \eqref{eqn:mdeltaYT2D}. For 2D OPS-DRM, these descriptors are given by Eqs. \eqref{eqn:PdOPS2D} and \eqref{eqn:mdeltaOPS2D}, respectively. For 3D OPS-DRM, $P(\delta)$ is given by Eq. \eqref{eqn:PdOPS3D} and $\langle\delta\rangle$ was computed numerically. The open symbols in (a) are numerically sampled $P(\delta)$ for 2D CP-DRM, are color-coded by volume fraction, and have negligibly small error bars that cannot be distinguished on the scale of this figure. $\langle\delta\rangle$ for CP-DRM is given by Eq. \eqref{eqn:CP_dn} with $n=1$.} \label{fig:pores2DCOMP}
\end{figure}

\begin{figure}
	    \centering
	    \subfloat[\label{fig:Qa}]{\includegraphics[width=0.78\columnwidth]{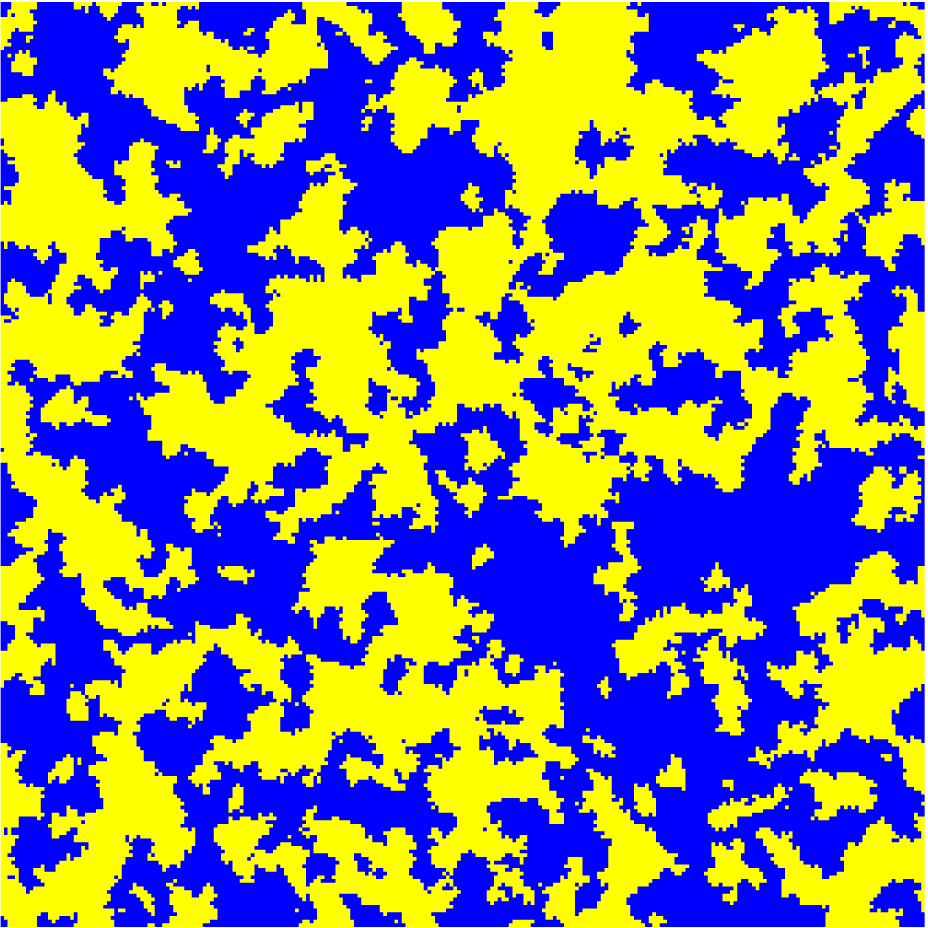}}
	    \hfill
	    \centering
	    \subfloat[\label{fig:Qb}]{\includegraphics[width=0.78\columnwidth]{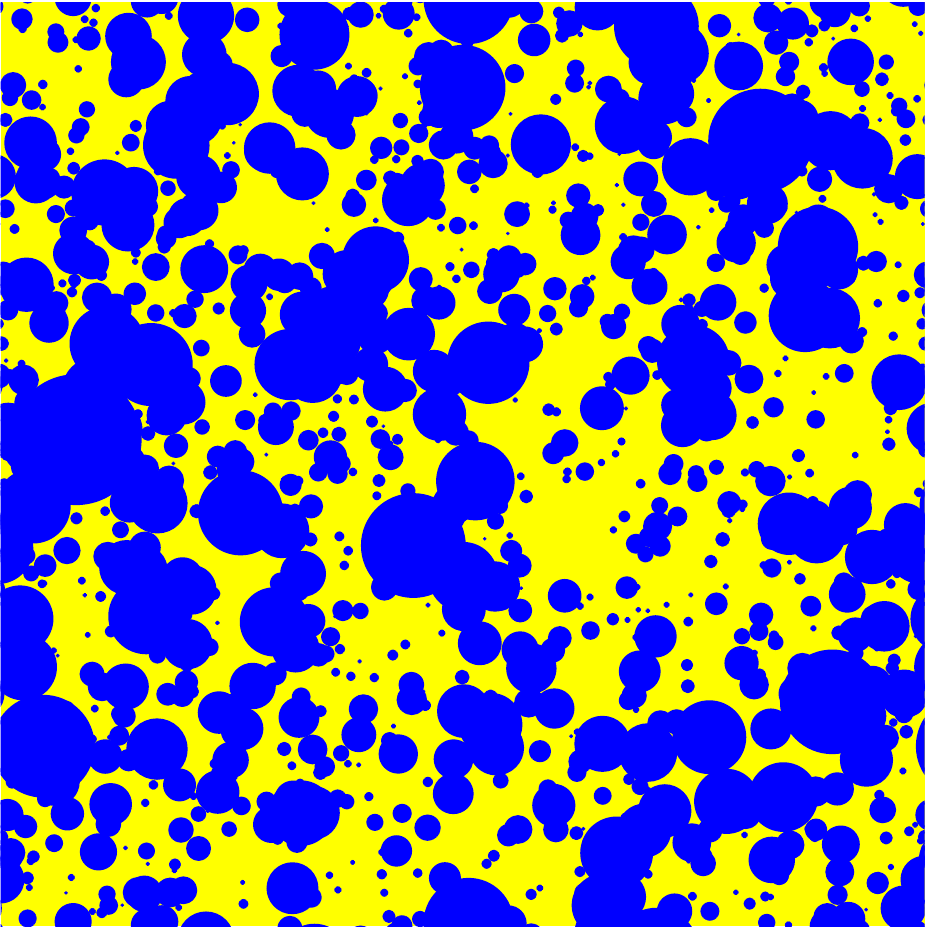}}
	    \hfill
	    \centering
	    \subfloat[\label{fig:Qc}]{\includegraphics[width=0.78\columnwidth]{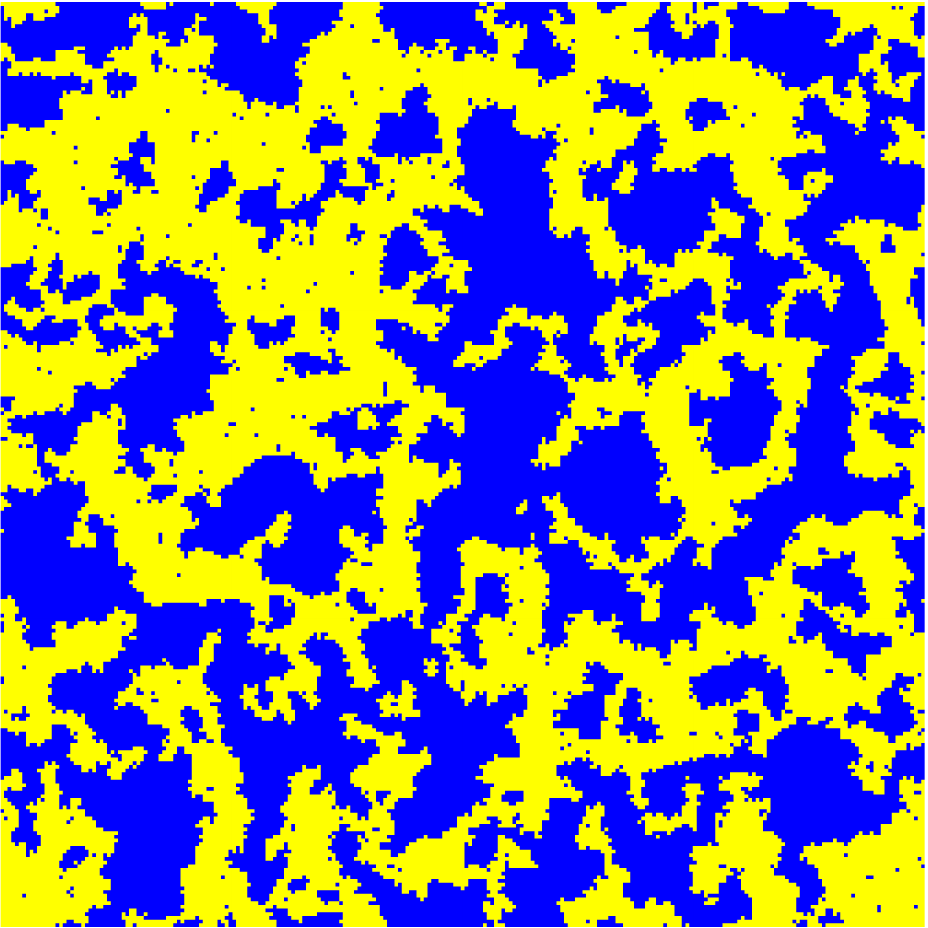}}
	    \caption{(a) YT-DRM with characteristic length $a=5$. (b) OPS-DRM with characteristic length $a=0.2$. (c) CP-DRM with characteristic length $a=5$. For all cases, $\phi_1=\phi_2=1/2$ and the characteristic length is chosen such that it is $1/100$ of the periodic system side length. Configurations (a) and (c) are both $251\times251$ pixels.}\label{fig:Q}
\end{figure}

In Fig. \ref{fig:pores2DCOMP}\subref{fig:pores2DCOMPa}, we show plots of $P(\delta)$ for all three classes of Debye random media in 2D and plots of $P(\delta)$ for 3D CP-DRM and OPS-DRM in Fig. \ref{fig:pores2DCOMP}\subref{fig:pores2DCOMPb}. Note that $P(\delta)$ is scaled by $\phi_1/s$ to bring its value at the origin to unity. Plots of $\langle \delta \rangle$ scaled by $a$ as a function of $\phi_1$ are shown in Fig. \ref{fig:pores2DCOMP}\subref{fig:pores2DCOMPc}. In 2D, these plots reveal that, for a given volume fraction, YT-DRM have the largest pores of the three classes. This difference in behavior can be explained by visual comparison of these three systems in Fig. \ref{fig:Q}. We see that OPS-DRM [Fig. \ref{fig:Q}\subref{fig:Qb}] have numerous islands of small disks embedded in the matrix phase which disrupt the pore space and collectively lower $\langle\delta\rangle$. Such islands are not present in Debye random media constructed with the Yeong-Torquato procedure due to the pixel refinement phase described in Sec. \ref{sec:YT}. The presence of these islands in the overlapping, polydisperse sphere systems is explained by examining the distribution of their radii: $f(R) = e^{-r/\langle R \rangle}/\langle R \rangle$, where smaller radii $R$ are clearly the most probable. 

These islands are also present in CP-DRM [Fig. \ref{fig:Q}\subref{fig:Qa}] where they similarly disrupt the pore space and, notably, have survived the pixel refinement phase of the Yeong-Torquato procedure. The persistence of these islands in CP-DRM indicates that they are critical to enforcing the strict-cutoff $\Lambda$ on the maximum pore radius. In 3D, we see that CP-DRM, on average, have larger pores than do OPS-DRM.

\subsection{\label{sec:comp_linealpath}Lineal-path function}

Here, we compare the lineal-path functions for the void phases of the three classes of Debye random media in 2D. Following Lu and Torquato \cite{Torq02,lu92II}, one will find that $L(z)$ for overlapping polydisperse disks is
\begin{equation}
    L(z) = \phi_1\exp\left( \ln\phi_1 \frac{2\langle R \rangle}{\pi\langle R^2 \rangle} z \right)\label{eqn:LzOPS}.
\end{equation}
Note that, from this expression, we can define the average \textit{lineal} size of these systems as
\begin{equation}
    L_w(\phi_1) = -\frac{\pi \langle R^2 \rangle}{2\ln\phi_1 \langle R \rangle}\label{eqn:avglinsize}.
\end{equation}
Specializing Eq. \eqref{eqn:LzOPS} for exponentially distributed radii, we find the lineal-path function for OPS-DRM to be
\begin{equation}
    L(z) = \phi_1^{1 + z/(\pi\langle R \rangle)}.
\end{equation}

\begin{figure}
    \centering
	\subfloat[\label{fig:lineal2DCOMPa}]{\includegraphics[width=1.0\columnwidth]{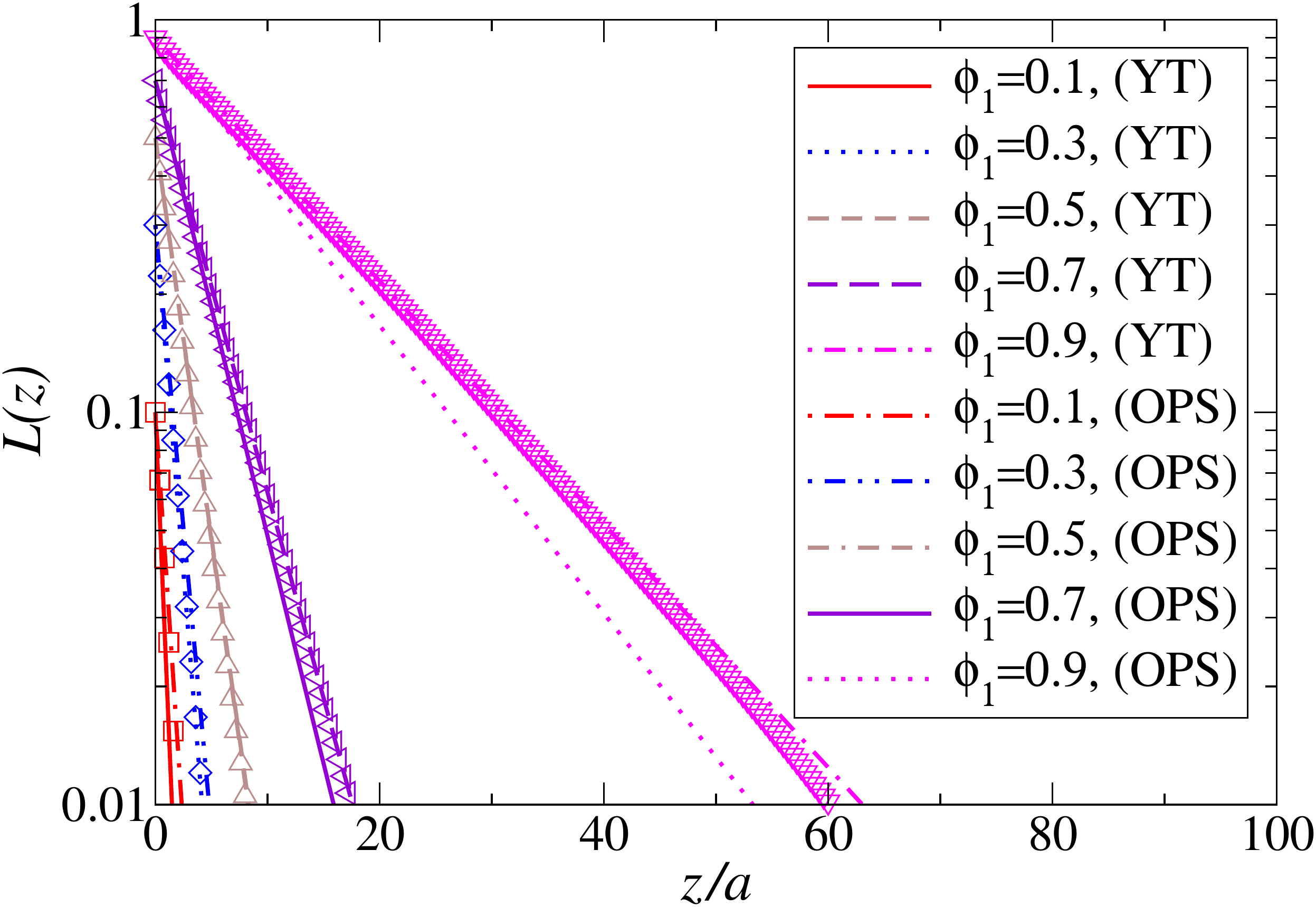}}
	\hfill
	\centering
	\subfloat[\label{fig:lineal2DCOMPb}]{\includegraphics[width=1.0\columnwidth]{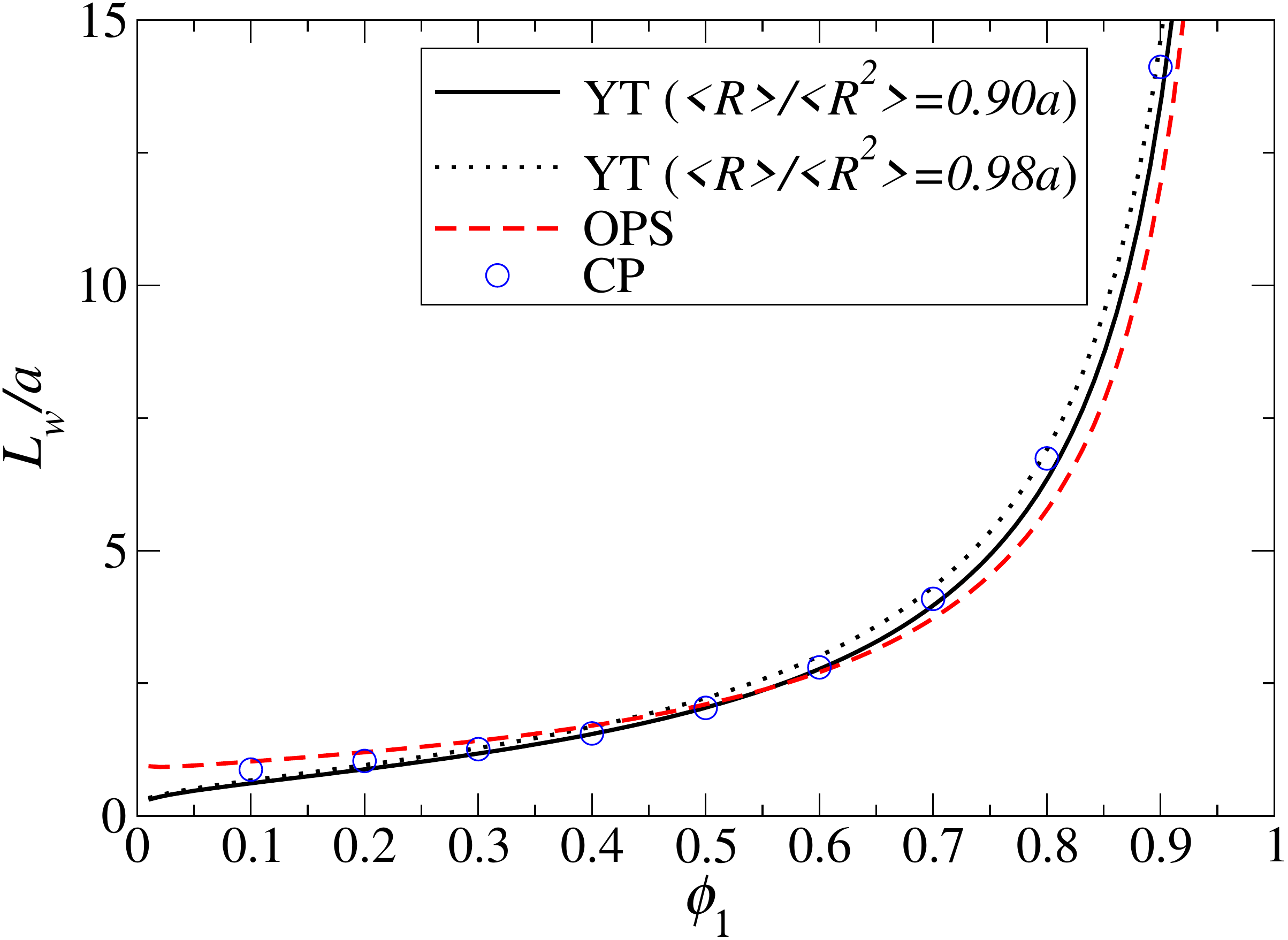}}
	\caption{Plots of the lineal-path function $L(z)$ (a) and the average lineal-size $L_w(\phi_1)$ (b) for the three classes of Debye random media in 2D. $L(z)$ for YT-DRM and OPS-DRM is given by Eq. \eqref{eqn:LzOPS} with their respective average lineal sizes. The open symbols in (a) are for numerically sampled $L(z)$ for CP-DRM and are color-coded by volume fraction. The open symbols in (b) are for numerically sampled $L_w$ for CP-DRM. For both sets of scatter plots, the error bars are too small to be distinguished on the scale of this figure. Note that the non linearity in the sampled $L(z)$ is caused by the cutoff $l_c$ used in the accelerated Yeong-Torquato procedure.}\label{fig:lineal2DCOMP}
\end{figure}

For 2D Debye random media constructed using the Yeong-Torquato procedure, Ma and Torquato found that the lineal-path function also exhibits an exponential decay. As such, they fit their data for $L(z)$ to Eq. \eqref{eqn:LzOPS} and found that the ratio $\langle R \rangle/\langle R^2 \rangle$ fell in the range $(0.94\pm0.04)a$ and was largely insensitive to changes in volume fraction. Here, we found that CP-DRM exhibit roughly the same $L(z)$ that YT-DRM do. This behavior contrasts that of OPS-DRM systems for which the ratio $\langle R \rangle/\langle R^2 \rangle$ is equal to $1/2\langle R \rangle$ and thus depends on the volume fraction [see Eq. \eqref{eqn:eplw}]. The lineal-path functions for the three classes of Debye random media are plotted in Fig. \ref{fig:lineal2DCOMP}\subref{fig:lineal2DCOMPa}, and the average lineal sizes of these structures are plotted in Fig. \ref{fig:lineal2DCOMP}\subref{fig:lineal2DCOMPb}. Interestingly, OPS-DRM have the largest $L_w$ for $\phi_1<1/2$.

\subsection{\label{sec:comp_chordlength}Chord-length probability density function}

\begin{figure}
    \centering
	\subfloat{\label{fig:pz2DCOMPa}}{\includegraphics[width=1.0\columnwidth]{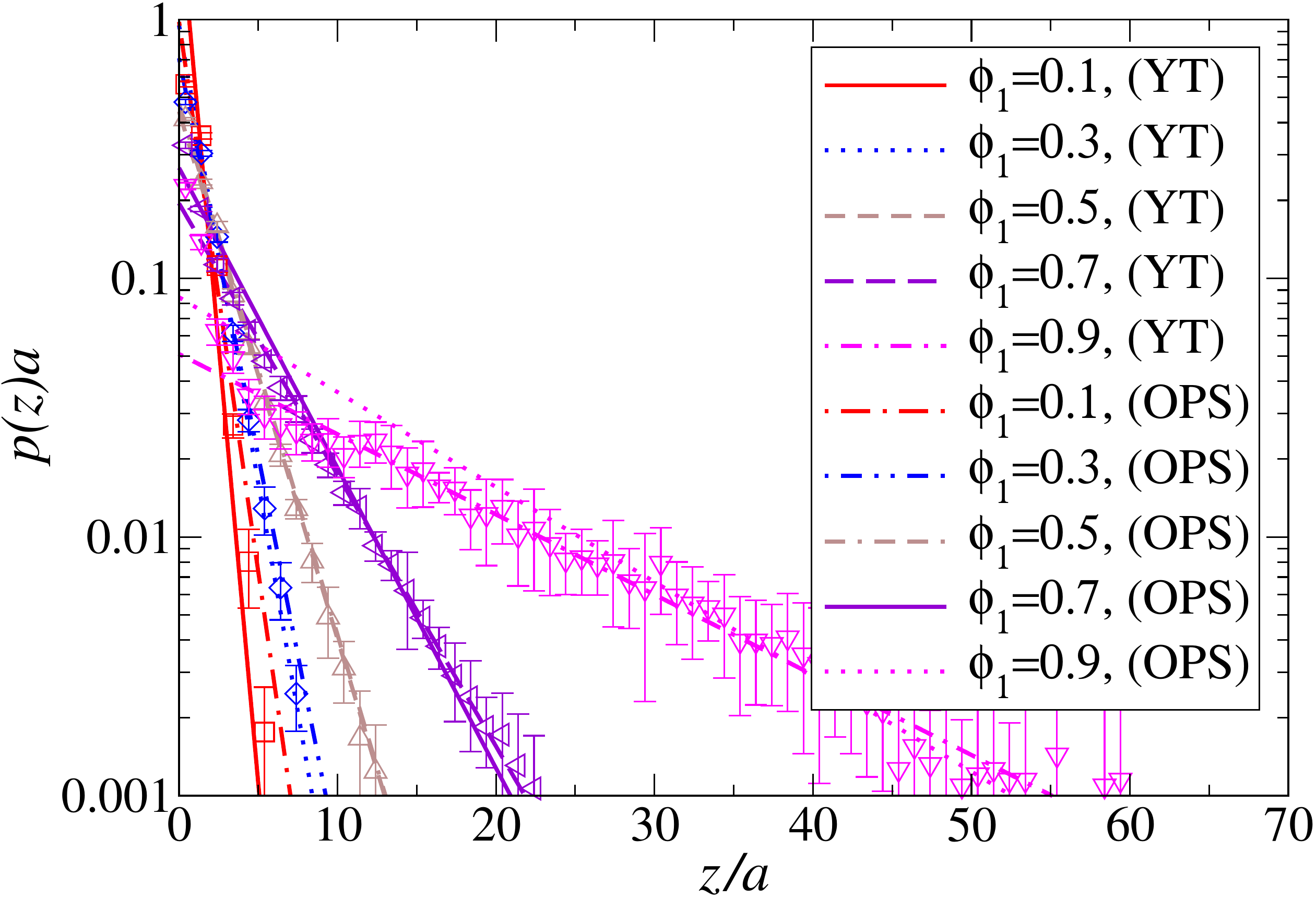}}
	\caption{Plots of the matrix chord-length probability density function $p(z)$ for the three classes of Debye random media in 2D. For YT-DRM and OPS-DRM, $p(z)$ is given by Eq. \eqref{eqn:pzOPS}. The open symbols are for numerically sampled $p(z)$ for CP-DRM and are color-coded by volume fraction.}\label{fig:pz2DCOMP}
\end{figure}

Using Eq. \eqref{eqn:pzdef}, it is trivial to obtain the matrix chord-length probability density function $p(z)$ from the lineal-path function \eqref{eqn:LzOPS}. Given that all three classes of Debye random media considered in this paper exhibit the same exponentially-decaying form for $L(z)$ [e.g., Eq. \eqref{eqn:LzOPS}], we find that
\begin{equation}
    p(z) = \frac{2\eta\langle R \rangle}{\pi\langle R^2 \rangle}\phi_1^{2\langle R \rangle z/ (\pi\langle R^2 \rangle)}\label{eqn:pzOPS}
\end{equation}
via relation \eqref{eqn:pzdef}. The matrix chord-length probability density functions for the three classes of Debye random are plotted in Fig. \ref{fig:pz2DCOMP}. Given that the three classes of degenerate Debye random media have similar $L(z)$, it is not surprising that they share similar $p(z)$ as well.

\section{\label{sec:percolation}Comparison of Percolation Thresholds}

In their study on Debye random media realized with the Yeong-Torquato procedure, Ma and Torquato \cite{drm2020} conjectured that the percolation threshold of the inclusion phase $\phi_2^c$ in $d=2$ is $1/2$. This prediction was based on the phase-inversion symmetry that is manifest in Eq. \eqref{eqn:debS2First} as well as visual inspection of their relatively large reconstructed samples. Additionally, using relation \eqref{eqn:euler}, Ma and Torquato found the specific Euler characteristic for this class of Debye random media to be
\begin{equation}
    \chi = \frac{\pi(\phi_1-\phi_2)\phi_1\phi_2}{4a^2}\label{eqn:eulerYT}.
\end{equation}
Prior work suggests that the zeros of the Euler characteristic can be used to estimate the percolation threshold of a two-phase system \cite{Mecke1991,prim20,neher08,Klatt2017}. We see from Eq. \eqref{eqn:eulerYT} that $\chi$ will vanish for $\phi_1=\phi_2=1/2$. We numerically estimated the percolation threshold of 2D YT-DRM to be $\phi_1^c\approx1/2$ by using a ``burning algorithm" \cite{stauffer} to detect percolating clusters in ten $501\times501$ pixel samples of YT-DRM at various volume fractions.

While Debye random media approximated by overlapping, polydisperse spheres has effective phase inversion symmetry [see Figs. \ref{fig:S2plots}\subref{fig:S2plotsb}, \ref{fig:S2plots}\subref{fig:S2plotsd}, and \ref{fig:S2plots}\subref{fig:S2plotsf}], we expect that the percolation threshold for the matrix phase will be lower than $1/2$. This expectation is motivated by analysis of the Euler characteristic of these systems. Using relation \eqref{eqn:euler} and Eq. \eqref{eqn:opsdrmfsv2D} we find that
\begin{equation}
    \chi = \frac{-\ln\phi_1}{2\pi\langle R \rangle^2} \left( \phi_1 + \phi_1\ln\phi_1 \right),
\end{equation} 
which has a nontrivial zero for $\phi_1=1/e\approx0.368$. Using the ``rescaled particle method", a Monte Carlo simulation method developed by Torquato and Jiao \cite{rescaleI12,rescaleII12}, we numerically estimated the percolation threshold as $\phi_1^c\approx0.303$. This value of $\phi_1^c$ is lower than the zero of $\chi$ and is closer to the percolation threshold found for overlapping disks with uniformly distributed radii which is $\phi_1^c\approx0.314$ \cite{perco01}. Our finding is also consistent with the observation of Klatt \textit{et al.} that the zero of $\chi$ was always an upper bound on the percolation threshold of overlapping squares \cite{Klatt2017}.

Given that the pore statistics of CP-DRM are distinct from those of YT-DRM, we expect that the void phase percolation threshold for this class of Debye random media will not be equal to $1/2$. Notably, using a procedure adapted from Ref. \cite{Klatt2020}, we numerically determined that the Euler characteristic for CP-DRM is negative for $\phi_1\in[0.05,0.9]$, strongly suggesting that it is only trivially equal to zero for $\phi_1=0,1$. Once again using the ``burning algorithm", we numerically estimated the percolation threshold for CP-DRM to be $\phi_1^c\approx0.39$. Interestingly, the Euler characteristic is only an accurate predictor of the percolation threshold of YT-DRM. It is likely that the additional constraints placed on the microstructures of OPS-DRM and CP-DRM alter the ability of the Euler characteristic to accurately predict the percolation thresholds of these systems.

\section{\label{sec:properties}Comparison of Effective Diffusion and Transport Properties of YT-DRM, OPS-DRM, and CP-DRM}

In this section, we treat YT-DRM, OPS-DRM, and CP-DRM as porous media (with phase 2 being solid and phase 1 being void space) and compare their diffusion and fluid permeability properties in 2D and 3D.

\subsection{\label{sec:diffusion}Bounds on Mean Survival and Principal Diffusion Relaxation Times}

Consider a porous medium in which a species diffuses throughout the pore space with diffusion coefficient $\mathpzc{D}$ and can react at the pore-solid interface via a surface with reaction rate $\kappa$. The diffusion-controlled limit is obtained when $\kappa\to\infty$, while taking $\kappa\to0$ corresponds to a perfectly reflective interface. A quantity of central interest in such diffusion and reaction problems is the mean survival time $\tau$, which is the average lifetime of the diffusing species before it gets trapped. Another important quantity, which is also pertinent to the description of viscous flow in porous media \cite{avellaneda}, is the principal relaxation time $T_1$ associated with the time-dependent decay of the initially uniform concentration field of the diffusing particles \cite{Torq02}.

\begin{figure}
    \centering
	\subfloat[\label{fig:tauT1UBa}]{\includegraphics[width=1.0\columnwidth]{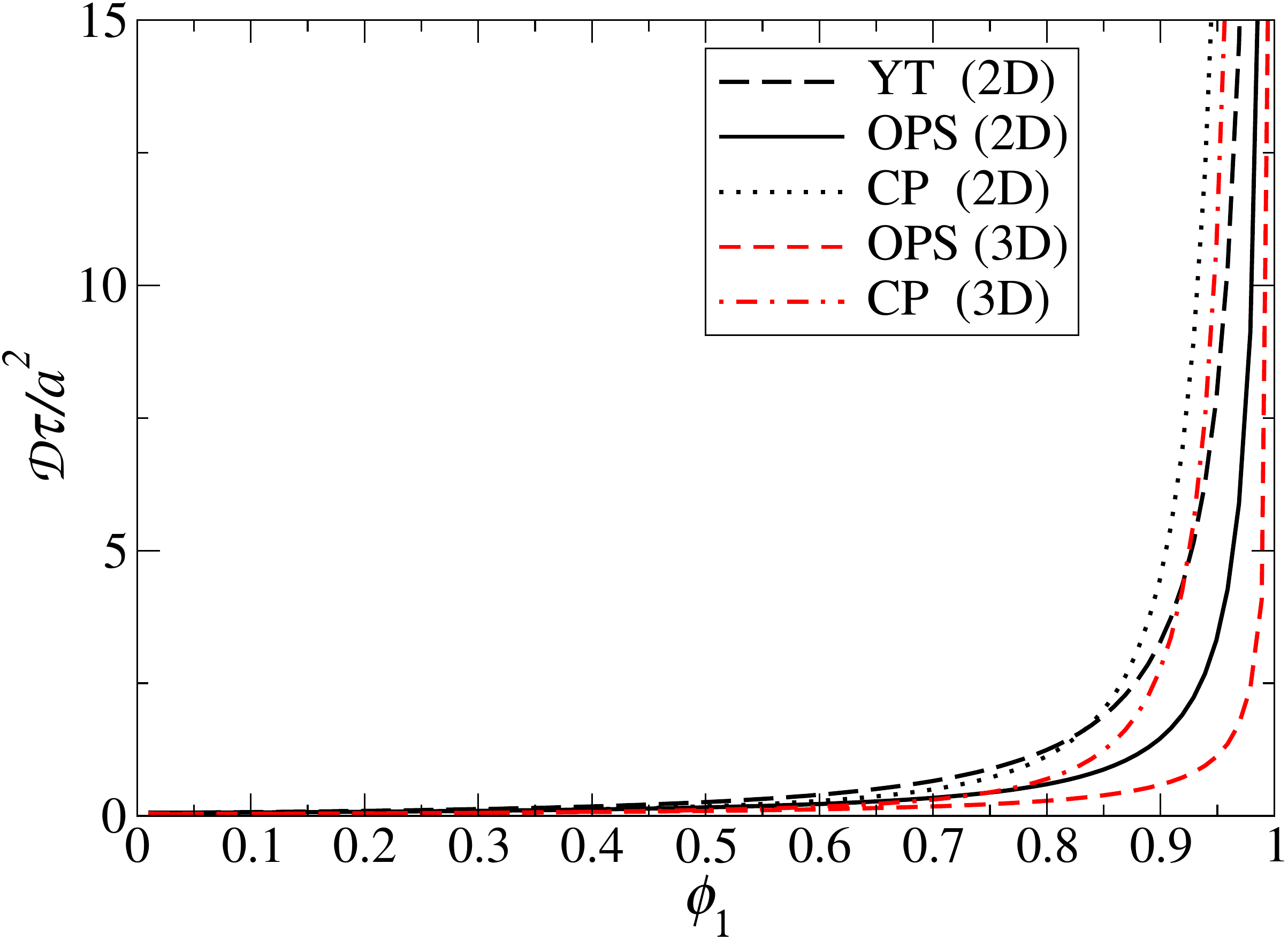}}
	\hfill
	\centering
	\subfloat[\label{fig:tauT1UBb}]{\includegraphics[width=1.0\columnwidth]{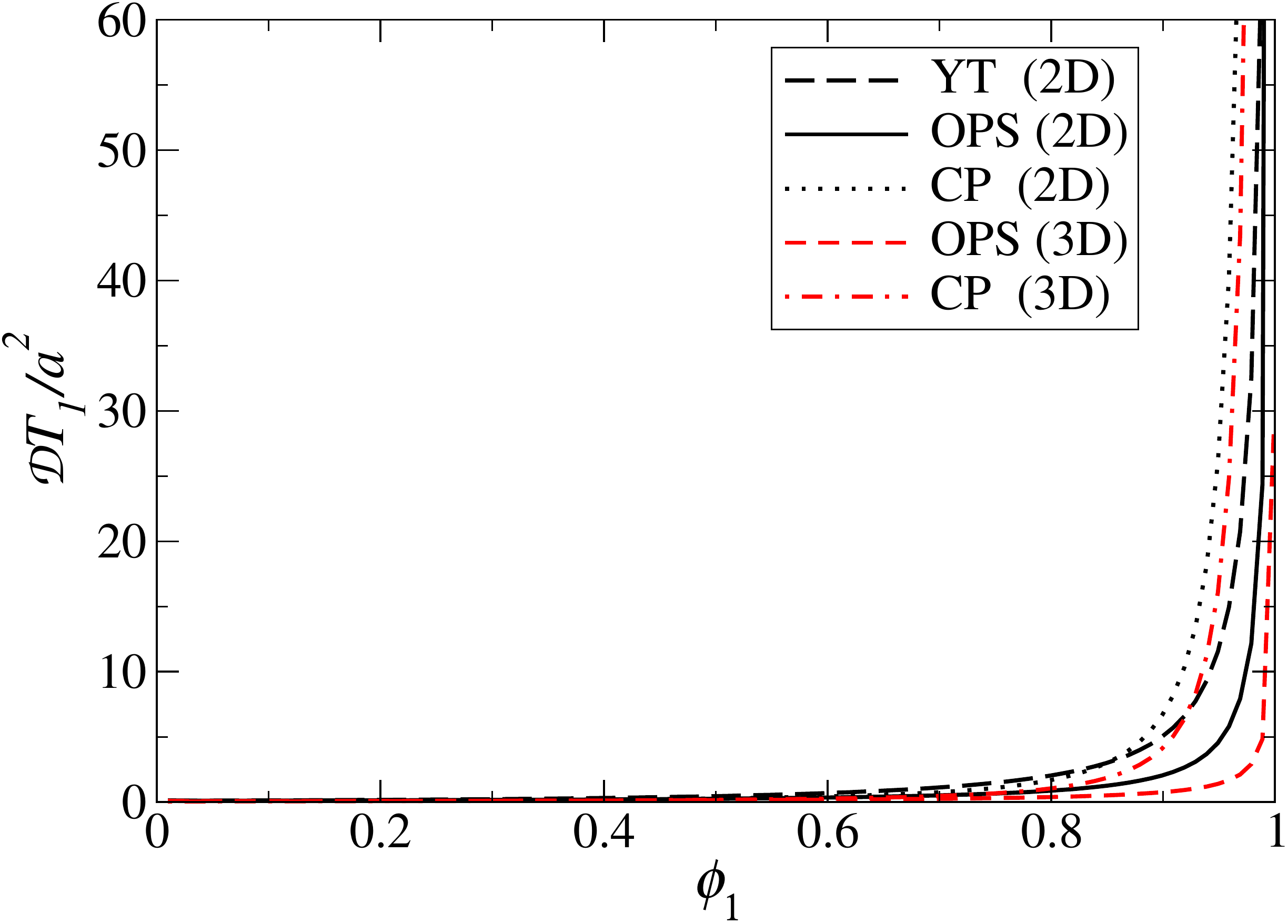}}
	\caption{Plots of upper bounds on the scaled mean survival time $\mathpzc{D}\tau/a^2$ (a) and scaled principal diffusion relaxation time $\mathpzc{D}T_1/a^2$ (b) as functions of $\phi_1$ for the three classes of Debye random media in 2D and 3D in the diffusion-controlled regime (e.g., $\kappa\to\infty$). $\tau$ and $T_1$ are obtained from inequalities \eqref{eqn:tauUB} and \eqref{eqn:T1UB}, respectively, and the length scale $a$ is defined in Eq. \eqref{eqn:debS2First}.}\label{fig:tauT1UB}
\end{figure}

Using the pore-size function $P(\delta)$ and variational principles, Torquato and Avellaneda \cite{avellaneda} derived the following upper bound on $\tau$:
\begin{equation}
    \tau \leq \frac{\langle\delta\rangle^2}{\mathpzc{D}} + \frac{\phi_1}{\kappa s}.\label{eqn:tauUB}
\end{equation}
They also computed the following upper bound on $T_1$ using a similar approach:
\begin{equation}
    T_1 \leq \frac{\langle\delta^2\rangle}{\mathpzc{D}} + \frac{3\phi_1\langle\delta\rangle^2}{4\kappa s \langle\delta^2\rangle}\label{eqn:T1UB}.
\end{equation}
Upper bounds on the mean survival time are plotted in Fig. \ref{fig:tauT1UB}\subref{fig:tauT1UBa} and those on the principal diffusion relaxation time in Fig. \ref{fig:tauT1UB}\subref{fig:tauT1UBb} for perfectly absorbing traps (i.e., $\kappa\to\infty$). In both 2D and 3D, we see that OPS-DRM has the lowest upper bounds on $\tau$ and $T_1$ which is consistent with our prior observation that this class of Debye random media has smaller pores on average than do YT-DRM and CP-DRM (see Sec. \ref{sec:comp_porefuncs}). Interestingly, 2D CP-DRM have slightly higher bounds for $\tau$ and $T_1$ than YT-DRM do for $\phi_1\approx0.83$.

\subsection{\label{sec:fluidpermeability}Bounds on Fluid Permeability}

Here, we present upper bounds on the fluid permeability $k$, which is defined in Darcy's law which describes slow, viscous flow through a porous medium \cite{perm92}, for YT-DRM and OPS-DRM. We also estimate $k$ for CP-DRM and OPS-DRM using an approximation that was recently suggested by Torquato \cite{awr20}. Using variational principles, Doi \cite{doi76}, and subsequently Rubinstein and Torquato \cite{rube89}, derived the following upper bound on the fluid permeability $k$ of statistically isotropic porous media:
\begin{equation}
    k \le k^{(2)}_U = \frac{2}{3}\int_0^{\infty} r \left[F_{vv}(r) - \frac{2\phi_1}{s}F_{sv}(r) + \frac{\phi_1^2}{s^2}F_{ss}(r) \right]dr\label{eqn:kbound}.
\end{equation}
Here, $\phi_1$ is the porosity and the void-void correlation function $F_{vv}(r)$ is the same as the two-point correlation function for phase 1, e.g., $S_2^{(1)}(r)$. Following Rubinstein and Torquato \cite{rube89}, we refer to Eq. \eqref{eqn:kbound} as a two-point ``interfacial-surface" upper bound.

Values of $k^{(2)}_U$ as a function of porosity for the two different classes of Debye random media are computed using their respective two-point and surface correlation functions (see Secs. \ref{sec:s2pds} and \ref{sec:comp_svfuncs}). Note that $k^{(2)}_U$ for overlapping, polydisperse spheres with various distributions of radii were computed in Ref. \cite{perm92}. For Debye random media realized via the Yeong-Torquato method, one finds that the two-point interfacial-surface upper bound on permeability to be
\begin{eqnarray}
    k^{(2)}_U = &&\frac{a^2}{576\phi_2^2} \Big[ 16\phi_1\phi_2\left(3 + 4\phi_2[6\phi_2 + \pi^2(\phi_1-\phi_2)]\right) \nonumber \\
    &&+ \pi^2|\phi_1-\phi_2|\Big]\label{eqn:k2YT}.
\end{eqnarray}
For OPS-DRM, the integral in Eq. \eqref{eqn:kbound} must be computed numerically. 

Torquato derived the following approximation for the fluid permeability \cite{awr20}: 
\begin{equation}
    k\approx\frac{\langle\delta^2\rangle}{\mathpzc{F}}\label{eqn:kest},
\end{equation}
which describes porous media with well-connected pore spaces. Note that $\mathpzc{F}$ is the formation factor, which is a measure of the tortuosity or ``windiness" of the entire pore space and is a monotonically decreasing function of the porosity \cite{awr20}. Notably, Eq. \eqref{eqn:kest} was recently confirmed by Klatt \textit{et al.} \cite{Klatt2021} to be highly accurate for models of porous media derived from overlapping spheres as well as various packings of spheres.

\begin{figure}
    \centering
	\subfloat[\label{fig:k2COMPa}]{\includegraphics[width=1.0\columnwidth]{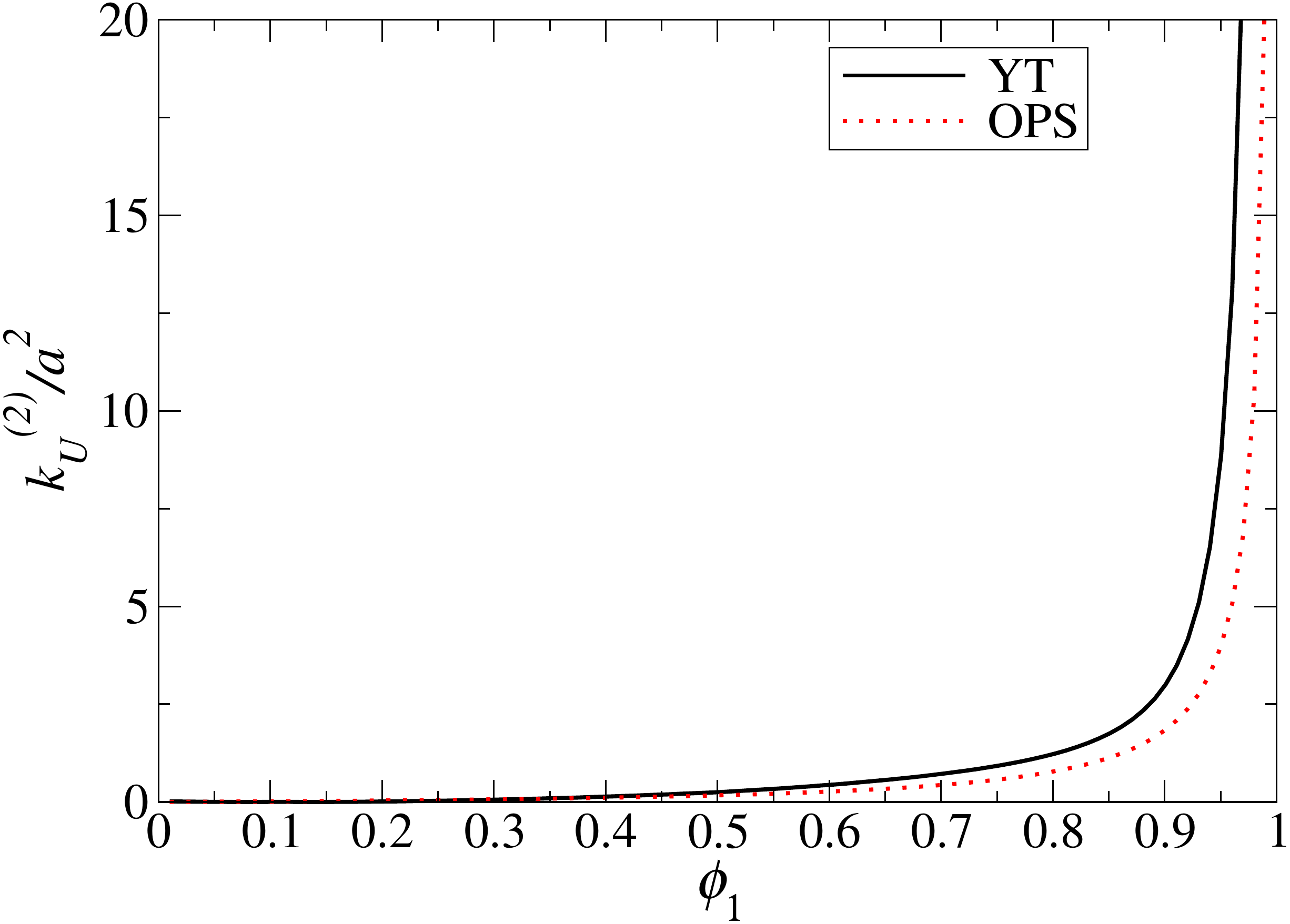}}
	\hfill
	\centering
	\subfloat[\label{fig:k2COMPb}]{\includegraphics[width=1.0\columnwidth]{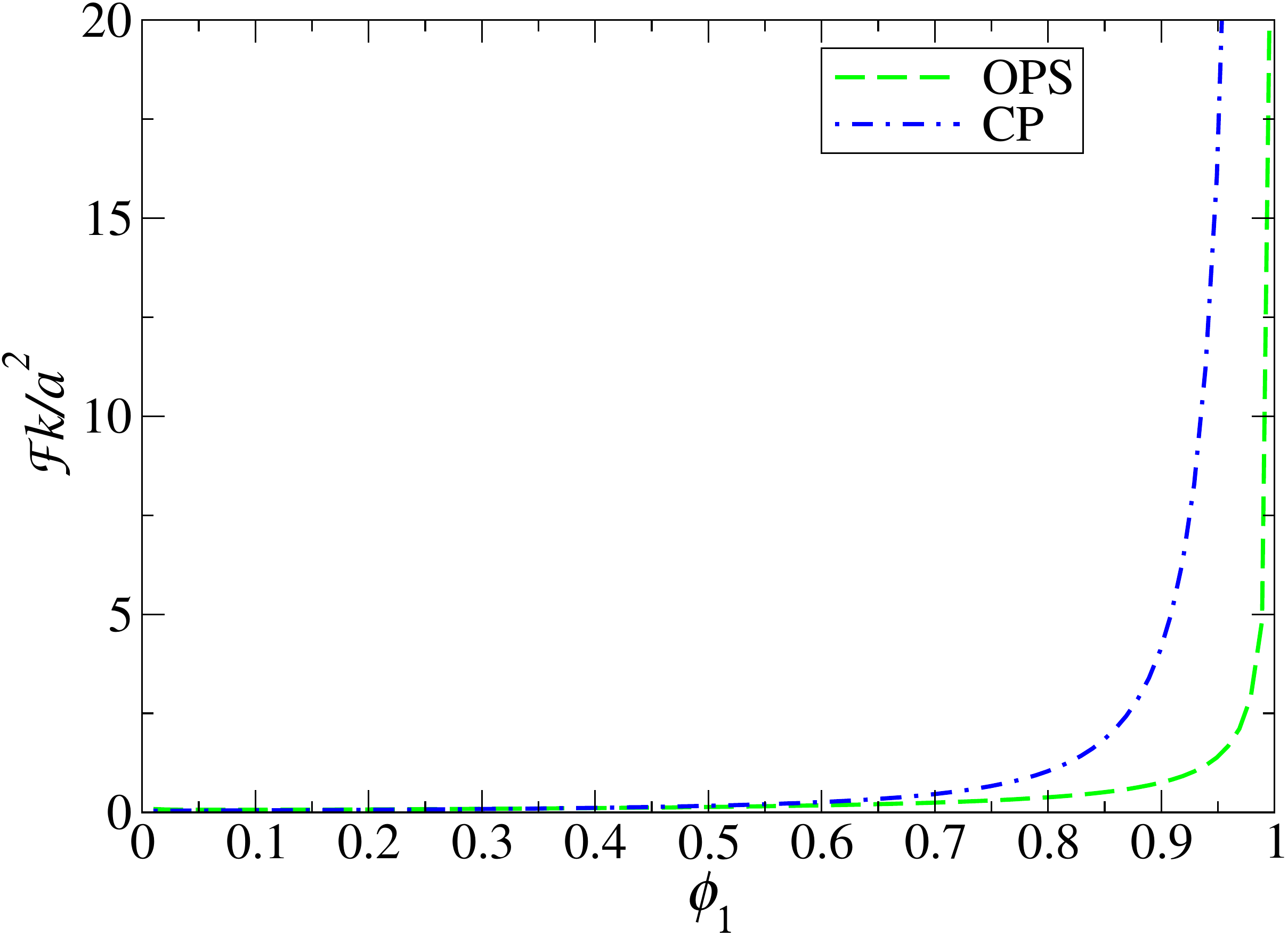}}
	\caption{(a) Plots of the two-point interfacial-surface upper bound on the scaled fluid permeability $k^{(2)}_U/a^2$ for OPS-DRM and YT-DRM, where $a$ is the length scale defined in Eq. \eqref{eqn:debS2First}. For YT-DRM, $k^{(2)}_U$ is given by Eq. \eqref{eqn:k2YT}. For OPS-DRM, the integral $\eqref{eqn:kbound}$ is computed numerically for each value of $\phi_1$. (b) Plots of the scaled fluid permeability $\mathpzc{F}k/a^2$ from the approximation \eqref{eqn:kest} for OPS-DRM and CP-DRM. Here, $\langle\delta^2\rangle$ is computed numerically from pore-size function \eqref{eqn:PdOPS3D} for OPS-DRM, and via Eq. \eqref{eqn:CP_dn} for CP-DRM.}\label{fig:k2COMP}
\end{figure}

Results for $k^{(2)}_U$ are plotted in Fig. \ref{fig:k2COMP}\subref{fig:k2COMPa}. We see that the upper bound on $k$ for YT-DRM and OPS-DRM are similar for low porosity, but that the bound for YT-DRM is larger than that of OPS-DRM for $\phi_1>0.4$ which is consistent with our observation that, on average, the pores of YT-DRM are larger than those of OPS-DRM in 2D [see Fig. \ref{fig:pores2DCOMP}\subref{fig:pores2DCOMPb}]. Additionally, our results agree with Torquato's observation that 3D Debye random media constructed with the Yeong-Torquato procedure have ``substantially large pore regions" \cite{awr20}. In the absence of estimates of the formation factor $\mathpzc{F}$ for our models, predictions of approximation \eqref{eqn:kest} of the product $\mathpzc{F}k$ are plotted in Fig. \ref{fig:k2COMP}\subref{fig:k2COMPb}. Similarly, we see that the the fluid permeabilities for OPS-DRM and CP-DRM are similar for $\phi_1<0.4$, while the latter becomes increasingly more permeable than the former as the porosity is increased. This result is consistent with our result where, in 3D, CP-DRM have larger pores than OPS-DRM do for $\phi_1>0.4$ [see Fig. \ref{fig:pores2DCOMP}\subref{fig:pores2DCOMPb}].

\section{\label{sec:conclusions}Conclusions and Discussions}

In this work, we have investigated and compared three classes of Debye random media to one another using a variety of descriptors in order to characterize how the microstructures of $S_2$-degenerate systems can vary. We specifically considered the ``most probable" class of Debye random media realized with the Yeong-Torquato procedure, as well as two other distinct classes of structures that we introduced in this work: Debye random media realized by certain systems of overlapping, polydisperse spheres with exponentially distributed radii, and Debye random media whose pore-size probability density function has compact support. To structurally discriminate these systems, we compared their surface correlation, pore-size, lineal-path, and chord-length distribution functions. In general, we found that these three classes of Debye random media are largely distinguished by these microstructural descriptors with the differences in their pore-size statistics and percolation thresholds being the most profound. Our results further support the well-known fact that the two-point correlation function is largely insufficient to determine the effective physical properties of two-phase random media.

Our analysis of the statistical descriptors of these degenerate Debye random media also revealed that OPS-DRM are only phase-inversion symmetric with respect to $S_2(r)$, while CP-DRM are only phase-inversion symmetric with respect to $S_2(r)$ and $F_{ss}(r)$. For OPS-DRM, this lack of symmetry is to be expected as particle models of two-phase media are generally not phase-inversion symmetric \cite{Torq02}. Conversely, by the nature of their construction, YT-DRM are likely truly phase-inversion symmetric, satisfying condition \eqref{eqn:phaseinv}. Furthermore, the additional constraint on the pore-size statistics in CP-DRM destroys such higher-order phase-inversion symmetry. We also determined that the percolation thresholds of these three classes of Debye random media are quite different which indicates that disordered, $S_2$-degenerate two-phase random media can exhibit a variety of topologies. Interestingly, we found that the Euler characteristic did not accurately predict the percolation thresholds of CP-DRM and OPS-DRM for reasons indicated in Sec. \ref{sec:percolation}.

Lastly, we found that the bounds on the effective mean survival times, principal diffusion relaxation times, and fluid permeabilities as well as the approximated fluid permeabilities of these degenerate Debye random media are distinct to varying degrees; with OPS-DRM having the lowest bounds in 2D and 3D for all three physical properties, as seen in Figs. \ref{fig:tauT1UB} and \ref{fig:k2COMP}. Moreover, these differences are largely due to the distinct pore spaces of YT-DRM, OPS-DRM, and CP-DRM. While we were able to compare statistical descriptors, percolation properties, and physical properties of the three classes of Debye random media in 2D, our analysis of 3D YT-DRM and CP-DRM was limited by the high computational cost of generating sufficiently large (e.g., $501^3$ voxel) samples of these structures with the Yeong-Torquato procedure. Hence, an outstanding problem for future research is to further accelerate the Yeong-Torquato procedure to efficiently (re)construct large samples of two-phase media with targeted statistical descriptors in 3D.

The large computational cost of generating Debye random media with the Yeong-Torquato procedure underscores an advantage of being able to effectively realize Debye random media with overlapping, polydisperse spheres, since the cost to generate samples of OPS-DRM does not scale appreciably with system size or dimension. For example, we note that the sample of YT-DRM in Fig. \ref{fig:Q}\subref{fig:Qa} took about 15 minutes to generate whereas over 2 million samples of OPS-DRM, such as the one in Fig. \ref{fig:Q}\subref{fig:Qb}, can be constructed in that time. Moreover, recall that any microstructural descriptor for OPS systems can be determined analytically via the canonical correlation function formalism \cite{torquato86}. Given these computational advantages of overlapping, polydisperse sphere models of random media, future work could consider using such systems with different distributions of radii $f(R)$ to realize microstructures with prescribed statistical descriptors.

An intriguing extension of the present work is to apply similar methodologies to study the degeneracies of disordered hyperuniform two-phase media, which are defined by a spectral density $\tilde{\chi}_{_V}(\mathbf{k})$ that tends to zero as the wave number $\mathbf{k}$ goes to zero \cite{zachary09}. As a result, hyperuniform media are characterized by an anomalous suppression of large-scale volume-fraction fluctuations relative to typical disordered two-phase media. For this purpose, one can employ the procedure of Chen and Torquato, which is a Fourier space analog of the Yeong-Torquato procedure to realize disordered two-phase media with general functional forms corresponding to hyperuniform spectral densities \cite{acta18}. Notably, it has been shown that disordered hyperuniform media are endowed with a variety of novel physical properties \cite{florescu09,derosa15,leseur16,skoro16,zhang16,florescu17,scheffold17,acta18,Klatt2018,hao19,gorsky19,sheremet20,kim20,yu2021}. Therefore, characterizing how microstructures with a prescribed hyperuniform $\tilde{\chi}_{_V}(\mathbf{k})$ are degenerate can aid in the design of multifunctional composite materials \cite{kim20,torquato18a,torquato18b,donev04,donev02} with sets of targeted physical properties.

\begin{acknowledgments}
The authors thank Michael Klatt for helpful discussions and his code for computing the Euler characteristic. They also gratefully acknowledge the support of Air Force Office of Scientific Research Program on Mechanics of Multifunctional Materials and Microsystems under Grant No. FA9550-18-1-0514.
\end{acknowledgments}



\begin{thebibliography}{82}
\bibitem [{C2note()}]{C2note}%
  \BibitemOpen
  \bibinfo {note} {{$C_2^{(i)}(\mathbf{x}_1,\mathbf{x}_2)$ is defined to be the
  probability of finding two points $\mathbf{x}_1$ and $\mathbf{x}_2$ in the
  same cluster of phase $i$ \cite{Torq02,Beasley88}.}}\BibitemShut {Stop}%
\bibitem [{Fssnote()}]{Fssnote}%
  \BibitemOpen
  \bibinfo {note} {Note that $F_{ss}(r)$ is intrinsically much more difficult
  to sample than are $S_2(r)$ and $F_{sv}(r)$.}\BibitemShut {Stop}%
\end{thebibliography}
%

\providecommand{\noopsort}[1]{}\providecommand{\singleletter}[1]{#1}%

\end{document}